\documentclass[final]{biometrika}

\usepackage{amsbsy,amsfonts,amsmath,xspace}
\usepackage{mathrsfs}
\usepackage{graphicx}
\usepackage{caption}
\usepackage{setspace}
\usepackage{booktabs}
\usepackage{longtable}
\usepackage{tikz}
\makeatletter
\providecommand{\@LN}[2]{}
\makeatother
\usepackage{times}
\usepackage{bm}
\usepackage{natbib}
\usepackage{subcaption}
\usepackage{multirow}
\usepackage{url}
\usepackage[toc,page]{appendix}
\usepackage{float}
\usepackage[plain,noend]{algorithm2e}
\usepackage{color}
\usepackage{verbatim}
\usepackage{xr}
\newcommand{\V}[1]{\ensuremath{\boldsymbol{#1}}\xspace}

\newcommand{\F}[1]{\ensuremath{\mathrm{#1}}\xspace}

\newcommand{\bx}{\mathbf{x}}
\newcommand{\e}{\mathbb{E}}
\newcommand{\p}{\mathbb{P}}

\newcommand{\I}{\mathcal{I}}
\newcommand{\II}{\mathcal{II}}
\newcommand{\ccal}{\mathcal{C}}

\newcommand{\tcal}{\mathcal{T}}

\newcommand{\vcal}{\mathcal{V}}

\newcommand{\bR}{\mathbb{R}}

\newcommand{\argmin}{\mathrm{argmin}}
\newcommand{\tr}{\mathrm{tr}}
\newcommand{\mbone}{{\mathbf{1}}}
\newcommand{\Vnode}{\mathcal{V}}

\newcommand{\diag}{\F{diag}}
\newcommand{\ind}{{\mathbf{I}}}

\newcommand{\norm}[1]{\Vert{#1}\Vert}

\newcommand{\rank}{\mathrm{rank}}

\theoremstyle{definition}
\newtheorem{ass}{Assumption}

\makeatletter
\renewcommand{\algocf@captiontext}[2]{#1\algocf@typo. \AlCapFnt{}#2} 
\def\@algocf@capt@plain{top}
\renewcommand{\algocf@makecaption}[2]{%
  \addtolength{\hsize}{\algomargin}%
  \sbox\@tempboxa{\algocf@captiontext{#1}{#2}}%
  \ifdim\wd\@tempboxa >\hsize
    \hskip .5\algomargin%
    \parbox[t]{\hsize}{\algocf@captiontext{#1}{#2}}
  \else%
    \global\@minipagefalse%
    \hbox to\hsize{\box\@tempboxa}
  \fi%
  \addtolength{\hsize}{-\algomargin}%
}
\makeatother

\begin{document}

\jname{Biometrika}
\jyear{XXX}
\jvol{XXX}
\jnum{XX}


\markboth{T. Li, E. Levina \and Ji. Zhu}{Network cross-validation}

\title{Network cross-validation by edge sampling}

\author{Tianxi Li}
\affil{Department of Statistics, University of Virginia\\ Charlottesville, Virginia 22904, U.S.A. \email{tianxili@virginia.edu}}

\author{Elizaveta Levina \and Ji Zhu}
\affil{Department of Statistics, University of Michigan\\ Ann Arbor, Michigan 48105, U.S.A. \\\email{elevina@umich.edu} \email{jizhu@umich.edu}}

\maketitle

\begin{abstract}
While many statistical models and methods are now available for network analysis, resampling network data remains a challenging problem.   Cross-validation is a useful general tool for model selection and parameter tuning, but is not directly applicable to networks since splitting network nodes into groups requires deleting edges and destroys some of the network structure. Here we propose a new network resampling strategy based on splitting node pairs rather than nodes applicable to cross-validation for a wide range of network model selection tasks. We provide a theoretical justification for our method in a general setting and examples of how our method can be used in specific network model selection and parameter tuning tasks.   Numerical results on simulated networks and on a citation network of statisticians show that this cross-validation approach works well for model selection.  
\end{abstract}


\section{Introduction}

Statistical methods for analyzing networks have received a lot of attention because of their wide-ranging applications in areas such as sociology, physics, biology and medical sciences.  Statistical network models provide a principled approach to extracting salient information about the network structure while filtering out the noise.   Perhaps the simplest statistical network model is the famous Erd\"{o}s-Renyi model \citep{erds1960evolution}, which served as a building block for a large body of more complex models,  including the stochastic block model  \citep{holland1983stochastic}, the degree-corrected stochastic block model \citep{karrer2011stochastic}, the mixed membership block model \citep{airoldi2008mixed}, and the latent space model \citep{hoff2002latent}, 
 to name a few.    
 
 While there has been plenty of work on models for networks and algorithms for fitting them, inference for these models is commonly lacking, making it hard to take advantage of the full power of statistical modeling.   Data splitting methods provide a general, simple, and relatively model-free  inference framework and are commonly used in modern statistics, with cross-validation (CV) being the tool of choice for many model selection and parameter tuning tasks.   For networks, both tasks are important  -- while there are  plenty of models to choose from,  it is a lot less clear how to select the best model for the data, and how to choose tuning parameters for the selected model, which is often necessary in order to fit it. In classical settings where the data points are assumed to be an i.i.d.\ sample, cross-validation  works by splitting the data into multiple parts (folds), holding out one fold at a time as a test set, fitting the model on the remaining folds and computing its error on the held-out fold, and finally averaging the errors across all folds to obtain the cross-validation error.   The model or the tuning parameter is then chosen to minimize this error.   To explain the challenge of applying this idea to networks,  we first introduce a probabilistic framework.

Let $\Vnode = \{1,2, \cdots, n\} =: [n]$ denote the node set of a network, and let $A$ be its $n\times n$ adjacency matrix, where $A_{ij} = 1$ if there is an edge from node $i$ to node $j$ and $0$ otherwise.  We view the elements of $A$ as realizations of independent Bernoulli variables, with $\e A = M$, where $M$ is a matrix of probabilities.  For undirected networks, $A_{ji} = A_{ij}$, thus both $A$ and $M$ are symmetric matrices.  We further assume the unique edges $A_{ij}$, $i < j$  are independent Bernoulli variables. The general network analysis task is to estimate $M$ from the data $A$, under  various structural assumptions we might make to address the difficulty of having a single realization of $A$.

To perform cross-validation on networks, one has to decide how to split the data contained in $A$, and how to treat the resulting partial data which is no longer a complete network.     To the best of our knowledge, there is little work available on the topic.   Cross-validation was used by \cite{hoff2008modeling} under a particular latent space model, and \cite{chen2014network} proposed a novel cross-validation strategy for model selection under the stochastic block model and its variants.   In this paper, we do not assume a specific model for the network, but instead make a more general structural assumption of $M$ being approximately low rank, which holds for most popular network models.    We  propose a new general edge cross-validation (ECV) strategy for networks, splitting node pairs rather than nodes into different folds, a natural yet crucial choice.   Treating the network after removing the entries of $A$ for some node pairs as a partially observed network, we apply low rank matrix completion to ``complete'' the network and then fit the relevant model.      This reconstructed network has the same rate of concentration around the true model as the full network adjacency matrix, allowing for valid analysis.     Our method is valid for directed and undirected, binary and weighted networks.      As concrete examples, we show how ECV can be applied to determine the latent space dimension of random dot product graph models, select between block model variants, tune regularization for spectral clustering,  and tune neighborhood smoothing for graphon models.

\section{The edge cross-validation  (ECV) algorithm}\label{sec:method}
\subsection{Notation and model}
For simplicity of presentation, we derive everything for binary networks, but it will be clear that  our framework is directly applicable to weighted networks, which are prevalent in practice, and in fact the application in Section~\ref{sec:app} is to a weighted network.   

Recall $n$ is the number of nodes and $A$ is the $n\times n$ adjacency matrix.   
Let $D = \F{diag}(d_1, \cdots, d_n)$ be the diagonal matrix with node degrees $d_i = \sum_{j}A_{ij}$ on the diagonal. The (normalized) Laplacian of a network is defined as $L  = D^{-1/2}AD^{-1/2}.$
Finally, we write $I_n$ for the $n\times n$ identity matrix and $\mbone_n$  for $n\times 1$ column vector of ones, suppressing the dependence on $n$ when it is clear from the context.   For any matrix $M$, we use $\norm{M}$ to denote its spectral norm and $\norm{M}_F$ to denote its Frobenius norm. 

Throughout the paper, we work with the widely used   inhomogeneous Erd\"{o}s-Renyi model for networks, defined by an $n\times n$ matrix of probabilities $M$, with (unique) edges $A_{ij}$ drawn as independent Bernoulli variables with $\p(A_{ij}=1) = M_{ij}$.    All the information about the structure of the network is thus contained in $M$.  While all $M_{ij}$'s can be different, with no additional assumptions on $M$  inference is impossible,  since we only have one observation.   On the other hand, we would like to avoid assuming a specific parametric model, since choosing the (type of) model is one of the primary applications of cross-validation.  As a compromise, we make a generic structural assumption on $M$, assuming it is low rank, which holds for many popular network models.    We describe three classes of examples below:
\vspace{-0.9cm}
\begin{enumerate}
\item The stochastic block model and its generalizations. The stochastic block model is perhaps the most widely used undirected network model with communities.   The model assumes that $M = ZBZ^T$ where $B \in [0,1]^{K\times K}$ is a symmetric probability matrix and $Z\in \{0,1\}^{n\times K}$ has exactly one ``1'' in each row, with $Z_{ik} = 1$ if node $i$ belongs to community $k$.  Let $\V{c} = (c_1,\cdots, c_n)$ be the vector of node membership labels with $c_i$ taking values in $\{1, \dots, K\}$. In particular, it is assumed that $P(A_{ij} = 1) =  B_{c_ic_j}$, that is, the probability of edge between two nodes depends only on the communities they belong to.    One of the commonly pointed out limitations of the stochastic block model is that it forces equal expected degrees for all the nodes in the same community, therefore ruling out ``hubs''.     The degree corrected stochastic block model corrects this by allowing nodes to have individual ``degree parameters''  $\theta_i$ associated with each node $i$, and models $P(A_{ij} = 1) =  \theta_i \theta_j B_{c_ic_j}$.  The degree corrected model needs a constraint to ensure identifiability, and here we use the constraint $\sum_{c_i=k}\theta_i=1$ for each $k$, proposed in the original paper \citep{karrer2011stochastic}.    The popular configuration model \citep{chung2002average} can be viewed as a special of the degree corrected model, and both these models have a probability matrix $M$ of rank $K$.   There are multiple other low rank variants of the stochastic block model, for example,  the mixed membership block model  \cite{airoldi2008mixed} and the popularity adjusted model recently proposed by \cite{sengupta2018block}.  For a review of recent developments on this class of models, see \cite{abbe2017community}. 

\item The random dot product graph model. The random dot product graph model \citep{young2007random}  is a  general low-rank network model.  It assumes each node of the network is associated with a latent $K$-dimensional vector $Z_i \in \bR^K$, and $M_{ij} = Z_i^TZ_j$. This model has been successfully applied to a number of network problems  \citep{sussman2014consistent,tang2017nonparametric} and its limiting behaviors can also be studied \citep{tang2016limit}. More details can be found in the review paper \cite{athreya2017statistical}. The random dot product graph model can include the stochastic block model as a special case, but only if the probability matrix $M$ of the stochastic block model is positive semi-definite.  

\item Latent space model and graphon models.  The latent space model \citep{hoff2002latent} is another popular inhomogeneous Erd\"os-R\'enyi model.  Similarly to the random dot product graph, it assumes the nodes correspond to $n$ latent positions $Z_i\in \bR^K$, and the probability matrix is some function of the latent positions, for example, the distance model $f(M_{ij}) = \alpha - \norm{Z_i - Z_j}$, or the projection model $f(M_{ij}) = \alpha - Z_i^TZ_j/(\norm{Z_i}\norm{Z_j})$ where $f$ is a known function, such as the logit function.    More generally, the Aldous-Hoover representation \citep{aldous1981representations, diaconis2007graph} says that the probability matrix of any exchangeable random graph can be written as $M_{ij} = f(\xi_i, \xi_j)$
for $\xi_i, i\in [n]$  independent uniform random variables on $[0,1]$ and a function $f:[0,1]\times [0,1] \to [0,1]$ symmetric in its two arguments, determined up to a measure-preserving transformation.   There is a substantial literature on estimating the function $f$, called the graphon, under various assumptions  \citep{wolfe2013nonparametric,choi2014co, gao2015rate}.  Under this framework $M$ is random, but the network follows an inhomogeneous  Erd\"{o}s-Renyi model conditional y on $M$, and thus our method is applicable conditionally.    The latent space models and the more general graphon models typically do not assume that $M$ is low rank, enforcing certain smoothness assumptions on the function $f$ instead.  Fortunately, when these smoothness assumptions apply, the corresponding matrix $M$ can typically be approximated reasonably well by a low rank matrix \citep{chatterjee2015matrix,zhang2015estimating}.   In this setting, the ECV procedure works with the best low rank approximation to the model;  see details in Section~\ref{secsec:graphon}).

\end{enumerate}

\subsection{The ECV procedure}\label{secsec:ECValgo}
For notational simplicity, we only present the algorithm for directed networks;  the only modification needed for undirected networks is treating node pairs $(i,j)$ and $(j,i)$ as one pair.   The key insight of ECV is to split node pairs rather than nodes, resulting in a partially observed network.  We randomly sample node pairs (regardless of the value of $A_{ij}$) with a fixed probability $1-p$ to be in the held-out set.  By exchangeable model assumption, the values of $A$ corresponding to held-out node pairs are independent of those corresponding to the rest.   The leftover training network now has missing edge values, which means many models and methods cannot be applied to it directly.    Our next step is to reconstruct a ``complete" network $\hat{A}$ from the training node pairs.   Fortunately, the missing entries are missing completely at random by construction, and this is the classic setting for matrix completion.    Any low-rank based matrix completion algorithm can now be used to fill in the missing entries, for example \cite{candes2010matrix}, \cite{davenport20141}.  We postpone the algorithm details to Section~\ref{secsec:matrixcompletion}. 

Once we complete $\hat{A}$ through matrix completion, we can fit the candidate models on $\hat{A}$ and evaluate the relevant loss on the held-out entries of $A$, just as in standard cross-validation.    There may be more than one way to evaluate the loss on the held-out set if the loss function itself is designed for binary input;  we will elaborate on this in examples in Section~\ref{sec:example}.    The general algorithm is summarized as  Algorithm~\ref{algo:genericECV} below.    We present the version with many random splits into training and test pairs, but it is obviously applicable to $K$-fold cross-validation if the computational cost of many random splits is prohibitive.   
 
\begin{algo}[The general ECV procedure]\label{algo:genericECV}
Input:  an adjacency matrix $A$, a loss function $L$, a set $\ccal$ of $Q$ candidate models or tuning parameter values to select from, the training proportion $p$, and the number of replications $N$.
\begin{enumerate}
\item Select rank $\hat{K}$ for matrix completion, either from prior knowledge or using the model-free cross-validation procedure in Section~\ref{secsec:ECV-AUC}.
\item For $m=1, \dots, N$
\begin{enumerate}
\item Randomly choose a subset of node pairs $\Omega \subset \Vnode\times\Vnode$, by selecting each pair independently with probability $p$. 
\item Apply a low-rank matrix completion algorithm to  $(A, \Omega)$ to obtain $\hat{A}$ with rank $\hat{K}$.  
\item For each of the candidate models $q =1, \dots, Q$, fit the model on $\hat{A}$, and evaluate its loss $L^{(m)}_q$ by averaging the loss function $L$ with the estimated parameters over the held-out set $A_{ij}, (i,j) \in \Omega^c$.
\end{enumerate}
\item Let $L_q = \sum_{m=1}^N L^{(m)}_q/N$ and return $\hat{q} = \argmin_q L_q$ (the best model from set $\ccal$).  
\end{enumerate}

\end{algo}

The two crucial parts of ECV are splitting node pairs at random and applying low-rank matrix completion to obtain a full matrix $\hat{A}$.  The two internal parameters we need to set for the ECV are the
selection probability $p$ and the number of repetitions $N$.  Our
numerical experiments suggest (see Supplementary Material~\ref{appendix:robustness})
that the accuracy is stable for $p \in (0.85,1)$ and the choice of $N$
does not have much effect after applying stability selection.  In all
of our examples, we take $p=0.9$ and $N=3$.

\subsection{Network recovery by matrix completion}\label{secsec:matrixcompletion}

There are many algorithms that can be used to recover $\hat{A}$ from the training pairs. 
 Define operator $P_\Omega : \bR^{n\times n} \to \bR^{n\times n}$ by $(P_\Omega A)_{ij} = A_{ij} \ind \{ (i,j)\in \Omega \},$
replacing held-out entries by zeros.    A generic low-rank matrix completion procedure solves the problem
\begin{align}\label{eq:genericMC}
\min_W~~  F(P_{\Omega}W, P_{\Omega}A)~~~~~~~~\text{subject to}~~ \rank(W) \le \hat{K} 
\end{align}
where $\hat{K}$ is the rank constraint and $F$ is a loss function measuring the discrepancy between $W$ and $A$ on entries in $\Omega$, for example, sum of squared errors or binomial deviance. Since the problem is non-convex due to the rank constraint, many computationally feasible variants of \eqref{eq:genericMC} have been proposed for use in practice, obtained via convex relaxation and/or problem reformulation.   While any such  method can be used in ECV, for concreteness we follow the  singular value thresholding procedure to construct a low rank approximation
\begin{equation}\label{eq:simpleHS}
\hat{A} = S_H \left( \frac{1}{p}P_{\Omega}A, \hat{K} \right),
\end{equation}
where $S_H(P_{\Omega}A,\hat{K})$ denotes rank $\hat{K}$ truncated SVD of a matrix $P_{\Omega}A$.  That is, if the SVD of $P_{\Omega}A$ is $P_{\Omega}A=UDV^T$ where $D=\F{diag}(\sigma_1, \cdots, \sigma_n)$, $\sigma_1 \ge \sigma_2 \cdots \ge \sigma_n \ge 0$, then $S_H(P_{\Omega}A,\hat{K}) = UD_{\hat{K}}V^T$,
where $D_{\hat{K}}=\F{diag}(\sigma_1, \cdots, \sigma_{\hat{K}}, 0, \cdots, 0)$. 

This matrix completion procedure is similar to the universal singular value thresholding method of \cite{chatterjee2015matrix}, except we fix $K$ and always use top $K$ eigenvalues instead of using a universal constant to threshold $\sigma$'s. This method is computationally efficient as it only requires a partial SVD  of the adjacency matrix with held-out entries replaced by zeros, which is typically sparse. It runs easily on a network of size $10^4-10^5$  on a laptop.    In principle, one can choose any matrix completion algorithm satisfying a bound similar to the one used in Theorem~\ref{thm:simplebound}.  One can choose a more sophisticated method such as, for example, \cite{keshavan2009matrix} and \cite{mazumder2010spectral}  if the size of the network allows, but since cross-validation is already computationally intensive, we prioritized low computational cost.   Additionally, imputation accuracy is not the primary goal;   we expect and in fact need noisy versions of $A$.  As a small-scale illustration, we have compared our SVD method to the iterative hardImpute algorithm of \cite{mazumder2010spectral} (see Supplementary Material~\ref{secsec:compareSoftImpute})  in the context of ECV.   
 We found that while it improves the accuracy of matrix completion by itself, it takes longer to compute and does not provide any tangible improvement in model selection, which is our ultimate goal here.

\begin{remark}
In some situations, the rank of $M$ itself is directly associated with the model to be selected;  see examples in Sections~\ref{secsec:ECV-AUC} and \ref{secsec:tuneRSC}.  In these situations, matrix completion rank $\hat{K}$ should be selected as part of model selection, omitting Step 1 in Algorithm~\ref{algo:genericECV} and instead merging  Steps  2(b) and 2(c) and using a value of $\hat{K}$ corresponding to the model being evaluated.   See Sections~\ref{secsec:ECV-AUC} and \ref{sec:blockmodels} for details. \end{remark}

\begin{remark}
If an upper bound on $\norm{M}_{\infty}$ is available, say $\norm{M}_{\infty} \le \bar{d}/n$, where $\norm{M}_{\infty} = \max_{ij} |M_{ij}|$, an improved estimator $\tilde{A}$ can be obtained by truncating the entries of $\hat{A}$ onto the interval $[0,\bar{d}/n]$, as in \cite{chatterjee2015matrix}.  A trivial option of truncating to the interval $[0,1]$ is always available, ensuring $\tilde{A}$ is a better estimator of $M$ in Frobenius norm than $\hat{A}$. We did not observe any substantial improvement in model selection performance from truncation, however.   In some applications, a binary adjacency matrix may be required for subsequent model fitting;  if that is the case, a binary matrix can be obtained from $\tilde{A}$ by using one of the standard link prediction methods, for example, by thresholding at 0.5   
\end{remark}

\begin{remark}
An alternative to matrix completion is to simply replace all of the held-out entries by zeros and use the resulting matrix $A^0$ for model estimation.    The resulting model estimate $\hat{M}^0$ of the probability matrix $\e A^0$ is a biased estimator of $M$, but since we know the sampling probability $p$, we can remove this bias by setting  $\hat{M}^* = \hat{M}^0/p$ as in \cite{chatterjee2015matrix} and \cite{gao2016optimal}, then use $\hat{M}^*$  for prediction and calculating the cross-validation error. This method is valid as long as the adjacency matrix is binary and probably the simplest of all  (though, surprisingly, we did not find any explicit references to this in the literature).    In particular, for the stochastic block model it is equivalent to our general ECV procedure when using \eqref{eq:simpleHS} for matrix completion.  However, in applications beyond block models these two approaches will give different results, and we have empirically observed that ECV with matrix completion works better and is much more robust to the choice of $p$.  Moreover, filling in zeros instead of doing matrix completion does not work for weighted networks, since that would  clearly change the weight distribution which cannot be fixed by a simple rescaling by $p$.   We do not pursue this version further.   
\end{remark}

\begin{remark}
Another matrix completion option is the 1-bit matrix completion \citep{davenport20141,cai2013max,bhaskar20151}, which uses binomial deviance instead of the least squares loss and assume that some smooth transformation of $M$ is low rank.   In particular, the special case of projection latent space model matches this framework.  However, 1-bit matrix completion methods are generally much more computationally demanding than the Frobenius norm-based completion, and given that computational cost is paramount for cross-validation whereas accurate matrix imputation is secondary, we do not pursue 1-bit matrix completion further.  
\end{remark}

\subsection{Theoretical justification}

Intuitively, ECV should work well if $\hat{A}$ reflects relevant structural properties of the true underlying model.   The following theorem formalizes this intuition.
All our results will be expressed as a function of the number of nodes $n$, the sampling probability $p$ which controls the size of the training set, the rank $K$ of the true matrix $M$, and an upper bound on the expected node degree  $d$, defined to be any value satisfying $\max_{ij}M_{ij}\le  d/n$, a crucial quantity for network concentration results.   We can always trivially set $d =  n$, but we will also consider the  sparse networks case with  $d = o(n)$.

\begin{theorem}\label{thm:simplebound}
Let $M$ be a probability matrix of rank $K$ and $d$ as defined above.  Let $A$ be an adjacency matrix with edges sampled independently and $\e(A) = M$.   Let $\Omega$ be an index matrix for a set of node pairs selected  independently with probability $p\ge C_1 \log n/ n$ for some absolute constant $C_1$, with $\Omega_{ij} = 1$ if the node pair $(i,j)$ are selected and 0 otherwise.  If $d \ge C_2\log(n)$ for some absolute constant $C_2$, then with probability at least $1-3n^{-\delta}$ for some $\delta>0$, the completed matrix $\hat{A}$ defined in \eqref{eq:simpleHS} with $\hat{K} = K$ satisfies
\begin{equation}\label{eq:generalerr}
\norm{\hat{A}-M} \le \tilde{C}\max\left(\sqrt{\frac{Kd^2}{np}}, \sqrt{\frac{d}{p}}, \frac{\sqrt{\log n}}{p}  \right)
\end{equation}
where $\tilde{C} = \tilde{C}(\delta, C_1, C_2)$ is a constant that only depends on $C_1, C_2$ and $\delta$. This also implies 
\begin{equation}\label{eq:generalerr_Frob}
\frac{\norm{\hat{A}-M}_F^2}{n^2} \le \frac{\tilde{C}^2}{2}\max\left(\frac{K^2d^2}{n^3p}, \frac{Kd}{n^2p}, \frac{K\log n}{n^2p^2}  \right).
\end{equation}
\end{theorem}

This theorem holds for both directed and undirected networks;  it can also be equivalently written in terms of the size of the set $|\Omega|$ since $|\Omega| \sim n^2p$.  From now on, we treat $p$ as a constant for simplicity, considering it is a user-chosen parameter.  We first compare Theorem~\ref{thm:simplebound} with known rates for previously studied network problems. In this case, the spectral norm error bound  \eqref{eq:generalerr},  taking into account the assumption $d\ge C_2 \log n$, becomes
 \begin{equation}\label{eq:pconstant}
 \norm{\hat{A}-M} \le \tilde{C}\max\left (\sqrt{\frac{Kd}{n}}, 1\right) \sqrt{d} \ .
 \end{equation}
 The bound \eqref{eq:pconstant} implies the rate of concentration of $\hat A$ around $M$ is the same as the  concentration of the full adjacency matrix $A$ around its expectation \citep{lei2014consistency,chin2015stochastic, le2017concentration}, as long as $Kd/n \le 1$. The sparser the network, the weaker our requirement for $K$. For instance, when the network is moderately sparse with $d = O(\log n)$, we only need $K \le (n/\log{n})$.   This may seem counter-intuitive but this happens because the dependence on $K$ in the bound comes entirely from $M$ itself.   A sparse network means that most entries of $M$ are very small, thus replacing the missing entries in $A$ with zeros does not contribute much to the overall error and the requirement on $K$ can be less stringent.    While for sparse networks the estimator is noisier, the noise bounds have the same order for the complete and the incomplete networks (when $p$ is a constant), and thus the two concentration bounds still match.  
 
Theorem~\ref{thm:simplebound} essentially indicates $\norm{\hat{A}-M} \approx \norm{A-M}$
if we assume $Kd \le n$. Thus in the sense of concentration in spectral norm, we can treat $\hat{A}$ as a network sampled from the same model.    Under many specific models, such concentration of $\hat{A}$ is sufficient to ensure model estimation consistency at the same rate as can be obtained from using the original matrix $A$ and also gives good properties about model selection (see Theorem~\ref{thm:RDPG-consistency} and Theorem~\ref{thm:K-Consistency}.)

\section{Examples of ECV for model selection}\label{sec:example}

\subsection{Model-free rank estimators} 
\label{secsec:ECV-AUC}

The rank constraint for the matrix completion problem for Algorithm~\ref{algo:genericECV} may be unknown, and we need to choose or estimate it in order to apply ECV. When the true model is a generic low-rank model such as the random dot product graph model, selecting $\hat{K}$ is essentially selecting its latent space dimension.   Rank selection for a general low-rank matrix (not a network) by cross-validation has been studied by \cite{owen2009bi} and \cite{kanagal2010rank}.    They split the matrix by blocks or multiple blocks instead of by individual entries, and evaluated performance on the task of non-negative matrix factorization, a completely different setting from ours.   More generally, selection of $\hat{K}$  can itself be treated as a model selection problem, since the completed matrix $\hat{A}$ itself is a low rank approximation to the unknown underlying probability matrix $M$.

To find a suitable value of $\hat{K}$, one has to compare $\hat{A}$ (completions for each candidate rank) to $A$ in some way.   
We take the natural approach of directly comparing the values of $\hat{A}$ and $A$ on the held-out set.   We can use the sum of squared errors on the held-out entries, $ = \sum_{(i,j)\in \Omega^c}(A_{ij}-\hat{A}_{ij})^2, $ or, when $A$ is binary,  the binomial deviance as the loss function to optimize.  Another possibility is to evaluate how well $\hat A$ predicts links (for unweighted networks).  We can predict $\tilde{A}_{ij} = \ind \{ \hat A_{ij} > c \} $  for all entries in the held-out set $\Omega^c$ for a threshold $c$, and vary $c$ to obtain a sequence of link prediction results.    A common measure of prediction performance is the area under the ROC curve (AUC), which compares false positive rates to true positive rates for all values of $c$, with perfect prediction corresponding to AUC of 1, and random guessing to 0.5. Therefore, the negative AUC can also be used as the loss function.   In summary, the completion rank $K$ can be chosen as follows:   
\begin{algo}[Model-free ECV for rank selection]\label{algo:rank}
Input:  an adjacency matrix $A$, the training proportion $p$, maximum possible rank $K_{\max}$ and the number of replications $N$.
\begin{enumerate}
\item For $m=1, \dots, N$
\begin{enumerate}
\item Randomly choose a subset of node pairs $\Omega \subset \Vnode\times\Vnode$, by selecting each pair independently with probability $p$. 
\item Apply a low-rank matrix completion algorithm to  $(A, \Omega)$ to obtain $\hat{A}$ with rank $\hat{K}$.  
\item For each of the candidate models $k =1, \dots, K_{\max}$, apply the low-rank matrix completion to  $(A, \Omega)$ with rank $k$ to obtain $\hat{A}$;  calculate the value of the loss function for using $\hat{A}$ to predict $A$ on $\Omega^c$, denoted by $L_k^{(m)}$
\end{enumerate}
\item Let $L_k = \sum_{m=1}^N L^{(m)}_k$/N and return $\hat{K} = \min_k L_k$.  
\end{enumerate}
\end{algo}
If the loss is the sum of squared errors, this algorithm can be viewed as a  network analogue of the tuning strategy of \cite{mazumder2010spectral}. In practice, we have observed that both the imputation error and the AUC work well in general rank estimation tasks. For block models, they perform comparably to likelihood-based methods most of the time.

From a theoretical perspective, the model of rank $K$ is a special case of the model with rank $K+1$, and while the former is preferable for parsimony, they will give very similar model fits (unless overfitting occurs). 
A reasonable goal for model selection in this situation is to guarantee the selected rank is not under-selected;  the same guarantee was provided by \cite{chen2014network}.
 The lack of protection against over-selection is a known issue for cross-validation in many problems and has been rigorously shown for regression \citep{shao1993linear,zhang1993model,yang2007consistency, lei2017cross} whenever the training proportion is non-vanishing.

\begin{assumption}\label{ass:RDPG-parameter}
Assume $M = \rho_n M^0$ where $M^0 = U\Sigma^0U^T$ is a probability matrix, $\Sigma^0 = \diag(\lambda_1^0, \cdots, \lambda_K^0)$ is the diagonal matrix of non-increasing eigenvalues, and  $U = (U_1,\cdots, U_K)$ contains the corresponding eigenvectors. Assume there exists a positive constant $\psi_1$  such that $n\psi_1^{-1} \le \lambda_K^0 \le \lambda_1^0 \le \psi_1 n$ and the minimum gap between any two distinct eigenvalues is at least $n/(2\psi_1)$.   Also, assume $\max_{i\in [n]}\sum_{j\in [n]}M^0_{ij} \ge \psi_2 n\max_{ij}M^0_{ij}$ for some positive constant $\psi_2$, i.e.,  the values of $M^0$ are all of similar magnitude.  With this parameterization, the expected node degree  is bounded by $\lambda_n = n\rho_n$. 
\end{assumption}

Another quantity we need is matrix coherence, introduced by \cite{candes2009exact}.    Under the parameterization of Assumption \ref{ass:RDPG-parameter},  coherence of $P$ is defined as
$$\mu(M) = \max_{i \in [n]}\frac{n}{K}\norm{U^T\V{e}_i}^2 = \frac{n}{K}\norm{U}_{2,\infty}^2.$$

To control prediction errors on the held-out entries in ECV, we need matrix completion to work well for most entries for which that it is generally believed in matrix literature that the matrix incoherence is necessary  \citep{chen2015completing,chi2019matrix}.    We will follow this literature and assume $\mu(M)$ is bounded, although in our context this assumption can be relaxed at the cost of a stronger condition on the network density. 

\begin{assumption}[Incoherent matrix]\label{ass:incoherence}
Under Assumption~\ref{ass:RDPG-parameter}, assume the coherence of $P^0$ is bounded with $\mu(P^0)  \le a$ for some constant $a > 1$.
\end{assumption}
Intuitively, Assumption \ref{ass:incoherence} says that the mass of eigenvectors of $P$ is not concentrated on a small number of coordinates. There is a large class of matrices satisfying the above bounded incoherence \citep{candes2009exact, candes2010power}. In the context of networks, it is easy to verify that, for example, the stochastic block model with a positive semi-definite probability matrix and non-vanishing communities satisfies both Assumptions \ref{ass:RDPG-parameter} and \ref{ass:incoherence}. In the special case of a fixed $K$ and $B =  \rho \cdot [(1-\beta)  I+ \beta \mbone \mbone^T]$, a sufficient condition for positive semi-definitiveness is $\beta \le 1/K$, implying a certain degree of assortativity in the network.     The degree corrected stochastic block model and the configuration model also satisfy these assumptions with similar restrictions on parameters, as long as the variability in the node degrees is of the same order for all nodes.   In general, any model that does not have ``spiky'' nodes that are very different from other nodes should satisfy these assumptions, possibly with some additional constraints on the parameter space.  

We next state a result on model selection, the primary task of ECV. 
\begin{theorem}[Consistency under the random dot product graph model]\label{thm:RDPG-consistency}
Assume $A$ is generated from a random dot product graph model satisfying \ref{ass:RDPG-parameter} and \ref{ass:incoherence},  with latent space dimension $K$.  Let $\hat{K}$ be the output of  Algorithm~\ref{algo:rank}.  If the sum of squared errors is used as the loss and the expected degree satisfies $\lambda_n/(n^{1/3}\log^{4/3}n) \to \infty$, 
$$\p(\hat{K} < K) \to 0.$$
\end{theorem}
To the best of our knowledge, Theorem~\ref{thm:RDPG-consistency} gives the first model selection guarantee under the random dot product graph model.

\color{black}

\subsection{Model selection for block models}
\label{sec:blockmodels}

Next we apply ECV to model selection under the stochastic block model and the degree corrected version (referred to as block models for conciseness). The choice of fitting method is not crucial for model selection, and many consistent methods are now available for fitting both models \citep{karrer2011stochastic, zhao2012consistency, bickel2013asymptotic, amini2013pseudo}.   Here we use one of the simplest, fastest, and most common methods, spectral clustering on the Laplacian $L = D^{1/2}AD^{1/2}$, where $D$ is the diagonal matrix of node degrees.   For the stochastic block model, spectral clustering takes $K$ leading eigenvectors of $L$, arranged in a $n \times K$ matrix $U$, and applies the $K$-means clustering algorithm to the rows of $U$ to obtain cluster assignments for the $n$ nodes.   For the degree corrected block model, the rows are normalized first, and then the same algorithm applies.

Spectral clustering enjoys asymptotic consistency under the block models when the average degree grows at least as fast as $\log n$ \citep{rohe2011spectral,lei2014consistency, sarkar2015role}. The possibility of strong consistency for spectral clustering was recently discussed by \cite{eldridge2017unperturbed}, \cite{abbe2017entrywise} and \cite{su2017strong}.  Variants of spectral clustering are consistent under the degree corrected model, for example, spherical spectral clustering \citep{qin2013regularized, lei2014consistency} 
and the SCORE method \citep{jin2015fast}. 

 Since both the stochastic block model and the degree corrected model are undirected network models, we use the undirected ECV, selecting edges at random from pairs $(i, j)$ with $i < j$, and including the pair $(j,i)$ whenever $(i,j)$ is selected.  Once node memberships are estimated, the other parameters are easy to estimate by conditioning on node labels, for example by the MLE evaluated on the available node pairs.  Let  $\hat{C}_k = \{i:  (i,j) \in \Omega, \hat c_i = k \}  $ be the estimated member sets for each group $k = 1, \dots, K$.   Then we can estimate the entries of the probability matrix $B$ as  
\begin{equation} \label{eq:estimation-SBM}
\hat{B}_{kl} = \frac{\sum_{(i,j) \in \Omega}A_{ij} 1(\hat c_i = k, \hat c_j = l)}{\hat{n}^{\Omega}_{kl}}
\end{equation}
where
$$\hat{n}^{\Omega}_{kl} = 
\begin{cases}   |(i,j) \in \Omega: \hat{c}_i=k, \hat{c}_j=l| & \quad  \text{if } k \ne l \\
                         |(i,j) \in \Omega: i<j, \hat{c}_i=\hat{c}_j=k|                   & \quad  \text{if } k = l. 
                   \end{cases}
$$

Under the degree corrected model, the probability matrix can be estimated similarly as in \cite{karrer2011stochastic, zhao2012consistency} and \cite{joseph2016impact} via the Poisson approximation, letting 
$$ \hat{O}^*_{kl} = \sum_{ (i,j)\in \Omega}A_{ij} 1(\hat c_i = k, \hat c_j = l)
\text{~~and setting~~} 
\hat{\theta}_i= \frac{\sum_{j:(i,j)\in \Omega}A_{ij}}{\sum_{k=1}^K\hat{O}^*_{\hat{c}_i,k}} \ , \ \ 
\hat{P}_{ij} = \hat{\theta}_i\hat{\theta}_j\hat{O}^*_{\hat{c}_i\hat{c}_j}/p \ . 
$$
The probability estimate $\hat{P}$ is scaled by $p$ to reflect missing edges, which makes it slightly different from the estimator for the fully observed degree corrected model  \citep{karrer2011stochastic}.   This rescaling happens automatically in the estimator \eqref{eq:estimation-SBM} since the sums in both the numerator and the denominator range over $\Omega$ only.    Finally, the loss function can again be the sum of squared errors or  binomial deviance; we found that the $L_2$ loss  works slightly better in practice for the block models.

The model selection task here includes the choice of the stochastic block model vs.\ the degree corrected model and the choice of $K$.  Suppose we consider the number of communities ranging from $1$ to $K_{\max}$.    The candidate set of models in Algorithm~\ref{algo:genericECV} is then both of the two block models with $K$ varying from 1 to $K_{\max}$.  The ECV algorithm for this task is presented next, as Algorithm~\ref{algo:ECV}.

\begin{algo}\label{algo:ECV}
Input:  an adjacency matrix $A$, the largest number of communities to consider $K_{\max}$, the training proportion $p$, and the number of replications $N$. 
\begin{enumerate}
\item For $m=1, \dots, N$
\begin{enumerate}
\item  Randomly choose a subset of node pairs $\Omega$, selecting each pair $(i,j)$, $i < j$  independently with probability $p$, and adding $(j,i)$ if $(i,j)$ is selected.  
\item For $k = 1, \dots, K_{\max}$, 
\begin{enumerate}
\item Apply matrix completion to $(A,\Omega)$ with rank constraint $k$ to obtain $\hat{A}_k$. 
\item Run spectral clustering on $\hat{A}_k$ to obtain the estimated stochastic block model membership vector $\hat{\V{c}}^{(m)}_{1,k}$, and spherical spectral clustering to obtain the estimated  degree corrected model $\hat{\V{c}}^{(m)}_{2,k}$.    
\item Estimate the two models' probability matrices $\hat{M}^{(m)}_{1,k}$,  $\hat{M}^{(m)}_{2,k}$ based on  $\hat{\V{c}}^{(m)}_{1,k}$,  $\hat{\V{c}}^{(m)}_{2,k}$ and evaluate the corresponding losses $L^{(m)}_{q,k}$, $q = 1, 2$ by applying the loss function $L$ with the estimated parameters to $A_{ij}, (i,j) \in \Omega^c$.  
\end{enumerate}
\end{enumerate}
\item Let $L_{q,k} = \sum_{m=1}^N L^{(m)}_{q,k}/N$.   Return $(\hat{q}, \hat{K}) = \arg\min_{q=1,2}\min_{k = 1, \dots, K_{\max}} L_{q,k}$ as the best model (with $\hat q = 1$ indicating no degree correction and $\hat q=2$ indicating degree correction).   
\end{enumerate}
\end{algo}

As a special case, one can also consider the task of just choosing $K$ under a specific model (the stochastic block model or the degree corrected model), for which there are many methods \citep{latouche2012variational, mcdaid2013improved, bickel2013Hypothesis,lei2016goodness, saldana2014how, wang2015likelihood, chen2014network, le2015estimating}. In particular,  Theorem~\ref{thm:simplebound} can be modified (see Proposition~\ref{prop:SBMperf} and \ref{prop:DCSBMperf} in Supplementary Material) to show that the parametric ECV (Algorithm~\ref{algo:ECV}) achieves  one-sided consistency of choosing $K$ under the stochastic block model, under the following standard assumption \citep{lei2014consistency}:
\begin{assumption}\label{ass:SBM_basic}
The probability matrix $B^{(n)} = \rho_n B_0$, where $B_0$ is a fixed $K\times K$ symmetric nonsingular matrix with all entries in $[0,1]$ and $K$ is fixed (and therefore the expected node degree is $\lambda_n  = n \rho_n$).    There exists a constant $\gamma >0$ such that $\min_kn_k >  \gamma n$ where $n_k = |\{i:c_i=k\}|$.
\end{assumption}

\begin{theorem}[Consistency under the stochastic block model]\label{thm:K-Consistency}
Let $A$ be the adjacency matrix of a network generated from the stochastic block model  satisfying Assumption~\ref{ass:SBM_basic}  and suppose the model is known to be the stochastic block model as a prior knowledge but the number of communities $K$ is to be estimated. Assume $\lambda_n / \log n \rightarrow \infty$. Let $\hat{K}$ be the selected number of communities by using Algorithm~\ref{algo:ECV} with the $L_2$ loss.  Then we have
$$\p(\hat{K} < K) \to 0.$$
If we assume $\lambda_n n^{-2/3} \rightarrow \infty$ and all entries of $B_0$ are positive, then the same result also holds for the binomial deviance loss.
\end{theorem}

The theorem requires a stronger assumption for the binomial deviance result than it does for the $L_2$ loss.   While these conditions may not be tight, empirically the $L_2$ loss performs better as well (see Section~\ref{sec:sim} and Supplementary Material~\ref{appendix:deviance}), which may intuitively be explained by the instability of binomial deviance near 0.    Just like in our random dot product graph result and the result  of \cite{chen2014network}, we have a one-sided guarantee, but the assumption on the expected degree is much weaker than that of Theorem~\ref{thm:RDPG-consistency}. This is a natural trade-off of better rates under a parametric version of the ECV against making additional model assumptions.  
\color{black}

\subsection{Parameter tuning in graphon estimation}\label{secsec:graphonModel}

Graphon (or probability matrix) estimation is another general  task which often relies on tuning parameters that can be determined by cross-validation.    \cite{zhang2015estimating} proposed a method called ``neighborhood smoothing" to estimate $M$ instead of $f$ under the assumption that $f$ is a piecewise Lipschitz function, avoiding the measure-preserving transformation ambiguity. They showed their method achieves a nearly optimal rate while requiring only polynomial complexity for computation (optimal methods are exponential).    The method depends on a tuning parameter $h$ which controls the degree of smoothing.  The theory suggests $h =  \tau \left({\log n}/{n}\right)^{1/2}$ for some $\tau$.  

This is a setting where we have no reason to assume a known rank of  the true probability matrix and $M$ does not have to be low rank. However, for a smooth graphon function a low rank matrix can approximate $M$ reasonably well \citep{chatterjee2015matrix}. The ECV procedure under the graphon model now has to select the best rank for its internal matrix completion step.  Specifically, in each split, we can run the rank estimation procedure discussed in Section~\ref{secsec:ECV-AUC} to estimate the best rank for approximation and the corresponding $\hat{A}$ as the input for the neighborhood smoothing algorithm. The selected tuning parameter is the one minimizing the average prediction error. 

The ECV algorithm can also be used to other tuning parameter selection problems. In Supplementary Material~\ref{secsec:tuneRSC}, we show its application in tuning network regularization in spectral clustering.

\subsection{Stability selection}\label{secsec:stability}
Stability selection \citep{meinshausen2010stability} was proposed as a general method to reduce noise by repeating model selection many times over random splits of the data and keeping only the features that are selected in the majority of splits;  any cross-validation procedure can benefit from stability selection since it relies on random data splits.  An additional benefit of stability selection in our context is increased robustness to the choice of $p$ and $N$ (see Supplementary Material~\ref{appendix:robustness}).  \cite{chen2014network} applied this idea as well,  repeating the procedure multiple times and choosing the most frequently selected model. We use the same strategy for ECV (and the CV method of \cite{chen2014network}), choosing the model selected most frequently out of 20 replications.   When we need to select a numerical parameter rather than a model, we can also average the values selected over the 20 replications (and round to an integer if needed, say for the number of communities).    Overall, picking the most frequent selection is more robust to different tasks, though picking the average may work better in some situations.  More details are given in Section~\ref{sec:sim}. 
  
\section{Numerical performance evaluation}\label{sec:sim}

\subsection{Model selection under block models}\label{secsec:sim:DCSBM}\label{secsec:sim-blockmodel}

Following \cite{chen2014network}, we evaluate performance under the block models on choosing both the model (with/without degree correction) and the number of communities $K$ simultaneously.  The setting for all simulated networks in this section is as
follows.   For the degree corrected block model, we first sample 300 values from the
power law distribution with the lower bound $1$ and scaling parameter
$5$, and then set the node degree parameters $\theta_i$, $i=1,\cdots,
n$ by randomly and independently choosing one of these 300 values.
For the stochastic block model, we set $\theta_i =  1$ for all $i$. We set the communities to have equal
sizes. The imbalanced community situation is given in the Supplementary Material.  Let $B_0 = (1-\beta)  I+ \beta \mbone \mbone^T$ and $B
\propto \Theta B_0 \Theta$, so that $\beta$ is the out-in ratio (the
ratio of between-block probability and within-block probability of
edge).  The scaling is selected so that the average node degree is
$\lambda$.   We consider several combinations of size and the number of communities: $(n=600, K=3)$, $(n=600, K=5)$ and $(n=1200, K=5)$.   For each configuration, we then vary two aspects of the model:
\begin{enumerate}
\item Sparsity:  set the expected average degree $\lambda$ to $15$, $20$, $30$, or $40$, fixing $t=0$ and $\beta = 0.2$.
\item Out-in ratio: set $\beta$ to $0$, $0.25$, or $0.5$,  fixing $\lambda = 40$ and $t = 0$.
\end{enumerate}

\begin{table}
\def~{\hphantom{0}}
\tbl{Overall model selection by two cross-validation methods (fraction correct out of 200 replications). The true model is the degree corrected block model.}{%
\begin{tabular}{ccccrrrr}
 \\
  \multicolumn{4}{c}{Configurations} 
&  \multicolumn{2}{c}{Proposed method } &  \multicolumn{2}{c}{\citet{chen2014network}  } \\
\hline
$K$ & $n$ & $\lambda$ & $\beta$ & $L_2$ loss & $L_2$ loss+stability  & $L_2$ loss & $L_2$ loss+stability  \\ 
  \hline
\multirow{4}{*}{3} &\multirow{4}{*}{600} &
15 &  0.2 & 0.73&  0.87  &  0.00 &  0.00 \\ 
&&20 & 0.2 &0.97 &  0.99 & 0.02 &  0.00 \\    
&&30 & 0.2 &1.00 &  1.00 & 0.43 &  0.40 \\    
&&40  & 0.2 &1.00 &  1.00  &  0.88 &  0.98 \\   
\hline 
\multirow{4}{*}{5} &\multirow{4}{*}{600} &
15  & 0.2 & 0.49 &  0.58  &  0.00 &  0.00 \\ 
&&20 & 0.2  &0.90 &  0.95 & 0.00 &  0.00 \\    
&&30 & 0.2  &0.99 &  1.00 & 0.05 &  0.01 \\    
&&40  & 0.2  &0.99 &  1.00  &  0.27 &  0.24 \\    
\hline
\multirow{4}{*}{5} &\multirow{4}{*}{1200} &
15 & 0.2 & 0.67 &  0.76  &  0.00 &  0.00 \\ 
&&20 & 0.2  &0.99 &  0.99 & 0.00 &  0.00 \\    
&&30 & 0.2  &1.00 &  1.00 & 0.04 &  0.00\\    
&&40 & 0.2  &1.00 &  1.00  &  0.41 &  0.33 \\  
\hline  
\multirow{3}{*}{3} &\multirow{3}{*}{600} &
40&0.1  & 1.00  &1.00&   0.99 &  1.00 \\    
&&40 &0.2 & 1.00 &  1.00  &  0.88 &  0.98\\    
&&40 & 0.5  &0.95 &  0.97  &  0.00 &  0.00 \\    
\hline
\multirow{3}{*}{5} &\multirow{3}{*}{600} &
40&0.1  & 1.00  &1.00&   0.79 &  0.96 \\    
&&40 &0.2 & 0.99 &  1.00  &  0.27 &  0.24\\    
&&40 & 0.5  &0.00 &  0.00  &  0.00 &  0.00 \\    
\hline
\multirow{3}{*}{5} &\multirow{3}{*}{1200} &
40&0.1  & 1.00  &1.00&   0.90 &  0.99 \\    
&&40 &0.2 & 1.00 &  1.00  &  0.41 &  0.33\\    
&&40 & 0.5  &0.00 &  0.00  &  0.00 &  0.00 \\    
\hline
\end{tabular}}
\label{tab:DCSBM-compareNCV}
\end{table}

All results are based on 200 replications. The four methods compared on this task are the ECV (Algorithm~\ref{algo:ECV}) with $L_2$ loss and its stable version where the most frequent selection of 20 independent repetitions is returned, and the corresponding versions of the procedure from \citet{chen2014network}. We only show the results from using the $L_2$ loss for model selection since we observed it works better than binomial deviance for both methods. The performance using binomial deviance as loss can be found in Supplementary Material~\ref{appendix:deviance}.

Table~\ref{tab:DCSBM-compareNCV} shows the fraction of times the correct model was selected when the true
model is the degree corrected model.    Over all settings, stability selection improves
performance as long as the single cross-validation is working reasonably well to
start with.   This is expected, since stability selection is only a
variance reduction step, and it cannot help if the original procedure
is not working.  Thought the method of   \citet{chen2014network} works well in easier settings (smaller number
of communities, denser networks, smaller
out-in ratio), it quickly loses
accuracy on model selection as the problem becomes harder.    In contrast, the ECV gives better selection in all cases,  and in harder settings the difference is very large.   In the Supplementary Material, we include the result from the experiment with the stochastic block model as the underlying truth and the message is still the same. 

Another popular model selection problem under the block models is the selection of number of communities, assuming the true model (with/without degree correction) is known. We have extensive simulation experiments on this task by comparing the two cross-validation methods above and a few other model-based methods. The details are included in the Supplementary Material. Between the two cross-validation methods, the ECV is again a clear winner. However, the model-based methods are overall more effective than cross-validation methods as expected.

\subsection{Tuning nonparametric graphon estimation}\label{secsec:graphon}

We now demonstrate the performance of ECV in tuning $\tau$ in the
neighborhood smoothing estimation for a graphon model used in   \cite{zhang2015estimating}. 

\begin{figure}
\centering
\begin{subfigure}{.328\textwidth}
  \centering
  \includegraphics[width=\linewidth]{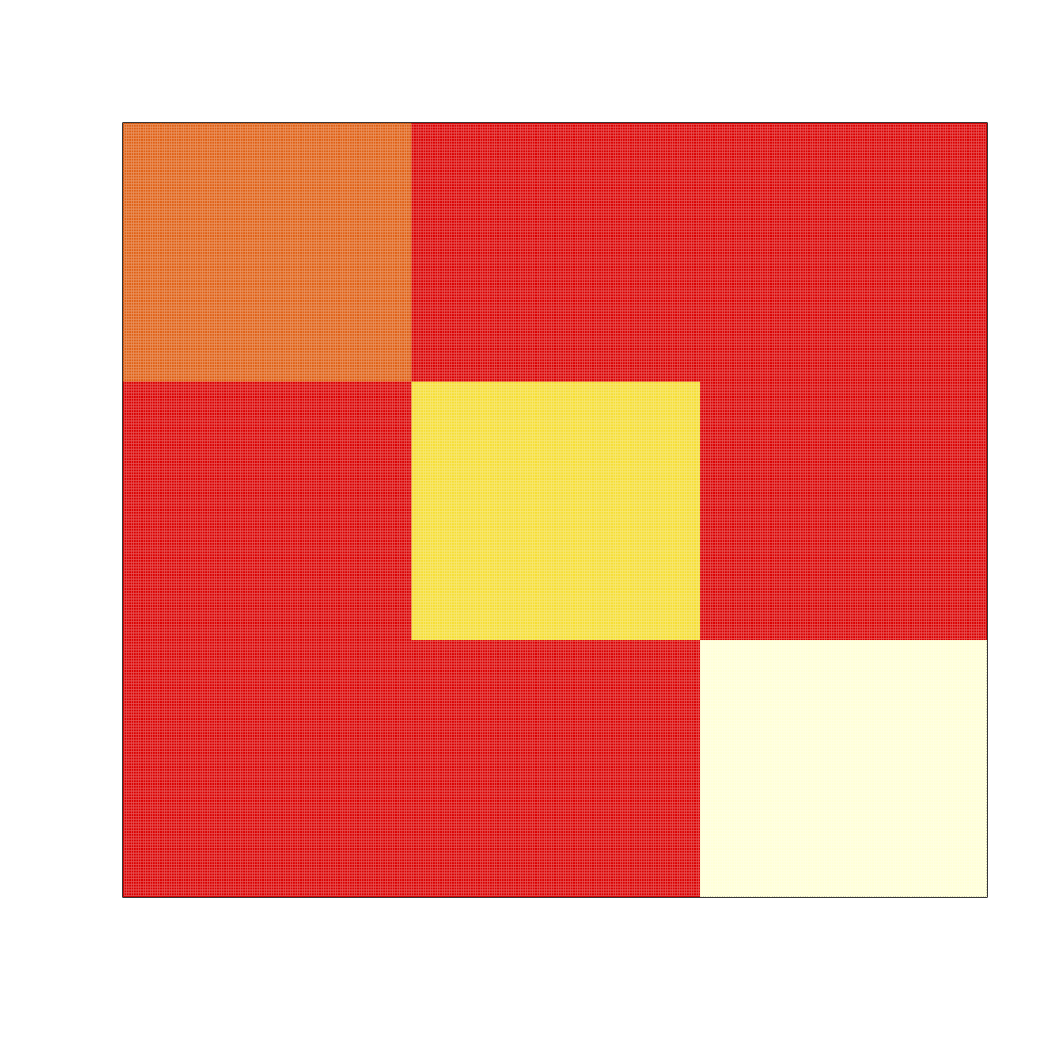}
  \vspace{-1.2cm}
  \caption{Graphon 1 heatmap}
  \label{fig:BlockGraphonMatrix}
\end{subfigure}%
\hspace{2cm}
\begin{subfigure}{.328\textwidth}
  \centering
  \includegraphics[width=\linewidth]{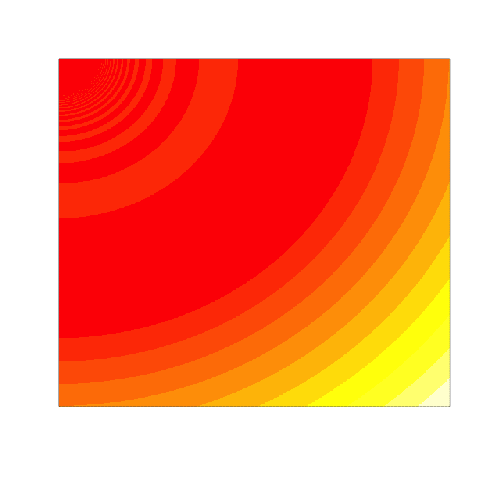}
  \vspace{-1.2cm}
  \caption{Graphon 2 heatmap}
\end{subfigure}%
\\
\begin{subfigure}{.448\textwidth}
  \centering
  \includegraphics[width=\linewidth]{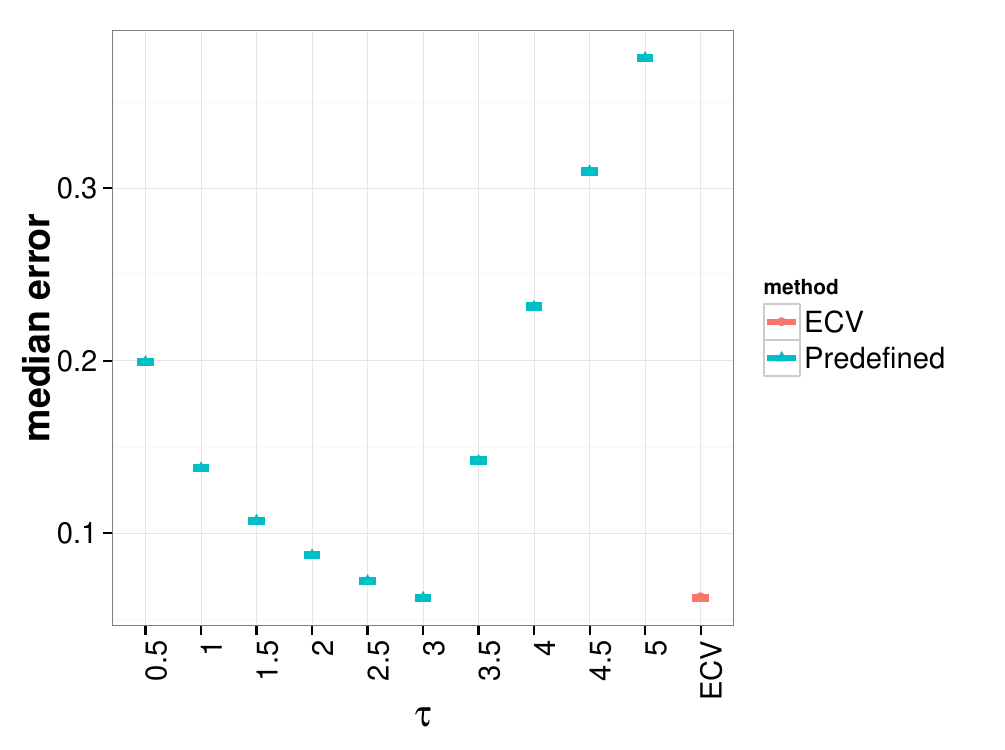}
  \caption{Graphon 1 errors}
  \label{fig:BlockGraphonTune}
\end{subfigure}%
\begin{subfigure}{.348\textwidth}
  \centering
  \includegraphics[width=\linewidth]{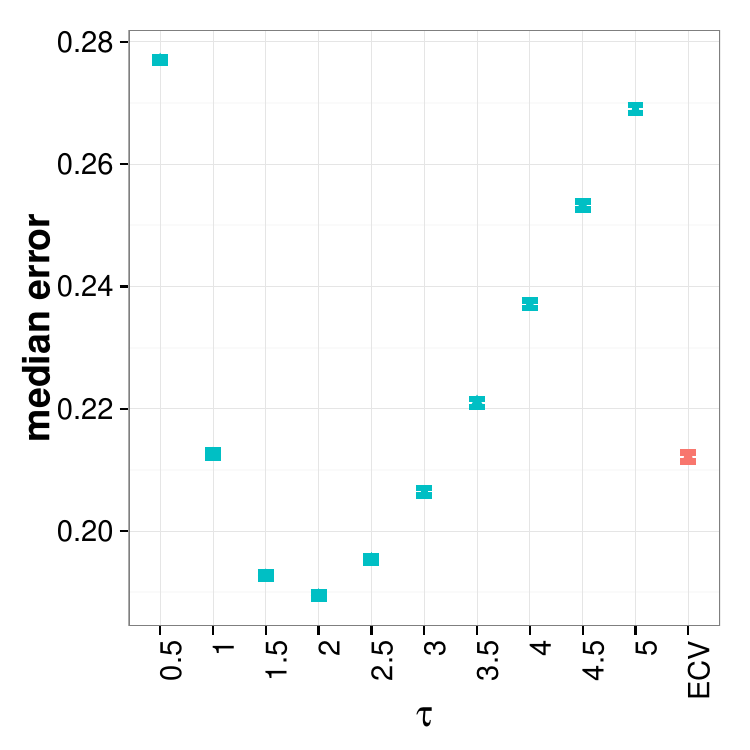}
  \caption{Graphon 2 errors}
  \label{fig:RainbowGraphonTune}
\end{subfigure}
\caption{Parameter tuning for piecewise constant graphon estimation.  }
\label{fig:Graphon}
\end{figure}

The tuning procedure is very stable for the graphon problem and
stability selection is unnecessary. Figure~\ref{fig:Graphon} shows the
tuning results for two graphon examples taken from
\cite{zhang2015estimating}, both for networks with $n=500$ nodes.
Graphon 1 is a block model (though this information is never used),
which is a piecewise constant function, and $M$ is low rank.   Graphon
2 is a smoothly varying function which is not low rank;  see
\cite{zhang2015estimating} for more details.  The errors are pictured
as the median over 200 replications with a 95\% confidence interval
(calculated by bootstrap) of the normalized Frobenius error
$\norm{\hat{M}-M}_F / \norm{M}_F$.   For Graphon 1, which is low rank,
the ECV works extremely well and picks the best $\tau$ from the
candidate set most of the time.  For Graphon 2, which is not low rank
and therefore more challenging for a procedure based on a low-rank
approximation, the ECV does not always choose the very best $\tau$,
but still achieves a fairly competitive error rate by successfully
avoiding the ``bad'' range of $\tau$.   This example illustrates that
the choice of constant can lead to a big difference in estimation error, and the ECV is successful at choosing it.  

\section{Community detection in a statistician citation network}\label{sec:app}

In this section, we demonstrate model selection on a publicly available dataset  compiled by \cite{ji2016coauthorship}.   This dataset contains information (title, author, year, citations and DOI)  about all papers published between 2003 and 2012 in four top statistics journals  (Annals of Statistics, Biometrika,  Journal of the American Statistical Association -- Theory and Methods, and Journal of the Royal Statistical Society Series B), which involves 3607 authors and 3248 papers in total.   This dataset was carefully curated  by \cite{ji2016coauthorship} to resolve name ambiguities and is relatively interpretable, at least to statisticians.

\begin{figure}
\begin{center}
\vspace{-1cm}
\includegraphics[width=0.8\textwidth]{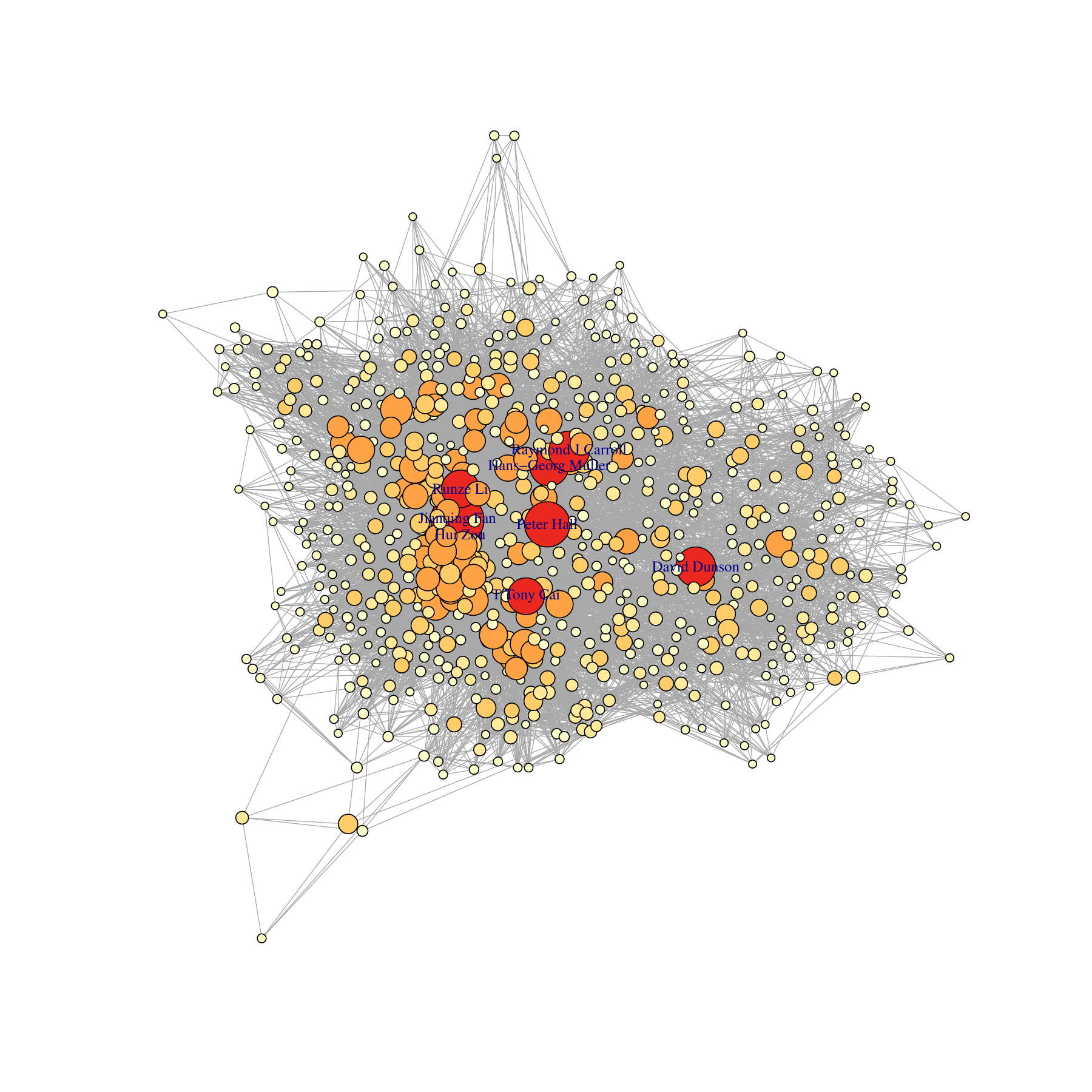}
\vspace{-2cm}
\caption{The core of statistician citation network. The network has 706 nodes with node citation count (ignoring directions) ranging from 15 to 703.  The nodes sizes and colors indicate the citation counts and the nodes with larger citation counts are larger and darker.}
\label{fig:StatisticianNet}
 \end{center}
\end{figure}

The citations of all the papers are available so we can construct the citation network between authors (as well as papers, but here we focus on authors as we are looking for research communities of people).     We thus construct a weighted undirected network between authors, where the weight is the total number of their mutual citations.  The largest connected component of the network contains 2654 authors.       Thresholding the weight to binary resulted in all methods for estimating $K$ selecting an unrealistically large and uninterpretable value, suggesting the network is too complex to be adequately described by a binary block model.    Since the weights are available and contain much more information than just the presence of an edge, we analyze the weighted network instead;   seamlessly switching between binary and weighted networks is a strength of the ECV.  
Many real world networks display a core-periphery structure, and citation networks especially are likely to have this form.    We focus on analyzing the core of the citation network, extracting it following the procedure proposed  by \cite{wang2016discussion}:   delete nodes with less than 15 mutual citations and their corresponding edges, and repeat until the network no longer changes.   This results in a network with 706 authors shown in Figure~\ref{fig:StatisticianNet}. The individual node citation count ranges from 15 to 703 with a median 30.

Block models are not defined for weighted networks, but the Laplacian is still well-defined and so the spectral clustering algorithm for community detection can be applied.   The model-free version Algorithm~\ref{algo:rank} can be used to determine the number of communities.  We apply the ECV with sum of squared error loss and repeat it 200 times, with the candidate values for $K$ from 1 to 50.  The stable version ECV selects $K=20$. We also used the ECV to tune the regularization parameter for spectral clustering, as described in Section~\ref{secsec:tuneRSC}. It turns out the regularization does make  the result more interpretable. We list the 20 communities in Table~\ref{tab:StatisticianCommunityTop10}, with each community represented by 10 authors with the largest number of citations, along with subjective and tentative names we assigned to these communities.  The names are assigned based on the majority of authors' interests or area of contributions, and that it is based exclusively on data collected in the period 2003-2012, so people who have worked on many topics over many years tend to appear under the topic they devoted the most attention to in that time period.   Many communities can be easily identified by their common research interests;   high-dimensional inference, a topic that many people published on in that period of time, is subdivided into several sub-communities that are in themselves interpretable (communities 1, 2, 4, 5, 10, 12, 15).    Overall, these groups are fairly easily interpretable to those familiar with the statistics literature of this decade.

\section{Discussion}\label{sec:con}

The general scheme of leaving out entries at random followed by matrix completion may be useful for other resampling-based methods.   In particular, an interesting future direction we plan to investigate is whether this strategy can be used to create something akin to bootstrap samples from a single network realization.     Another direction we did not explore in this paper is cross-validation under alternatives to the inhomogeneous Erd\"{o}s-Renyi model, such as \cite{crane2015framework} or \cite{lauritzen2018random}.   ECV may also be modified for the setting where additional node features are available  \citep{li2016prediction, newman2015structure}.    We leave these questions for future work.  

\begin{table}[H]
{\scriptsize
\centering
\caption{The 10 authors with largest total citation numbers (ignoring the direction) within 20 communities, as well as the community interpretations.  The communities are ordered by size and authors within a community are ordered by mutual citation count.}
\label{tab:StatisticianCommunityTop10}
\begin{tabular}{p{0.2cm}|p{4cm}|p{11cm}}
  \hline
 & Interpretation [size] & Authors \\ 
  \hline
  1 & high-dimensional inference (multiple testing, machine learning) [57]&T Tony Cai, Jiashun Jin, Larry Wasserman, Christopher Genovese, Bradley Efron, John D Storey, David L Donoho, Yoav Benjamini, Jonathan E Taylor, Joseph P Romano \\ 
  \hline
  2 & high-dimensional inference (sparse penalties) [53]& Hui Zou, Ming Yuan, Yi Lin, Trevor J Hastie, Robert J Tibshirani, Xiaotong Shen, Jinchi Lv, Gareth M James, Hongzhe Li, Peter Radchenko \\ 
  \hline
 3 & functional data analysis [52]& Hans-Georg Muller, Jane-Ling Wang, Fang Yao, Yehua Li, Ciprian M Crainiceanu, Jeng-Min Chiou, Alois Kneip, Hulin Wu, Piotr Kokoszka, Tailen Hsing \\ 
   \hline
  4 & high-dimensional inference (theory and sparsity)  [45]& Peter Buhlmann, Nicolai Meinshausen, Cun-Hui Zhang, Alexandre B Tsybakov, Emmanuel J Candes, Terence Tao, Marten H Wegkamp, Bin Yu, Florentina Bunea, Martin J Wainwright \\ 
  \hline
  5 & high-dimensional covariance estimation [43]&Peter J Bickel, Ji Zhu, Elizaveta Levina, Jianhua Z Huang, Mohsen Pourahmadi, Clifford Lam, Wei Biao Wu, Adam J Rothman, Weidong Liu, Linxu Liu \\ 
  \hline
  6 & Bayesian machine learning [41]& David Dunson, Alan E Gelfand, Abel Rodriguez, Michael I Jordan, Peter Muller, Gareth Roberts, Gary L Rosner, Omiros Papaspiliopoulos, Steven N MacEachern, Ju-Hyun Park \\ 
  \hline
  7 &  spatial statistics  [41]& Tilmann Gneiting, Marc G Genton, Sudipto Banerjee, Adrian E Raftery, Haavard Rue, Andrew O Finley, Bo Li, Michael L Stein, Nicolas Chopin, Hao Zhang \\ 
  \hline
  8 & biostatistics (machine learning) [40]&Donglin Zeng, Dan Yu Lin, Michael R Kosorok, Jason P Fine, Jing Qin, Guosheng Yin, Guang Cheng, Yi Li, Kani Chen, Yu Shen \\ 
  \hline
9 & sufficent dimension reduction [39] & Lixing Zhu, R Dennis Cook, Bing Li, Chih-Ling Tsai, Liping Zhu, Yingcun Xia, Lexin Li, Liqiang Ni, Francesca Chiaromonte, Liugen Xue \\ 
\hline
  10 & high-dimensional inference (penalized methods) [38]& Jianqing Fan, Runze Li, Hansheng Wang, Jian Huang, Heng Peng, Song Xi Chen, Chenlei Leng, Shuangge Ma, Xuming He, Wenyang Zhang \\ 
  \hline
  11 & Bayesian (general) [33] & Jeffrey S Morris, James O Berger, Carlos M Carvalho, James G Scott, Hemant Ishwaran, Marina Vannucci, Philip J Brown, J Sunil Rao, Mike West, Nicholas G Polson \\ 
  \hline
  12 & high-dimensional theory and wavelets [33]& Iain M Johnstone, Bernard W Silverman, Felix Abramovich, Ian L Dryden, Dominique Picard, Richard Nickl, Holger Dette, Marianna Pensky, Piotr Fryzlewicz, Theofanis Sapatinas \\ 
  \hline
  13 &  mixed (causality + theory + Bayesian) [32]& James R Robins, Christian P Robert, Paul Fearnhead, Gilles Blanchard, Zhiqiang Tan, Stijn Vansteelandt, Nancy Reid, Jae Kwang Kim, Tyler J VanderWeele, Scott A Sisson \\ 
  \hline
  14 & semiparametrics and nonparametrics [28]& Hua Liang, Naisyin Wang, Joel L Horowitz, Xihong Lin, Enno Mammen, Arnab Maity, Byeong U Park, Wolfgang Karl Hardle, Jianhui Zhou, Zongwu Cai \\ 
  \hline
  15 & high-dimensional inference (machine learning)  [27]&Hao Helen Zhang, J S Marron, Yufeng Liu, Yichao Wu, Jeongyoun Ahn, Wing Hung Wong, Peter L Bartlett, Michael J Todd, Amnon Neeman, Jon D McAuliffe \\ 
  \hline
  16 & semiparametrics [24]& Peter Hall, Raymond J Carroll, Yanyuan Ma, Aurore Delaigle, Gerda Claeskens, David Ruppert, Alexander Meister, Huixia Judy Wang, Nilanjan Chatterjee, Anastasios A Tsiatis \\ 
  \hline
 17 & mixed (causality + financial) [22]& Qiwei Yao, Paul R Rosenbaum, Yacine Ait-Sahalia, Yazhen Wang, Marc Hallin, Dylan S Small, Davy Paindaveine, Jian Zou, Per Aslak Mykland, Jean Jacod \\
  \hline 
   18 & biostatistics (survival, clinical trials) [22]& L J Wei, Lu Tian, Tianxi Cai, Zhiliang Ying, Zhezhen Jin, Peter X-K Song, Hui Li, Bin Nan, Hajime Uno, Jun S Liu \\ 
  \hline
  19 & biostatistics - genomics [21]& Joseph G Ibrahim, Hongtu Zhu, Jiahua Chen, Amy H Herring, Heping Zhang, Ming-Hui Chen, Stuart R Lipsitz, Denis Heng-Yan Leung, Weili Lin, Armin Schwartzman \\ 
  \hline
  20 & Bayesian (nonparametrics) [15]& Subhashis Ghosal, Igor Prunster, Antonio Lijoi, Stephen G Walker, Aad van der Vaart, Anindya Roy, Judith Rousseau, J H van Zanten, Richard Samworth, Aad W van der Vaart \\ 
  \hline
\end{tabular}
}
\end{table}

\section*{Acknowledgement}
This research was partially supported by NSF grant
DMS-1521551 and ONR grant N000141612910 (to E. Levina), and NSF grants DMS-1407698 and DMS-1821243 (to J. Zhu).  

\bibliographystyle{biometrika}
\bibliography{CommonBib}

\begin{thebibliography}{28}
\expandafter\ifx\csname natexlab\endcsname\relax\def\natexlab#1{#1}\fi

\bibitem[{Amini et~al.(2013)Amini, Chen, Bickel \& Levina}]{amini2013pseudo}
\textsc{Amini, A.~A.}, \textsc{Chen, A.}, \textsc{Bickel, P.~J.} \&
  \textsc{Levina, E.} (2013).
\newblock Pseudo-likelihood methods for community detection in large sparse
  networks.
\newblock \textit{The Annals of Statistics} \textbf{41}, 2097--2122.

\bibitem[{Athreya et~al.(2017)Athreya, Fishkind, Levin, Lyzinski, Park, Qin,
  Sussman, Tang, Vogelstein \& Priebe}]{athreya2017statistical}
\textsc{Athreya, A.}, \textsc{Fishkind, D.~E.}, \textsc{Levin, K.},
  \textsc{Lyzinski, V.}, \textsc{Park, Y.}, \textsc{Qin, Y.}, \textsc{Sussman,
  D.~L.}, \textsc{Tang, M.}, \textsc{Vogelstein, J.~T.} \& \textsc{Priebe,
  C.~E.} (2017).
\newblock Statistical inference on random dot product graphs: a survey.
\newblock \textit{arXiv preprint arXiv:1709.05454} .

\bibitem[{Bandeira \& van Handel(2016)}]{bandeira2016sharp}
\textsc{Bandeira, A.~S.} \& \textsc{van Handel, R.} (2016).
\newblock Sharp nonasymptotic bounds on the norm of random matrices with
  independent entries.
\newblock \textit{The Annals of Probability} \textbf{44}, 2479--2506.

\bibitem[{Bhaskar \& Javanmard(2015)}]{bhaskar20151}
\textsc{Bhaskar, S.~A.} \& \textsc{Javanmard, A.} (2015).
\newblock 1-bit matrix completion under exact low-rank constraint.
\newblock In \textit{Information Sciences and Systems (CISS), 2015 49th Annual
  Conference on}. IEEE.

\bibitem[{Bhojanapalli \& Jain(2014)}]{bhojanapalli2014universal}
\textsc{Bhojanapalli, S.} \& \textsc{Jain, P.} (2014).
\newblock Universal matrix completion.
\newblock In \textit{Proceedings of The 31st International Conference on
  Machine Learning}.

\bibitem[{Boyd \& Vandenberghe(2004)}]{boyd2004convex}
\textsc{Boyd, S.} \& \textsc{Vandenberghe, L.} (2004).
\newblock \textit{Convex optimization}.
\newblock Cambridge University Press.

\bibitem[{Chaudhuri et~al.(2012)Chaudhuri, Graham \&
  Tsiatas}]{chaudhuri2012spectral}
\textsc{Chaudhuri, K.}, \textsc{Graham, F.~C.} \& \textsc{Tsiatas, A.} (2012).
\newblock Spectral clustering of graphs with general degrees in the extended
  planted partition model.
\newblock In \textit{COLT}, vol.~23.

\bibitem[{Chen \& Lei(2018)}]{chen2014network}
\textsc{Chen, K.} \& \textsc{Lei, J.} (2018).
\newblock Network cross-validation for determining the number of communities in
  network data.
\newblock \textit{Journal of the American Statistical Association}
  \textbf{113}, 241--251.

\bibitem[{Chi \& Li(2019)}]{chi2019matrix}
\textsc{Chi, E.~C.} \& \textsc{Li, T.} (2019).
\newblock Matrix completion from a computational statistics perspective.
\newblock \textit{Wiley Interdisciplinary Reviews: Computational Statistics} ,
  e1469.

\bibitem[{Chin et~al.(2015)Chin, Rao \& Vu}]{chin2015stochastic}
\textsc{Chin, P.}, \textsc{Rao, A.} \& \textsc{Vu, V.} (2015).
\newblock Stochastic block model and community detection in sparse graphs: A
  spectral algorithm with optimal rate of recovery.
\newblock In \textit{Conference on Learning Theory}.

\bibitem[{Eldridge et~al.(2017)Eldridge, Belkin \&
  Wang}]{eldridge2017unperturbed}
\textsc{Eldridge, J.}, \textsc{Belkin, M.} \& \textsc{Wang, Y.} (2017).
\newblock Unperturbed: spectral analysis beyond {Davis-Kahan}.
\newblock \textit{arXiv preprint arXiv:1706.06516} .

\bibitem[{Gao et~al.(2017)Gao, Ma, Zhang \& Zhou}]{gao2015achieving}
\textsc{Gao, C.}, \textsc{Ma, Z.}, \textsc{Zhang, A.~Y.} \& \textsc{Zhou,
  H.~H.} (2017).
\newblock Achieving optimal misclassification proportion in stochastic block
  models.
\newblock \textit{The Journal of Machine Learning Research} \textbf{18},
  1980--2024.

\bibitem[{Hastie \& Mazumder(2015)}]{softImpute}
\textsc{Hastie, T.} \& \textsc{Mazumder, R.} (2015).
\newblock \textit{softImpute: Matrix Completion via Iterative Soft-Thresholded
  SVD}.
\newblock R package version 1.4.

\bibitem[{Jin(2015)}]{jin2015fast}
\textsc{Jin, J.} (2015).
\newblock Fast community detection by {SCORE}.
\newblock \textit{The Annals of Statistics} \textbf{43}, 57--89.

\bibitem[{Joseph \& Yu(2016)}]{joseph2016impact}
\textsc{Joseph, A.} \& \textsc{Yu, B.} (2016).
\newblock Impact of regularization on spectral clustering.
\newblock \textit{The Annals of Statistics} \textbf{44}, 1765--1791.

\bibitem[{Klopp(2015)}]{klopp2015matrix}
\textsc{Klopp, O.} (2015).
\newblock Matrix completion by singular value thresholding: sharp bounds.
\newblock \textit{Electronic Journal of Statistics} \textbf{9}, 2348--2369.

\bibitem[{Le \& Levina(2015)}]{le2015estimating}
\textsc{Le, C.~M.} \& \textsc{Levina, E.} (2015).
\newblock Estimating the number of communities in networks by spectral methods.
\newblock \textit{arXiv preprint arXiv:1507.00827} .

\bibitem[{Le et~al.(2017)Le, Levina \& Vershynin}]{le2017concentration}
\textsc{Le, C.~M.}, \textsc{Levina, E.} \& \textsc{Vershynin, R.} (2017).
\newblock Concentration and regularization of random graphs.
\newblock \textit{Random Structures \& Algorithms} .

\bibitem[{Lei \& Rinaldo(2014)}]{lei2014consistency}
\textsc{Lei, J.} \& \textsc{Rinaldo, A.} (2014).
\newblock Consistency of spectral clustering in stochastic block models.
\newblock \textit{The Annals of Statistics} \textbf{43}, 215--237.

\bibitem[{Mazumder et~al.(2010)Mazumder, Hastie \&
  Tibshirani}]{mazumder2010spectral}
\textsc{Mazumder, R.}, \textsc{Hastie, T.} \& \textsc{Tibshirani, R.} (2010).
\newblock Spectral regularization algorithms for learning large incomplete
  matrices.
\newblock \textit{The Journal of Machine Learning Research} \textbf{11},
  2287--2322.

\bibitem[{Saldana et~al.(2017)Saldana, Yu \& Feng}]{saldana2014how}
\textsc{Saldana, D.}, \textsc{Yu, Y.} \& \textsc{Feng, Y.} (2017).
\newblock How many communities are there?
\newblock \textit{Journal of Computational and Graphical Statistics}
  \textbf{26}, 171--181.

\bibitem[{Srebro \& Shraibman(2005)}]{srebro2005rank}
\textsc{Srebro, N.} \& \textsc{Shraibman, A.} (2005).
\newblock Rank, trace-norm and max-norm.
\newblock In \textit{International Conference on Computational Learning
  Theory}. Springer.

\bibitem[{Sugar \& James(2003)}]{sugar2003finding}
\textsc{Sugar, C.~A.} \& \textsc{James, G.~M.} (2003).
\newblock Finding the number of clusters in a dataset.
\newblock \textit{Journal of the American Statistical Association} \textbf{98}.

\bibitem[{Tibshirani \& Walther(2005)}]{tibshirani2005cluster}
\textsc{Tibshirani, R.} \& \textsc{Walther, G.} (2005).
\newblock Cluster validation by prediction strength.
\newblock \textit{Journal of Computational and Graphical Statistics}
  \textbf{14}, 511--528.

\bibitem[{Tibshirani et~al.(2001)Tibshirani, Walther \&
  Hastie}]{tibshirani2001estimating}
\textsc{Tibshirani, R.}, \textsc{Walther, G.} \& \textsc{Hastie, T.} (2001).
\newblock Estimating the number of clusters in a data set via the gap
  statistic.
\newblock \textit{Journal of the Royal Statistical Society: Series B
  (Statistical Methodology)} \textbf{63}, 411--423.

\bibitem[{Wang \& Bickel(2017)}]{wang2015likelihood}
\textsc{Wang, Y.~R.} \& \textsc{Bickel, P.~J.} (2017).
\newblock Likelihood-based model selection for stochastic block models.
\newblock \textit{The Annals of Statistics} \textbf{45}, 500--528.

\bibitem[{Yao(2003)}]{yao2003information}
\textsc{Yao, Y.} (2003).
\newblock Information-theoretic measures for knowledge discovery and data
  mining.
\newblock In \textit{Entropy Measures, Maximum Entropy Principle and Emerging
  Applications}. Springer, pp. 115--136.

\bibitem[{Young \& Scheinerman(2007)}]{young2007random}
\textsc{Young, S.~J.} \& \textsc{Scheinerman, E.~R.} (2007).
\newblock Random dot product graph models for social networks.
\newblock In \textit{International Workshop on Algorithms and Models for the
  Web-Graph}. Springer.

\end{thebibliography}


\begin{thebibliography}{79}
\expandafter\ifx\csname natexlab\endcsname\relax\def\natexlab#1{#1}\fi

\bibitem[{Abbe(2018)}]{abbe2017community}
\textsc{Abbe, E.} (2018).
\newblock Community detection and stochastic block models: Recent developments.
\newblock \textit{Journal of Machine Learning Research} \textbf{18}, 1--86.

\bibitem[{Abbe et~al.(2017)Abbe, Fan, Wang \& Zhong}]{abbe2017entrywise}
\textsc{Abbe, E.}, \textsc{Fan, J.}, \textsc{Wang, K.} \& \textsc{Zhong, Y.}
  (2017).
\newblock Entrywise eigenvector analysis of random matrices with low expected
  rank.
\newblock \textit{arXiv preprint arXiv:1709.09565} .

\bibitem[{Airoldi et~al.(2008)Airoldi, Blei, Fienberg \&
  Xing}]{airoldi2008mixed}
\textsc{Airoldi, E.~M.}, \textsc{Blei, D.~M.}, \textsc{Fienberg, S.~E.} \&
  \textsc{Xing, E.~P.} (2008).
\newblock Mixed membership stochastic blockmodels.
\newblock \textit{Journal of Machine Learning Research} \textbf{9}, 1981--2014.

\bibitem[{Aldous(1981)}]{aldous1981representations}
\textsc{Aldous, D.~J.} (1981).
\newblock Representations for partially exchangeable arrays of random
  variables.
\newblock \textit{Journal of Multivariate Analysis} \textbf{11}, 581--598.

\bibitem[{Amini et~al.(2013)Amini, Chen, Bickel \& Levina}]{amini2013pseudo}
\textsc{Amini, A.~A.}, \textsc{Chen, A.}, \textsc{Bickel, P.~J.} \&
  \textsc{Levina, E.} (2013).
\newblock Pseudo-likelihood methods for community detection in large sparse
  networks.
\newblock \textit{The Annals of Statistics} \textbf{41}, 2097--2122.

\bibitem[{Athreya et~al.(2017)Athreya, Fishkind, Levin, Lyzinski, Park, Qin,
  Sussman, Tang, Vogelstein \& Priebe}]{athreya2017statistical}
\textsc{Athreya, A.}, \textsc{Fishkind, D.~E.}, \textsc{Levin, K.},
  \textsc{Lyzinski, V.}, \textsc{Park, Y.}, \textsc{Qin, Y.}, \textsc{Sussman,
  D.~L.}, \textsc{Tang, M.}, \textsc{Vogelstein, J.~T.} \& \textsc{Priebe,
  C.~E.} (2017).
\newblock Statistical inference on random dot product graphs: a survey.
\newblock \textit{arXiv preprint arXiv:1709.05454} .

\bibitem[{Bandeira \& van Handel(2016)}]{bandeira2016sharp}
\textsc{Bandeira, A.~S.} \& \textsc{van Handel, R.} (2016).
\newblock Sharp nonasymptotic bounds on the norm of random matrices with
  independent entries.
\newblock \textit{The Annals of Probability} \textbf{44}, 2479--2506.

\bibitem[{Bhaskar \& Javanmard(2015)}]{bhaskar20151}
\textsc{Bhaskar, S.~A.} \& \textsc{Javanmard, A.} (2015).
\newblock 1-bit matrix completion under exact low-rank constraint.
\newblock In \textit{Information Sciences and Systems (CISS), 2015 49th Annual
  Conference on}. IEEE.

\bibitem[{Bhojanapalli \& Jain(2014)}]{bhojanapalli2014universal}
\textsc{Bhojanapalli, S.} \& \textsc{Jain, P.} (2014).
\newblock Universal matrix completion.
\newblock In \textit{Proceedings of The 31st International Conference on
  Machine Learning}.

\bibitem[{Bickel et~al.(2013)Bickel, Choi, Chang \&
  Zhang}]{bickel2013asymptotic}
\textsc{Bickel, P.}, \textsc{Choi, D.}, \textsc{Chang, X.} \& \textsc{Zhang,
  H.} (2013).
\newblock Asymptotic normality of maximum likelihood and its variational
  approximation for stochastic blockmodels.
\newblock \textit{The Annals of Statistics} \textbf{41}, 1922--1943.

\bibitem[{Bickel \& Sarkar(2016)}]{bickel2013Hypothesis}
\textsc{Bickel, P.~J.} \& \textsc{Sarkar, P.} (2016).
\newblock Hypothesis testing for automated community detection in networks.
\newblock \textit{Journal of the Royal Statistical Society: Series B
  (Statistical Methodology)} \textbf{78}, 253--273.

\bibitem[{Boyd \& Vandenberghe(2004)}]{boyd2004convex}
\textsc{Boyd, S.} \& \textsc{Vandenberghe, L.} (2004).
\newblock \textit{Convex optimization}.
\newblock Cambridge University Press.

\bibitem[{Cai \& Zhou(2013)}]{cai2013max}
\textsc{Cai, T.} \& \textsc{Zhou, W.-X.} (2013).
\newblock A max-norm constrained minimization approach to 1-bit matrix
  completion.
\newblock \textit{The Journal of Machine Learning Research} \textbf{14},
  3619--3647.

\bibitem[{Candes \& Plan(2010)}]{candes2010matrix}
\textsc{Candes, E.~J.} \& \textsc{Plan, Y.} (2010).
\newblock Matrix completion with noise.
\newblock \textit{Proceedings of the IEEE} \textbf{98}, 925--936.

\bibitem[{Cand{\`e}s \& Recht(2009)}]{candes2009exact}
\textsc{Cand{\`e}s, E.~J.} \& \textsc{Recht, B.} (2009).
\newblock Exact matrix completion via convex optimization.
\newblock \textit{Foundations of Computational mathematics} \textbf{9},
  717--772.

\bibitem[{Cand{\`e}s \& Tao(2010)}]{candes2010power}
\textsc{Cand{\`e}s, E.~J.} \& \textsc{Tao, T.} (2010).
\newblock The power of convex relaxation: Near-optimal matrix completion.
\newblock \textit{IEEE Transactions on Information Theory} \textbf{56},
  2053--2080.

\bibitem[{Chatterjee(2015)}]{chatterjee2015matrix}
\textsc{Chatterjee, S.} (2015).
\newblock Matrix estimation by universal singular value thresholding.
\newblock \textit{The Annals of Statistics} \textbf{43}, 177--214.

\bibitem[{Chaudhuri et~al.(2012)Chaudhuri, Graham \&
  Tsiatas}]{chaudhuri2012spectral}
\textsc{Chaudhuri, K.}, \textsc{Graham, F.~C.} \& \textsc{Tsiatas, A.} (2012).
\newblock Spectral clustering of graphs with general degrees in the extended
  planted partition model.
\newblock In \textit{COLT}, vol.~23.

\bibitem[{Chen \& Lei(2018)}]{chen2014network}
\textsc{Chen, K.} \& \textsc{Lei, J.} (2018).
\newblock Network cross-validation for determining the number of communities in
  network data.
\newblock \textit{Journal of the American Statistical Association}
  \textbf{113}, 241--251.

\bibitem[{Chen et~al.(2015)Chen, Bhojanapalli, Sanghavi \&
  Ward}]{chen2015completing}
\textsc{Chen, Y.}, \textsc{Bhojanapalli, S.}, \textsc{Sanghavi, S.} \&
  \textsc{Ward, R.} (2015).
\newblock Completing any low-rank matrix, provably.
\newblock \textit{The Journal of Machine Learning Research} \textbf{16},
  2999--3034.

\bibitem[{Chi \& Li(2019)}]{chi2019matrix}
\textsc{Chi, E.~C.} \& \textsc{Li, T.} (2019).
\newblock Matrix completion from a computational statistics perspective.
\newblock \textit{Wiley Interdisciplinary Reviews: Computational Statistics} ,
  e1469.

\bibitem[{Chin et~al.(2015)Chin, Rao \& Vu}]{chin2015stochastic}
\textsc{Chin, P.}, \textsc{Rao, A.} \& \textsc{Vu, V.} (2015).
\newblock Stochastic block model and community detection in sparse graphs: A
  spectral algorithm with optimal rate of recovery.
\newblock In \textit{Conference on Learning Theory}.

\bibitem[{Choi \& Wolfe(2014)}]{choi2014co}
\textsc{Choi, D.} \& \textsc{Wolfe, P.~J.} (2014).
\newblock Co-clustering separately exchangeable network data.
\newblock \textit{The Annals of Statistics} \textbf{42}, 29--63.

\bibitem[{Chung \& Lu(2002)}]{chung2002average}
\textsc{Chung, F.} \& \textsc{Lu, L.} (2002).
\newblock The average distances in random graphs with given expected degrees.
\newblock \textit{Proceedings of the National Academy of Sciences} \textbf{99},
  15879--15882.

\bibitem[{Crane \& Dempsey(2018)}]{crane2015framework}
\textsc{Crane, H.} \& \textsc{Dempsey, W.} (2018).
\newblock Edge exchangeable models for interaction networks.
\newblock \textit{Journal of the American Statistical Association}
  \textbf{113}, 1311--1326.

\bibitem[{Davenport et~al.(2014)Davenport, Plan, van~den Berg \&
  Wootters}]{davenport20141}
\textsc{Davenport, M.~A.}, \textsc{Plan, Y.}, \textsc{van~den Berg, E.} \&
  \textsc{Wootters, M.} (2014).
\newblock 1-bit matrix completion.
\newblock \textit{Information and Inference} \textbf{3}, 189--223.

\bibitem[{Diaconis \& Janson(2007)}]{diaconis2007graph}
\textsc{Diaconis, P.} \& \textsc{Janson, S.} (2007).
\newblock Graph limits and exchangeable random graphs.
\newblock \textit{arXiv preprint arXiv:0712.2749} .

\bibitem[{Eldridge et~al.(2017)Eldridge, Belkin \&
  Wang}]{eldridge2017unperturbed}
\textsc{Eldridge, J.}, \textsc{Belkin, M.} \& \textsc{Wang, Y.} (2017).
\newblock Unperturbed: spectral analysis beyond {Davis-Kahan}.
\newblock \textit{arXiv preprint arXiv:1706.06516} .

\bibitem[{Erd\"os \& R{\'e}nyi(1960)}]{erds1960evolution}
\textsc{Erd\"os, P.} \& \textsc{R{\'e}nyi, A.} (1960).
\newblock On the evolution of random graphs.
\newblock \textit{Publ. Math. Inst. Hung. Acad. Sci} \textbf{5}, 17--61.

\bibitem[{Gao et~al.(2016)Gao, Lu, Ma \& Zhou}]{gao2016optimal}
\textsc{Gao, C.}, \textsc{Lu, Y.}, \textsc{Ma, Z.} \& \textsc{Zhou, H.~H.}
  (2016).
\newblock Optimal estimation and completion of matrices with biclustering
  structures.
\newblock \textit{Journal of Machine Learning Research} \textbf{17}, 1--29.

\bibitem[{Gao et~al.(2015)Gao, Lu \& Zhou}]{gao2015rate}
\textsc{Gao, C.}, \textsc{Lu, Y.} \& \textsc{Zhou, H.~H.} (2015).
\newblock Rate-optimal graphon estimation.
\newblock \textit{The Annals of Statistics} \textbf{43}, 2624--2652.

\bibitem[{Gao et~al.(2017)Gao, Ma, Zhang \& Zhou}]{gao2015achieving}
\textsc{Gao, C.}, \textsc{Ma, Z.}, \textsc{Zhang, A.~Y.} \& \textsc{Zhou,
  H.~H.} (2017).
\newblock Achieving optimal misclassification proportion in stochastic block
  models.
\newblock \textit{The Journal of Machine Learning Research} \textbf{18},
  1980--2024.

\bibitem[{Hastie \& Mazumder(2015)}]{softImpute}
\textsc{Hastie, T.} \& \textsc{Mazumder, R.} (2015).
\newblock \textit{softImpute: Matrix Completion via Iterative Soft-Thresholded
  SVD}.
\newblock R package version 1.4.

\bibitem[{Hoff(2008)}]{hoff2008modeling}
\textsc{Hoff, P.} (2008).
\newblock Modeling homophily and stochastic equivalence in symmetric relational
  data.
\newblock In \textit{Advances in Neural Information Processing Systems}.

\bibitem[{Hoff et~al.(2002)Hoff, Raftery \& Handcock}]{hoff2002latent}
\textsc{Hoff, P.~D.}, \textsc{Raftery, A.~E.} \& \textsc{Handcock, M.~S.}
  (2002).
\newblock Latent space approaches to social network analysis.
\newblock \textit{Journal of the American Statistical Association} \textbf{97},
  1090--1098.

\bibitem[{Holland et~al.(1983)Holland, Laskey \&
  Leinhardt}]{holland1983stochastic}
\textsc{Holland, P.~W.}, \textsc{Laskey, K.~B.} \& \textsc{Leinhardt, S.}
  (1983).
\newblock Stochastic blockmodels: First steps.
\newblock \textit{Social Networks} \textbf{5}, 109--137.

\bibitem[{Ji \& Jin(2016)}]{ji2016coauthorship}
\textsc{Ji, P.} \& \textsc{Jin, J.} (2016).
\newblock Coauthorship and citation networks for statisticians.
\newblock \textit{The Annals of Applied Statistics} \textbf{10}, 1779--1812.

\bibitem[{Jin(2015)}]{jin2015fast}
\textsc{Jin, J.} (2015).
\newblock Fast community detection by {SCORE}.
\newblock \textit{The Annals of Statistics} \textbf{43}, 57--89.

\bibitem[{Joseph \& Yu(2016)}]{joseph2016impact}
\textsc{Joseph, A.} \& \textsc{Yu, B.} (2016).
\newblock Impact of regularization on spectral clustering.
\newblock \textit{The Annals of Statistics} \textbf{44}, 1765--1791.

\bibitem[{Kanagal \& Sindhwani(2010)}]{kanagal2010rank}
\textsc{Kanagal, B.} \& \textsc{Sindhwani, V.} (2010).
\newblock Rank selection in low-rank matrix approximations: A study of
  cross-validation for {NMFs}.
\newblock In \textit{Advances in Neural Information Processing Systems},
  vol.~1.

\bibitem[{Karrer \& Newman(2011)}]{karrer2011stochastic}
\textsc{Karrer, B.} \& \textsc{Newman, M.~E.} (2011).
\newblock Stochastic blockmodels and community structure in networks.
\newblock \textit{Physical Review E} \textbf{83}, 016107.

\bibitem[{Keshavan et~al.(2009)Keshavan, Montanari \& Oh}]{keshavan2009matrix}
\textsc{Keshavan, R.}, \textsc{Montanari, A.} \& \textsc{Oh, S.} (2009).
\newblock Matrix completion from noisy entries.
\newblock In \textit{Advances in Neural Information Processing Systems}.

\bibitem[{Klopp(2015)}]{klopp2015matrix}
\textsc{Klopp, O.} (2015).
\newblock Matrix completion by singular value thresholding: sharp bounds.
\newblock \textit{Electronic Journal of Statistics} \textbf{9}, 2348--2369.

\bibitem[{Latouche et~al.(2012)Latouche, Birmele \&
  Ambroise}]{latouche2012variational}
\textsc{Latouche, P.}, \textsc{Birmele, E.} \& \textsc{Ambroise, C.} (2012).
\newblock Variational {Bayesian} inference and complexity control for
  stochastic block models.
\newblock \textit{Statistical Modelling} \textbf{12}, 93--115.

\bibitem[{Lauritzen et~al.(2018)Lauritzen, Rinaldo \&
  Sadeghi}]{lauritzen2018random}
\textsc{Lauritzen, S.}, \textsc{Rinaldo, A.} \& \textsc{Sadeghi, K.} (2018).
\newblock Random networks, graphical models and exchangeability.
\newblock \textit{Journal of the Royal Statistical Society: Series B
  (Statistical Methodology)} \textbf{80}, 481--508.

\bibitem[{Le \& Levina(2015)}]{le2015estimating}
\textsc{Le, C.~M.} \& \textsc{Levina, E.} (2015).
\newblock Estimating the number of communities in networks by spectral methods.
\newblock \textit{arXiv preprint arXiv:1507.00827} .

\bibitem[{Le et~al.(2017)Le, Levina \& Vershynin}]{le2017concentration}
\textsc{Le, C.~M.}, \textsc{Levina, E.} \& \textsc{Vershynin, R.} (2017).
\newblock Concentration and regularization of random graphs.
\newblock \textit{Random Structures \& Algorithms} .

\bibitem[{Lei(2016)}]{lei2016goodness}
\textsc{Lei, J.} (2016).
\newblock A goodness-of-fit test for stochastic block models.
\newblock \textit{The Annals of Statistics} \textbf{44}, 401--424.

\bibitem[{Lei(2017)}]{lei2017cross}
\textsc{Lei, J.} (2017).
\newblock Cross-validation with confidence.
\newblock \textit{arXiv preprint arXiv:1703.07904} .

\bibitem[{Lei \& Rinaldo(2014)}]{lei2014consistency}
\textsc{Lei, J.} \& \textsc{Rinaldo, A.} (2014).
\newblock Consistency of spectral clustering in stochastic block models.
\newblock \textit{The Annals of Statistics} \textbf{43}, 215--237.

\bibitem[{Li et~al.(2019)Li, Levina \& Zhu}]{li2016prediction}
\textsc{Li, T.}, \textsc{Levina, E.} \& \textsc{Zhu, J.} (2019).
\newblock Prediction models for network-linked data.
\newblock \textit{The Annals of Applied Statistics} \textbf{13}, 132--164.

\bibitem[{Mazumder et~al.(2010)Mazumder, Hastie \&
  Tibshirani}]{mazumder2010spectral}
\textsc{Mazumder, R.}, \textsc{Hastie, T.} \& \textsc{Tibshirani, R.} (2010).
\newblock Spectral regularization algorithms for learning large incomplete
  matrices.
\newblock \textit{The Journal of Machine Learning Research} \textbf{11},
  2287--2322.

\bibitem[{McDaid et~al.(2013)McDaid, Murphy, Friel \&
  Hurley}]{mcdaid2013improved}
\textsc{McDaid, A.~F.}, \textsc{Murphy, T.~B.}, \textsc{Friel, N.} \&
  \textsc{Hurley, N.~J.} (2013).
\newblock Improved {Bayesian} inference for the stochastic block model with
  application to large networks.
\newblock \textit{Computational Statistics \& Data Analysis} \textbf{60},
  12--31.

\bibitem[{Meinshausen \& B{\"u}hlmann(2010)}]{meinshausen2010stability}
\textsc{Meinshausen, N.} \& \textsc{B{\"u}hlmann, P.} (2010).
\newblock Stability selection.
\newblock \textit{Journal of the Royal Statistical Society: Series B
  (Statistical Methodology)} \textbf{72}, 417--473.

\bibitem[{Newman \& Clauset(2016)}]{newman2015structure}
\textsc{Newman, M.~E.} \& \textsc{Clauset, A.} (2016).
\newblock Structure and inference in annotated networks.
\newblock \textit{Nature Communications} \textbf{7}.

\bibitem[{Owen \& Perry(2009)}]{owen2009bi}
\textsc{Owen, A.~B.} \& \textsc{Perry, P.} (2009).
\newblock Bi-cross-validation of the svd and the nonnegative matrix
  factorization.
\newblock \textit{The Annals of Applied Statistics} \textbf{3}, 564--594.

\bibitem[{Qin \& Rohe(2013)}]{qin2013regularized}
\textsc{Qin, T.} \& \textsc{Rohe, K.} (2013).
\newblock Regularized spectral clustering under the degree-corrected stochastic
  blockmodel.
\newblock In \textit{Advances in Neural Information Processing Systems}.

\bibitem[{Rohe et~al.(2011)Rohe, Chatterjee \& Yu}]{rohe2011spectral}
\textsc{Rohe, K.}, \textsc{Chatterjee, S.} \& \textsc{Yu, B.} (2011).
\newblock Spectral clustering and the high-dimensional stochastic blockmodel.
\newblock \textit{The Annals of Statistics} \textbf{39}, 1878--1915.

\bibitem[{Saldana et~al.(2017)Saldana, Yu \& Feng}]{saldana2014how}
\textsc{Saldana, D.}, \textsc{Yu, Y.} \& \textsc{Feng, Y.} (2017).
\newblock How many communities are there?
\newblock \textit{Journal of Computational and Graphical Statistics}
  \textbf{26}, 171--181.

\bibitem[{Sarkar \& Bickel(2015)}]{sarkar2015role}
\textsc{Sarkar, P.} \& \textsc{Bickel, P.~J.} (2015).
\newblock Role of normalization in spectral clustering for stochastic
  blockmodels.
\newblock \textit{The Annals of Statistics} \textbf{43}, 962--990.

\bibitem[{Sengupta \& Chen(2018)}]{sengupta2018block}
\textsc{Sengupta, S.} \& \textsc{Chen, Y.} (2018).
\newblock A block model for node popularity in networks with community
  structure.
\newblock \textit{Journal of the Royal Statistical Society: Series B
  (Statistical Methodology)} \textbf{80}, 365--386.

\bibitem[{Shao(1993)}]{shao1993linear}
\textsc{Shao, J.} (1993).
\newblock Linear model selection by cross-validation.
\newblock \textit{Journal of the American Statistical Association} \textbf{88},
  486--494.

\bibitem[{Srebro \& Shraibman(2005)}]{srebro2005rank}
\textsc{Srebro, N.} \& \textsc{Shraibman, A.} (2005).
\newblock Rank, trace-norm and max-norm.
\newblock In \textit{International Conference on Computational Learning
  Theory}. Springer.

\bibitem[{Su et~al.(2017)Su, Wang \& Zhang}]{su2017strong}
\textsc{Su, L.}, \textsc{Wang, W.} \& \textsc{Zhang, Y.} (2017).
\newblock Strong consistency of spectral clustering for stochastic block
  models.
\newblock \textit{arXiv preprint arXiv:1710.06191} .

\bibitem[{Sugar \& James(2003)}]{sugar2003finding}
\textsc{Sugar, C.~A.} \& \textsc{James, G.~M.} (2003).
\newblock Finding the number of clusters in a dataset.
\newblock \textit{Journal of the American Statistical Association} \textbf{98}.

\bibitem[{Sussman et~al.(2014)Sussman, Tang \& Priebe}]{sussman2014consistent}
\textsc{Sussman, D.~L.}, \textsc{Tang, M.} \& \textsc{Priebe, C.~E.} (2014).
\newblock Consistent latent position estimation and vertex classification for
  random dot product graphs.
\newblock \textit{IEEE transactions on pattern analysis and machine
  intelligence} \textbf{36}, 48--57.

\bibitem[{Tang et~al.(2017)Tang, Athreya, Sussman, Lyzinski \&
  Priebe}]{tang2017nonparametric}
\textsc{Tang, M.}, \textsc{Athreya, A.}, \textsc{Sussman, D.~L.},
  \textsc{Lyzinski, V.} \& \textsc{Priebe, C.~E.} (2017).
\newblock A nonparametric two-sample hypothesis testing problem for random
  graphs.
\newblock \textit{Bernoulli} \textbf{23}, 1599--1630.

\bibitem[{Tang \& Priebe(2018)}]{tang2016limit}
\textsc{Tang, M.} \& \textsc{Priebe, C.~E.} (2018).
\newblock Limit theorems for eigenvectors of the normalized {Laplacian} for
  random graphs.
\newblock \textit{The Annals of Statistics} \textbf{46}, 2360--2415.

\bibitem[{Tibshirani \& Walther(2005)}]{tibshirani2005cluster}
\textsc{Tibshirani, R.} \& \textsc{Walther, G.} (2005).
\newblock Cluster validation by prediction strength.
\newblock \textit{Journal of Computational and Graphical Statistics}
  \textbf{14}, 511--528.

\bibitem[{Tibshirani et~al.(2001)Tibshirani, Walther \&
  Hastie}]{tibshirani2001estimating}
\textsc{Tibshirani, R.}, \textsc{Walther, G.} \& \textsc{Hastie, T.} (2001).
\newblock Estimating the number of clusters in a data set via the gap
  statistic.
\newblock \textit{Journal of the Royal Statistical Society: Series B
  (Statistical Methodology)} \textbf{63}, 411--423.

\bibitem[{Wang \& Rohe(2016)}]{wang2016discussion}
\textsc{Wang, S.} \& \textsc{Rohe, K.} (2016).
\newblock Discussion of ``coauthorship and citation networks for
  statisticians".
\newblock \textit{The Annals of Applied Statistics} \textbf{10}, 1820--1826.

\bibitem[{Wang \& Bickel(2017)}]{wang2015likelihood}
\textsc{Wang, Y.~R.} \& \textsc{Bickel, P.~J.} (2017).
\newblock Likelihood-based model selection for stochastic block models.
\newblock \textit{The Annals of Statistics} \textbf{45}, 500--528.

\bibitem[{Wolfe \& Olhede(2013)}]{wolfe2013nonparametric}
\textsc{Wolfe, P.~J.} \& \textsc{Olhede, S.~C.} (2013).
\newblock Nonparametric graphon estimation.
\newblock \textit{arXiv preprint arXiv:1309.5936} .

\bibitem[{Yang(2007)}]{yang2007consistency}
\textsc{Yang, Y.} (2007).
\newblock Consistency of cross validation for comparing regression procedures.
\newblock \textit{The Annals of Statistics} \textbf{35}, 2450--2473.

\bibitem[{Yao(2003)}]{yao2003information}
\textsc{Yao, Y.} (2003).
\newblock Information-theoretic measures for knowledge discovery and data
  mining.
\newblock In \textit{Entropy Measures, Maximum Entropy Principle and Emerging
  Applications}. Springer, pp. 115--136.

\bibitem[{Young \& Scheinerman(2007)}]{young2007random}
\textsc{Young, S.~J.} \& \textsc{Scheinerman, E.~R.} (2007).
\newblock Random dot product graph models for social networks.
\newblock In \textit{International Workshop on Algorithms and Models for the
  Web-Graph}. Springer.

\bibitem[{Zhang(1993)}]{zhang1993model}
\textsc{Zhang, P.} (1993).
\newblock Model selection via multifold cross validation.
\newblock \textit{The Annals of Statistics} \textbf{21}, 299--313.

\bibitem[{Zhang et~al.(2017)Zhang, Levina \& Zhu}]{zhang2015estimating}
\textsc{Zhang, Y.}, \textsc{Levina, E.} \& \textsc{Zhu, J.} (2017).
\newblock Estimating network edge probabilities by neighbourhood smoothing.
\newblock \textit{Biometrika} \textbf{104}, 771--783.

\bibitem[{Zhao et~al.(2012)Zhao, Levina \& Zhu}]{zhao2012consistency}
\textsc{Zhao, Y.}, \textsc{Levina, E.} \& \textsc{Zhu, J.} (2012).
\newblock Consistency of community detection in networks under degree-corrected
  stochastic block models.
\newblock \textit{The Annals of Statistics} \textbf{40}, 2266--2292.

\end{thebibliography}

\begin{appendix}

\section{Proofs}\label{sec:proof}

We start with additional notation.   For any vector $\V{x}$, we use $\norm{\V{x}}$ to denote its Euclidean norm. We denote the singular values of a matrix $M$ by $\sigma_1(M)\ge\sigma_2(M)\ge \cdots \sigma_K(M) >  \sigma_{K+1}(M) = \sigma_{K+2}(M) \cdots \sigma_{n}(M) = 0$, where $K=\rank(M)$.   Recall the Frobenius norm $\norm{M}_F$ is defined by $\norm{M}_F^2 = \sum_{ij}M_{ij}^2 = \sum_i\sigma_i(M)^2$, the spectral norm $\norm{M} = \sigma_1(M)$, the infinity norm  $\norm{M}_{\infty} = \max_{ij}|M_{ij}|$,  and the nuclear norm $\norm{M}_* = \sum_i \sigma_{i}(M)$ be the nuclear norm. In addition, the max norm of $M$ \citep{srebro2005rank} is defined as
$$\norm{M}_{\max} = \min_{M = UV^T}\max(\norm{U}_{2,\infty}^2, \norm{V}_{2,\infty}^2),$$
where $\norm{U}_{2,\infty} = \max_i (\sum_{j}U_{ij}^2)^{1/2}$.

We will need the following well-known inequalities: 
\begin{align}
\norm{M} \le \norm{M}_F  \le \sqrt{K}\norm{M}, \label{eq:norms} \\
\norm{M}_F  \le \norm{M}_* \le \sqrt{K}\norm{M}_F \label{eq:nuclearFrobenius}\\
|\tr(M_1^TM_2)| \le \norm{M_1}\norm{M_2}_* \label{eq:duality}\\
\max(\norm{M^T}_{2,\infty}, \norm{M}_{2,\infty}) \le \norm{M} \label{eq:2toInfty}\\
\norm{M}_{\max} \le \sqrt{K}\norm{M}_{\infty}. \label{eq:maxnorm}
\end{align}
Relationship \eqref{eq:duality}, which holds for any two matrices $M_1$, $M_2$ with matching dimensions, is called norm duality for the spectral norm and the nuclear norm \citep{boyd2004convex}.  Relationship \eqref{eq:maxnorm} can be found in \cite{srebro2005rank}. The last one we need is the variational property of spectral norm:
\begin{equation}\label{eq:variational}
\norm{M} = \max_{\V{x}, \V{y} \in \bR^n: \norm{\V{x}}=\norm{\V{y}}=1}\V{y}^TM\V{x}.
\end{equation}

\subsection{Proof of Theorem~\ref{thm:simplebound}}

Our proof will rely on a concentration result for the adjacency matrix.   To the best of our knowledge, Lemma~\ref{lemma:concentration} stated next is the best concentration bound currently available, proved by \cite{lei2014consistency}. The same concentration was also obtained by \cite{chin2015stochastic} and \cite{le2017concentration}.

\begin{lemma}
\label{lemma:concentration}
Let $A$ be the adjacency matrix of a random graph on $n$ nodes with independent edges. Set $\e(A) = P = [p_{ij}]_{n\times n}$ and assume that $n\max_{ij}p_{ij}\le d$ for $d \ge C_0\log n$ and $C_0>0$. Then for any $\delta>0$, there exists a constant $C = C(\delta, C_0)$ such that 
$$\norm{A - P} \le C\sqrt{d}$$
with probability at least $1-n^{-\delta}$.
\end{lemma}

Another tool we need is the discrepancy between a bounded matrix and its partially observed version given in Lemma \ref{lemma:descrepancy}, which can be viewed as a generalization of Theorem 4.1 of \cite{bhojanapalli2014universal} and Lemma 6.4 of \cite{bhaskar20151} to the more realistic uniform missing mechanism in the matrix completion problem.  Let $G \in \bR^{n\times n}$ be the indicator matrix associated with the hold-out set $\Omega$, such that if  $(i,j) \in \Omega$,  $G_{ij}=0$ and otherwise $G_{ij}=1$. Note that under the uniform missing mechanism, $G$ can be viewed as an adjacency matrix of an Erd\"{o}s-Renyi random graph where all edges appear independently with probability $p$. Note that $P_{\Omega}A = A\circ G$ where $\circ$ is the Hadamard (element-wise) matrix product.
\begin{lemma}\label{lemma:descrepancy}
Let $G$ an adjacency matrix of an Erd\"{o}s-Renyi graph with the probability of edge  $p \ge C_1\log n / n$ for a constant $C_1$. Then for any $\delta > 0$, with probability at least $1-n^{-\delta}$, the following relationship holds for any $Z \in \bR^{n\times n}$ with $\rank(Z) \le K$ 
$$\left\| \frac{1}{p}Z \circ G - Z \right\|  \le 2C\sqrt{\frac{nK}{p}}\norm{Z}_{\infty} $$
where $C = C(\delta, C_1)$ is the constant from Lemma~\ref{lemma:concentration} that only depends on $\delta$ and $C_1$.
\end{lemma}
\begin{proof}[(Proof of Lemma~\ref{lemma:descrepancy})]
Let $Z= UV^T$, where $U \in \bR^{n\times K}$ and $V \in \bR^{n\times K}$ are the matrices that achieve the minimum in the definition of $\norm{Z}_{\max}$.    Denote the $\ell$th column of $U$ by $U_{\cdot \ell}$ and the $\ell$th row by $U_{\ell \cdot}$.

Given any unit vectors $\V{x}, \V{y} \in \bR^n$, we have
\begin{align}\label{eq:variational1}
\V{y}^T \left( \frac{1}{p}Z\circ G - Z \right)\V{x} 
& = \sum_{\ell}\bigg[\frac{1}{p}\V{y}^T(U_{\cdot \ell}V_{\cdot \ell}^T)\circ G \V{x} - (\V{y}^TU_{\cdot \ell})(\V{x}^TV_{\cdot \ell})\bigg]  \notag\\
&= \sum_{\ell}\bigg[\frac{1}{p}(\V{y}\circ U_{\cdot\ell})^T G  (\V{x}\circ V_{\cdot \ell}) - (\V{y}^TU_{\cdot \ell})(\V{x}^TV_{\cdot \ell})\bigg] . 
\end{align}

Let $\tilde{\mbone} = \mbone_n/\sqrt{n}$ be the constant unit vector. For any $1 \le \ell \le n $, let $\V{y}\circ U_{\cdot \ell} = \alpha_{\ell}\tilde{\mbone}  + \beta_{\ell}\tilde{\mbone}_{\perp}^{\ell}$ in which $\tilde{\mbone}_{\perp}^{\ell}$ is a vector that is orthogonal to $\tilde{\mbone}$.   It is easy to check that
$$\alpha_{\ell} = (\V{y}\circ U_{\cdot \ell})^T\tilde{\mbone} = \frac{1}{\sqrt{n}}\V{y}^TU_{\cdot \ell}.$$
Similarly, we also have
$$(\V{x}\circ V_{\cdot \ell})^T\tilde{\mbone} = \frac{1}{\sqrt{n}}\V{x}^TV_{\cdot \ell}.$$

Let  $\bar{G} = p\mbone\mbone^T$ be the expectation of $G$ with respect to the missing mechanism.   Then 
\begin{align}\label{eq:AI}
(&\V{y}\circ U_{\cdot\ell})^TG(\V{x}\circ V_{\cdot \ell}) = \frac{1}{\sqrt{n}}(\V{y}^TU_{\cdot \ell})\tilde{\mbone}^TG(\V{x}\circ  V_{\cdot \ell}) +\beta_{\ell}\tilde{\mbone}_{\perp}^{\ell~T}G(\V{x}\circ V_{\cdot \ell})\notag \\
& = \frac{1}{\sqrt{n}}(\V{y}^TU_{\cdot \ell})\tilde{\mbone}^T\bar{G}(\V{x}\circ V_{\cdot \ell}) + \frac{1}{\sqrt{n}}(\V{y}^TU_{\cdot \ell})\tilde{\mbone}^T(G-\bar{G})(\V{x}\circ V_{\cdot \ell}) +\beta_{\ell}\tilde{\mbone}_{\perp}^{\ell~T} G (\V{x}\circ V_{\cdot \ell})\ . 
\end{align}
Notice that  $\tilde{\mbone}^T\bar{G} = np\tilde{\mbone}^T$, and therefore 
$$ \frac{1}{\sqrt{n}}(\V{y}^TU_{\cdot \ell})\tilde{\mbone}^T\bar{G}(\V{x}\circ V_{\cdot \ell}) 
= \frac{np}{\sqrt{n}}(\V{y}^TU_{\cdot \ell})\tilde{\mbone}^T(\V{x}\circ V_{\cdot \ell}) \ . 
  = p(\V{y}^TU_{\cdot \ell})(\V{x}^TV_{\cdot \ell}) 
$$

%
Further, since $\bar{G}\tilde{\mbone}_{\perp}^{\ell} = 0$ for any $\ell$, we can rewrite \eqref{eq:AI} as 
\begin{align}
(\V{y}\circ U_{\cdot\ell})^TG(\V{x}\circ V_{\cdot \ell}) &= p(\V{y}^TU_{\cdot \ell})(\V{x}^TV_{\cdot \ell})\notag \\
 &~ + \frac{1}{\sqrt{n}}(\V{y}^TU_{\cdot \ell})\tilde{\mbone}^T(G-\bar{G})(\V{x}\circ V_{\cdot \ell}) +\beta_{\ell}\tilde{\mbone}_{\perp}^{\ell~T}(G-\bar{G})(\V{x}\circ V_{\cdot \ell}) \ . \label{eq:AI2}
\end{align}

%
%
%
%
Substituting \eqref{eq:AI2} into \eqref{eq:variational1}  and applying \eqref{eq:variational} and the Cauchy-Schwarz inequality leads to
\begin{align}\label{eq:variational2}
\V{y}^T & (\frac{1}{p}P_{\Omega}Z - Z)\V{x} =\frac{1}{p}\sum_{\ell}\bigg[\frac{1}{\sqrt{n}}(\V{y}^TU_{\cdot \ell})\tilde{\mbone}^T(G-\bar{G})(\V{x}\circ V_{\cdot \ell}) +\beta_{\ell}\tilde{\mbone}_{\perp}^{\ell~T}(G-\bar{G})(\V{x}\circ V_{\cdot \ell})\bigg] \notag \\
& \le \frac{1}{p}\norm{G-\bar{G}}\bigg[\sum_{\ell}\frac{1}{\sqrt{n}}|\V{y}^TU_{\cdot \ell}|\norm{\V{x}\circ V_{\cdot \ell}} + \sum_{\ell}|\beta_{\ell}|\norm{\V{x}\circ V_{\cdot \ell}} \bigg]\notag \\
& \le \frac{1}{p}\norm{G-\bar{G}}\bigg[\frac{1}{\sqrt{n}}\sqrt{\sum_{\ell}(\V{y}^TU_{\cdot \ell})^2}\sqrt{\sum_{\ell}\norm{\V{x}\circ V_{\cdot \ell}}^2} + \sqrt{\sum_{\ell}\beta_{\ell}^2}\sqrt{\sum_{\ell}\norm{\V{x}\circ V_{\cdot \ell}}^2}\bigg].
\end{align}
Using Cauchy-Schwarz inequality, the definition of max norm and the relationship \eqref{eq:maxnorm}, we get
\begin{align}\label{eq:termbound1}
\sum_{\ell}(\V{y}^TU_{\cdot \ell})^2   \le \sum_{\ell}\norm{\V{y}}^2\norm{U_{\cdot \ell}}^2 
& = \norm{U}_F^2 \le n\norm{U}_{2,\infty}^2  \le n\norm{Z}_{\max}  \le n\sqrt{K}\norm{Z}_{\infty} \ . 
\end{align}
Similarly, 
\begin{align}\label{eq:termbound2}
\sum_{\ell}\beta_{\ell}^2 &= \sum_{\ell}(\tilde{\mbone}_{\perp}^{\ell~T}(\V{y}\circ U_{\cdot \ell}))^2  \le \sum_{\ell}\norm{\V{y}\circ U_{\cdot \ell}}^2 \notag\\
&= \sum_{\ell}\sum_iy_i^2U_{il}^2 \le \norm{U}_{2,\infty}^2\sum_i{y_i^2} \le \norm{Z}_{\max}\le \sqrt{K}\norm{Z}_{\infty}.
\end{align}
 We also have
\begin{align}\label{eq:termbound3}
\sum_{\ell}\norm{\V{x}\circ V_{\cdot \ell}}^2 \le \sqrt{K}\norm{Z}_{\infty}.
\end{align}
Combining \eqref{eq:termbound1}, \eqref{eq:termbound2} and \eqref{eq:termbound3} with \eqref{eq:variational2}, we get
\begin{equation}\label{eq:variational3}
\V{y}^T(\frac{1}{p}P_{\Omega}Z - Z)\V{x} \le \frac{2\sqrt{K}}{p}\norm{G-\bar{G}}\norm{Z}_{\infty}.
\end{equation}
From \eqref{eq:variational}, we have
$$\norm{\frac{1}{p}P_{\Omega}Z - Z} = \sup_{\norm{\V{x}}=\norm{\V{y}}=1}\V{y}^T(\frac{1}{p}P_{\Omega}Z - Z)\V{x} \le \frac{2\sqrt{K}}{p}\norm{G-\bar{G}}\norm{Z}_{\infty}.$$
Finally, Lemma~\ref{lemma:concentration} implies
\begin{equation}\label{eq:edconcentration}
\norm{G-\bar{G}} \le C(\delta, C_1)\sqrt{pn} 
\end{equation}
with probability at least $1-n^{-\delta}$ defined in Lemma~\ref{lemma:concentration}.  Therefore, with probability at least $1-n^{-\delta}$, 
$$\norm{\frac{1}{p}P_{\Omega}Z - Z} \le 2C(\delta, C_1)\sqrt{\frac{nK}{p}}\norm{Z}_{\infty}. $$
\end{proof}

The following lemma is from \cite{klopp2015matrix}; see also Corollary 3.3 of \cite{bandeira2016sharp} for a more general result. 
\begin{lemma}[Proposition 13 of \citep{klopp2015matrix}]\label{lemma:klopp-spectral}
Let $X$ be an $n\times n$ matrix with each entry $X_{ij}$ being independent and bounded random variables, such that $\max_{ij}|X_{ij}| \le \sigma$ with probability 1. Then for any $\delta>0$, 
$$\norm{X} \le C'\max(\sigma_1, \sigma_2, \sqrt{\log n})$$
in which $C' = C'(\sigma, \delta)$ is a constant that only depends on $\delta$ and $\sigma$, 
$$\sigma_1 = \max_i\sqrt{\e \sum_j X_{ij}^2} \ , \ \sigma_2 = \max_j\sqrt{\e \sum_i X_{ij}^2}.$$
\end{lemma}

\begin{proof}[(Proof of Theorem~\ref{thm:simplebound})]
Our proof is valid regardless of whether the network is directed, as Lemma~\ref{lemma:concentration} holds for both directed and undirected networks. So we ignore the fact that $M$ can be symmetric.
Let $W=A-M$, so $\e W = 0$. It is known that
\begin{equation}\label{eq:SVD-spectral}
S_H\left(\frac{1}{p}P_{\Omega}A,K\right) = \argmin_{M: \rank(M)\le K} \norm{\frac{1}{p}P_{\Omega}A-M}.
\end{equation}
Therefore, we have
\begin{align*}
\norm{\hat{A}-M} & = \norm{\hat{A}-\frac{1}{p}P_{\Omega}A+\frac{1}{p}P_{\Omega}A - M}\\
&\le \norm{\frac{1}{p}P_{\Omega}A-\hat{A}} + \norm{\frac{1}{p}P_{\Omega}A-M} \le 2\norm{\frac{1}{p}P_{\Omega}A-M}\\
&\le 2\norm{\frac{1}{p}P_{\Omega}M-M+\frac{1}{p}P_{\Omega}W}  \le 2\norm{\frac{1}{p}P_{\Omega}M-M} + \frac{2}{p}\norm{P_{\Omega}W}\\
&= 2\norm{\frac{1}{p}G\circ M-M} + \frac{2}{p}\norm{G\circ W} := \I + \II.
\end{align*}
Since $\rank(M)\le K$, by Lemma~\ref{lemma:descrepancy}, we have
\begin{equation}\label{eq:meanError}
\I \le 4C(\delta, C_1)\sqrt{\frac{nK}{p}}\norm{Z}_{\infty} \le 4C(\delta, C_1)\sqrt{\frac{Kd^2}{np}}
\end{equation}
with probability at least $1-n^{-\delta}$ for any $\delta>0$.

We want to apply the result of Lemma~\ref{lemma:klopp-spectral} to control $\II$, by conditioning on $W$. Notice that $(G\circ W)_{ij} = \eta_{ij}W_{ij}$ where $\eta_{ij} \sim B(p)$. Clearly we can set $\sigma = 1$ in the lemma. Also, 
\begin{align*}
\sigma_1 & = \max_i\sqrt{\e(\sum_j \eta_{ij}^2W_{ij}^2|W)} = \max_i\sqrt{ \sum_j W_{ij}^2\e(\eta_{ij}^2|W)}\\
& = \max_i\sqrt{p}\sqrt{ \sum_j W_{ij}^2} = \max_i\sqrt{p}\sqrt{ \norm{W_{i\cdot}}_2^2}\\
&= \sqrt{p}\sqrt{\norm{W}_{2,\infty}^2} \le \sqrt{p}\norm{W}
\end{align*}
in which the last inequality comes from \eqref{eq:2toInfty}. Similarly, we have
$$\sigma_2 = \max_j\sqrt{\e(\sum_i \eta_{ij}^2W_{ij}^2|W)} \le \sqrt{p}\norm{W}.$$
Now by  Lemma~\ref{lemma:klopp-spectral}, we know that given $W$,
\begin{equation}\label{eq:ConditionalResidualProjection}
\II = \frac{2}{p}\norm{G\circ W} \le \frac{2}{p}C'(\delta)(\sqrt{p}\norm{W} \vee \sqrt{\log n})
\end{equation}
with probability at least $1-n^{-\delta}$ where $C'(\delta)$ is the $C'(1,\delta)$ in Lemma~\ref{lemma:klopp-spectral}. 

Finally, applying Lemma~\ref{lemma:concentration} to \eqref{eq:ConditionalResidualProjection}, we have for any $\delta_2, \delta_3>0$
\begin{equation}\label{eq:ResidualProjection}
\II \le \frac{2}{p}C'(\delta)\max(C(\delta, C_2)\sqrt{p}\sqrt{d}, \sqrt{\log n}) \le C''(\delta, C_2)\frac{\max(\sqrt{pd}, \sqrt{\log n})}{p}
\end{equation}
with probability at least $1-2n^{-\delta}$ where $C''(\delta, C_2) = 2C'(\delta)\max(C(\delta, C_2),1)$.

Combining \eqref{eq:meanError} and $\eqref{eq:ResidualProjection}$ gives

$$
\norm{\hat{A}-M} \le \I+\II \le  \tilde{C}\max(\sqrt{\frac{Kd^2}{np}}, \sqrt{\frac{d}{p}}, \frac{\sqrt{\log n}}{p}  )
$$
with probability at least $1-3n^{-\delta}$ where $\tilde{C}(\delta, C_1, C_2) = 4C(\delta, C_1) + C''(\delta, C_2)$.

The bound about Frobenius norm \eqref{eq:generalerr_Frob} directly comes from \eqref{eq:norms} since $\rank(\hat{A}-M) \le 2K$.

\end{proof}

\subsection{Proofs for block models}

We first proceed to show the community detection result based on each ECV split of Algorithm~\ref{algo:ECV}. Many different versions of $K$-means can be used in spectral clustering. Here we state the result for the version of $K$-means used by \cite{lei2014consistency}.

\begin{proposition}[Community recovery for each ECV split under the stochastic block model]\label{prop:SBMperf}
Let $A$ be the adjacency matrix of a network generated from a stochastic block model  satisfying  Assumption~\ref{ass:SBM_basic} with $K$ blocks, and $M = \e A$. Let $\hat{A}$ be the recovered adjacency matrix in \eqref{eq:simpleHS}. Assume the expected node degree $\lambda_n \ge C\log(n)$. Let $\hat{\V{c}}$ be the output of spectral clustering on $\hat{A}$. Then $\hat{\V{c}}$ coincides with the true $\V{c}$ on all but $O(n\lambda_n^{-1})$  nodes within each of the $K$ communities (up to a permutation of block labels), with probability tending to one. 
\end{proposition}

To state an analogous result for the degree corrected model, we need one more standard assumption on the degree parameters, similar to \cite{jin2015fast, lei2014consistency, chen2014network}.
\begin{ass}\label{ass:DCSBM-Degree}
$\min_i \theta_i \ge \theta_0$ for some constant $\theta_0 >0$ and $\sum_{i:c_i=k}\theta_i = 1$ for all $k \in [K]$.
\end{ass}

\begin{proposition}[Community recovery for each ECV split under the degree corrected block model]\label{prop:DCSBMperf}
Let $A$ be an adjacency matrix from a degree corrected block model satisfying Assumption~\ref{ass:SBM_basic} and \ref{ass:DCSBM-Degree}  with $K$ blocks, and $M = \e A$. Let $\hat{A}$ be the recovered adjacency matrix in \eqref{eq:simpleHS}.  Assume the expected node degree $\lambda_n \ge C\log(n)$.  Let $\hat{\V{c}}$ be the output of spherical spectral clustering on $\hat{A}$. Then $\hat{\V{c}}$ coincides with the true $\V{c}$ on all but $O(n\lambda_n^{-1/2})$  nodes  within each of the $K$ communities (up to a permutation of block labels), with probability tending to one.

\end{proposition}

\begin{proof}[(Proof of Proposition~\ref{prop:SBMperf} and \ref{prop:DCSBMperf})]
A direct consequence of Theorem~\ref{thm:simplebound} is the concentration bound
$$\norm{\hat{A}-M} \le C\sqrt{d}$$
with high probability. Then the conclusion of Proposition~\ref{prop:SBMperf}  can be proved following the strategy of Corollary 3.2 of \cite{lei2014consistency}. The same concentration bound also holds for the degree corrected model. To prove Proposition~\ref{prop:DCSBMperf},  recall that $n_k = |\{i: c_i=k\}|$.    Following \cite{lei2014consistency}, define $\V{\theta}_k = \{\theta_i\}_{c_i=k}$ and
$$\nu_k = \frac{1}{n_k^2}\sum_{i:c_i=k}\frac{\norm{\V{\theta}_k}^2}{\theta_i^2}.$$
Let $\tilde{n}_k = \norm{\V{\theta}_k}^2$ be the ``effective size" of the $k$th community. 
Under Assumption~\ref{ass:DCSBM-Degree}, we have
$$\nu_k \le \frac{1}{n_k^2}\sum_{i:c_i=k}\frac{n_k}{\theta_0^2} = \frac{1}{\theta_0^2}.$$
Furthermore, when Assumption~\ref{ass:SBM_basic} and \ref{ass:DCSBM-Degree} hold, we have
\begin{equation}\label{eq:dcorder}
\frac{\sum_kn_k^2\nu_k^2}{\min_k\tilde{n}_k^2}\le \frac{\sum_kn_k^2\nu_k^2}{\min_kn_k^2\theta_0^4} \le \frac{\sum_kn_k^2}{\gamma^2\theta_0^8} \le \frac{K}{\gamma^2\theta_0^8} = O(1).
\end{equation}
Proposition~\ref{prop:DCSBMperf} can then be proved by following the proof of Corollary 4.3 of \cite{lei2014consistency} and applying \eqref{eq:dcorder}.
\end{proof}

Next we prove model selection consistency.    For a true community label vector $c$, define $G_{k} = \{i: c_i = k\}$, and similarly let  $\hat{G}_k$ be communities corresponding to an estimated label vector $\hat{c}$.   For any $\hat{c}$ for which the number of communities is smaller than the true $K$, we have the following basic observation.
\begin{lemma}\label{lem:pigeon}
Assume the network is drawn from the stochastic block model with $K$ communities satisfying Assumption~\ref{ass:SBM_basic}, and consider one split of ECV.  Suppose we cluster the nodes into $K'$ communities, where $K' < K$. Define $I_{k_1k_2} = (G_{k_1}\times G_{k_2})\cap \Omega^c$ and $\hat{I}_{k_1k_2} = (\hat{G}_{k_1}\times \hat{G}_{k_2})\cap \Omega^c$. Then with probability tending to 1, there must exist $l_1, l_2, l_3\in [K]$ and $k_1, k_2 \in [K']$ such that
\begin{enumerate}
\item $|\hat{I}_{k_1k_2}\cap I_{l_1l_2}| \ge \tilde{c}n^2$
\item $|\hat{I}_{k_1k_2}\cap I_{l_1l_3}| \ge  \tilde{c}n^2$
\item $B_{0,(l_1l_2)} \ne B_{0,(l_1l_3)}$ where $B_{0,(ij)}$ denotes the $(i,j)$-th element of $B_0$.
\end{enumerate}
\end{lemma}
\begin{proof}
We first prove a uniform bound on the test sample size in any partition of communities with size at least $\frac{\gamma n}{K}$, where $\gamma$ is the constant in Assumption~\ref{ass:SBM_basic}.   Consider two subsets $S_i \subset G_i$ and $S_j \subset G_j$ with $|S_i| = n_{S_i} \ge \frac{\gamma n}{K}$ and $|S_j| = n_{S_j} \ge \frac{\gamma n}{K}$. Without loss of generality, assume $i\ne j$; otherwise the bound is only different by a factor of 2. We know that the cardinality of the test set within $S_i \times S_j$, given by $|(S_i\times S_j)\cap \Omega^c|$, is $\text{Binomial}(n_{S_i}n_{S_j},1-p).$

Thus by Hoeffding's inequality, we have
\begin{align*}
\p(|(S_i\times S_j)\cap \Omega^c| &\ge \frac{\gamma^2n^2(1-p)}{2K^2}) \ge \p\big(|(S_i\times S_j)\cap \Omega^c| \ge \frac{n_{S_i}n_{S_j}(1-p)}{2}\big)\\
& \ge 1-2\exp(-\frac{c_H}{4}n_{S_i}^2n_{S_j}^2(1-p)^2)  \ge 1-2\exp(-\frac{c_H\gamma^2n^4(1-p)^2}{4K^2})
\end{align*}
where $c_H$ is an absolute constant from Hoeffding's inequality. Taking $\tilde{c} = \frac{\gamma^2(1-p)}{2K^2}$ gives 
$$\p(|(S_i\times S_j)\cap \tcal| \ge \tilde{c}n^2) \ge 1-2\exp(-2c_H\tilde{c}n^4).$$
To obtain a uniform bound for all $i,j \in [K]$ and all subsets satisfying $|S_i| = n_{S_i} \ge \frac{\gamma n}{K}$ and $|S_j| = n_{S_j} \ge \frac{\gamma n}{K}$, we apply the union bound across all such sets. The number of terms in the sum is bounded above by
$$K^2\sum_{n_1, n_2 \in [\frac{\gamma n}{K},  (1-K\gamma) n]}{n \choose n_1}{n \choose n_2}.$$
By Sterling's approximation, when $n_1 = \omega(n)$ is true, ${n \choose n_1} = \Theta(2^{\ell_{dev}(n_1/n)n})$ where $\ell_{dev}$ is the binomial deviance. Thus the number of terms can be bounded by $2K^2n^22^{2c'n}$, 
where $c'$ is a constant depending on $K$ and $\gamma$.  Thus by the union bound, with probability at least $1-4K^2n^22^{2c'n}\exp(-2c_H\tilde{c}n^4) \to 1$, for any $i, j \in [K]$ and $S_i \subset G_i$ and $S_j \subset G_j$ with $|S_i| = n_{S_i} \ge \frac{\gamma n}{K}$ and $|S_j| = n_{S_j} \ge \frac{\gamma n}{K}$, we have
$$|(S_i\times S_j)\cap \Omega^c| \ge \tilde{c}n^2.$$

Now, for each $k \in [K]$, there must exist $\hat{k} \in [K']$ such that
$$|G_k \cap \hat{G}_{\hat{k}}| \ge \frac{|G_k|}{K'} \ge \frac{\gamma n}{K}.$$
By the pigeonhole principle, there must exist $l_2$, $l_3$ such that $\hat{l}_2=\hat{l}_3$, and in this case we set $k_2 = \hat{l}_2$.   Since no two rows of $B_0$ are identical, we know there exists $l_1$ such that $B_{0,l_1l_3} \ne B_{0, l_1l_2}$ and we set $k_1 = \hat{l}_1$. These $l_1, l_2, l_3$ and $k_1, k_2$ satisfy the claim of the Lemma.  
\end{proof}

\begin{proof}[(Proof of Theorem~\ref{thm:K-Consistency})]
We prove the result for one random ECV split, since the number of splits is finite and if the conclusion holds for each one with probability tending to 1, it trivially holds for the intersection of events as well. We follow a strategy similar to \cite{chen2014network}.   We  start from deriving an upper bound on the prediction loss for $K'=K$, with respect to an oracle.    Then we derive a lower bound on the prediction loss when $K' < K$ and show that asymptotically the latter bound is larger.  A key ingredient in the proof is independence of the test entries from the training entries (and thus $\hat{c}$), conditional on the data.   This allows us to condition on the event in Proposition~\ref{prop:SBMperf} for the test entries.  

Recall $B = \rho_n B_0$ (we suppress $n$ for clarity).  Consider the loss function
$$L(A,K) = \sum_{(i,j) \in \Omega^c}\ell(A_{ij},\hat{B}_{\hat{c}_i,\hat{c}_j})$$
and the oracle loss based on the true model, 
$$L_0(A,K) = \sum_{(i,j) \in \Omega^c}\ell(A_{ij},B_{ c_i,c_j})$$
where $\ell(x,y)$ is the entrywise loss function, with is either the $L_2$ loss, 
$$\ell(x,y) = (x-y)^2$$
or the binomial deviance loss (with $y>0$), 
$$\ell(x,y) = -x\log(y)-(1-x)\log(1-y).$$

Next, we define several sets of entries:
\begin{align*}
  T_{k_1, k_2, l_1, l_2} & = \{(i,j) \in \Omega^c: c_i = l_1, \hat{c}_i = k_1, c_j = l_2, \hat{c}_j = k_2 \} , \\
U_{k_1, k_2, l_1, l_2} & =  \{(i,j)  \in \Omega:  c_i = l_1, \hat{c}_i = k_1, c_j = l_2, \hat{c}_j = k_2   \} , \\
  Q_{k_1, k_2, l_1, l_2} & =  \{(i,j) :  c_i = l_1, \hat{c}_i = k_1, c_j = l_2, \hat{c}_j = k_2\}
  \end{align*}
 Let  $T_{k_1, k_2, \cdot, \cdot} = \cup_{l_1,l_2}T_{k_1, k_2, l_1, l_2}$ be the union taken over the first two indices, and similarly define $T_{\cdot, \cdot,l_1, l_2}$, $U_{k_1, k_2, \cdot, \cdot}$ and $Q_{k_1, k_2, \cdot, \cdot}$.  

 \noindent Case 1: $K' = K$.   First we bound the error in estimation of $B$.    Under the conditions of Proposition~\ref{prop:SBMperf}, the mislabeled proportion in each community is at most of the order of $\frac{1}{\lambda_n}$ where $\lambda_n = n\rho_n$.   The resulting relationship between  $Q_{k_1,k_2,\cdot, \cdot}$ and $Q_{\cdot, \cdot, k_1,k_2}$ is shown in Figure \ref{fig:Qintersect}.
 \begin{figure}
\begin{center}
\begin{tikzpicture}[thin,fill opacity=0.7]
\filldraw[black, very thick,fill=lightgray] (-3,3) rectangle (2,-2);
\filldraw[black, very thick,fill=lightgray] (3,-3) rectangle (-2,2);
\begin{scope}
	\clip (-3,3) rectangle (2,-2);
	\draw [draw=none, fill=white] (3,-3) rectangle (-2,2);
\end{scope} 
\node at (2.5,2.5) {$n/\lambda_n$};
\node at (-2.5,-2.5) {$n/\lambda_n$};
\node at (-0.5,3.2) {$n$};
\node at (0,0) {$Q_{k_1,k_2,k_1,k_2}$};
\node at (-2,2.5) {$Q_{k_1,k_2,\cdot, \cdot}$};
\node at (2,-2.5) {$Q_{\cdot, \cdot,k_1,k_2}$};
\end{tikzpicture}
\end{center}
\caption{Intersecting sets   $Q_{k_1,k_2,\cdot, \cdot}$ and $Q_{\cdot, \cdot, k_1,k_2}$.}
\label{fig:Qintersect}
\end{figure}
In particular, for each $k \in [K]$, we have $|\hat{G}_k \Delta G_k| = O_P(n/\lambda_n)$, and $|\hat{n}_k - n_k| = O_P(n/\lambda_n)$.   This implies that for any $k_1, k_2 \in [K]$, the symmetric difference 
$$(Q_{k_1, k_2, \cdot, \cdot}) \Delta (Q_{\cdot, \cdot,k_1, k_2}) = O_P\left(\frac{n^2}{\lambda_n}\right).$$
Finally, we have
\begin{align*}
U_{k_1, k_2, \cdot, \cdot}= Q_{k_1, k_2, \cdot, \cdot} \cap \Omega &=\left[\left(Q_{\cdot, \cdot,k_1, k_2}\cap \Omega   \right)/\left((Q_{\cdot, \cdot,k_1, k_2}/Q_{k_1, k_2, \cdot, \cdot})   \cap \Omega   \right)\right] \\
& ~~~~~~~~~~~~~~\cup \left[\left((Q_{k_1, k_2, \cdot, \cdot}/Q_{\cdot, \cdot,k_1, k_2})   \cap \Omega   \right)\right].
\end{align*}
Therefore 
\begin{align*}
|U_{k_1, k_2, \cdot, \cdot}| &\ge |\left(Q_{\cdot, \cdot,k_1, k_2}\cap \Omega   \right)/\left((Q_{\cdot, \cdot,k_1, k_2}/Q_{k_1, k_2, \cdot, \cdot})   \cap \Omega   \right)|\\
& \ge |Q_{\cdot, \cdot,k_1, k_2}\cap \Omega| - |Q_{\cdot, \cdot,k_1, k_2}/Q_{k_1, k_2, \cdot, \cdot}|  \ge c n^2.
\end{align*}
for some constant $c$.

Then by Bernstein inequality, for $k_1 \ne k_2$, we have
\begin{align}\label{eq:Bernstein}
|\hat{B}_{k_1k_2}-B_{k_1k_2}| &= |\frac{\sum_{U_{k_1, k_2, \cdot, \cdot}}A_{ij}}{|U_{k_1, k_2, \cdot, \cdot}|} - B_{k_1k_2}|\notag\\
& \le \frac{|U_{k_1, k_2, k_1, k_2}|}{|U_{k_1, k_2, \cdot, \cdot}|}|\frac{\sum_{U_{k_1, k_2, k_1, k_2}}A_{ij}}{|U_{k_1, k_2, k_1, k_2}|}-B_{k_1k_2}| + |(1-\frac{|U_{k_1, k_2, k_1, k_2}|}{|U_{k_1, k_2, \cdot, \cdot}|})B_{k_1k_2}|\notag\\
&~~~~~~~~~~~~~~~~~~ + 
\frac{|U_{k_1, k_2, \cdot, \cdot}/U_{k_1, k_2, k_1, k_2}|}{|U_{k_1, k_2, \cdot, \cdot}|}|\frac{\sum_{U_{k_1, k_2, \cdot, \cdot}/U_{k_1, k_2, k_1, k_2}}A_{ij}}{|U_{k_1, k_2, \cdot, \cdot}/U_{k_1, k_2, k_1, k_2}|}|\notag\\
&\le O_P(\sqrt{\frac{\rho_n}{n^2}}) + O_P(\frac{1}{\lambda_n})O_P(\rho_n) + O_P(\frac{1}{\lambda_n})O_P(\rho_n) = O_P(\frac{1}{n} ).
\end{align}
For $k_1 = k_2$, $\hat{B}_{k_1k_2}$ is the average over $U_{k_1, k_2, \cdot, \cdot} \cap \{(i,j): i<j\}$ which makes both its denominator and nominator half of those in \eqref{eq:Bernstein}, and the same concentration holds.

Now comparing $L$ and $L_0$, we have
\begin{align}\label{eq:Decomposition0}
L(A,K) - L_0(A,K) &= \sum_{k_1,k_2, l_1, l_2}\sum_{(i,j) \in T_{k_1, k_2, l_1, l_2}}[\ell(A_{ij}, \hat{B}_{k_1k_2}) - \ell(A_{ij}, B_{l_1l_2})] \notag\\
& =  \sum_{k_1,k_2}\sum_{(i,j) \in T_{k_1, k_2, k_1, k_2}}[\ell(A_{ij}, \hat{B}_{k_1k_2}) - \ell(A_{ij}, B_{k_1k_2})] \notag\\
&~~~~~ + \sum_{(k_1,k_2)\ne (l_1, l_2)}\sum_{(i,j) \in T_{k_1, k_2, l_1, l_2}}[\ell(A_{ij}, \hat{B}_{k_1k_2}) - \ell(A_{ij}, B_{l_1l_2})]\\
& := \mathcal{I} + \mathcal{II}. \notag
\end{align}

For the $L_2$ loss $\ell(x,y) = (x-y)^2$, we have
$$|\ell(x,y_1)-\ell(x,y_2)| = |(2x-y_1-y_2)(y_2-y_1)| \le 2(|x|+|y_2|+|y_1-y_2|)|y_1-y_2|.$$ 
Thus we have
\begin{align}
  \label{eq:Term1Bound-L2}
|\mathcal{I}| &\le \sum_{k_1,k_2}\sum_{(i,j) \in T_{k_1, k_2, k_1, k_2}}2|\hat{B}_{k_1k_2}-B_{k_1k_2}|(A_{ij}+B_{k_1k_2}+|\hat{B}_{k_1k_2}-B_{k_1k_2}|)\notag\\
& \le O_P\left(\frac{1}{n}n^2\rho_n\right) + O_P\left(\frac{1}{n}n^2\rho_n\right) + O_P\left((\frac{1}{n})^2n^2\right)  = O_P( n\rho_n) ; \\
|\mathcal{II}| &\le \sum_{(k_1,k_2)\ne (l_1, l_2)}\sum_{(i,j) \in T_{k_1, k_2, l_1, l_2}}(2A_{ij} + \hat{B}_{k_1k_2} + B_{l_1l_2})(\hat{B}_{k_1k_2} + B_{l_1l_2}) \notag \\
& =  \sum_{(k_1,k_2)\ne (l_1, l_2)}\sum_{(i,j) \in T_{k_1, k_2, l_1, l_2}}(2A_{ij} + B_{k_1k_2} + B_{l_1l_2} + (\hat{B}_{k_1k_2}-B_{k_1k_2}))(B_{k_1k_2} + B_{l_1l_2}+(\hat{B}_{k_1k_2}-B_{k_1k_2})) \notag \\
& = O_P(\frac{n^2}{n\rho_n}\rho_n^2) = O_P(n\rho_n). \notag
\end{align}

Combining these, we have, for the $L_2$ loss, 
\begin{equation}\label{seq:L2Bound2}
L(A,K) \le L_0(A,K) + O_P(n\rho_n).
\end{equation}
For the binomial deviance loss, we need an additional condition since $\log y$  is unbounded when $y$ is approaching 0.  We assume all entries of $B_0$ are positive and $n$ is sufficiently large so that for all $k_1, k_2$
\begin{equation}\label{eq:CurrentEvent}
|\hat{B}_{k_1k_2}-B_{k_1k_2}| < B_{k_1k_2}/2 < 1/6
\end{equation}
under the current event of \eqref{eq:Bernstein}, where the $1/6$ can be replaced by any other positive constant.  

Applying the inequality 
\begin{equation}\label{eq:basiclog}
|\log y_1-\log y_2| \le \frac{|y_1-y_2|}{\min(y_1,y_2)}
\end{equation}
we have
\begin{equation}\label{eq:dev-basic}
|\ell(x,y_1)-\ell(x,y_2)| \le  x\frac{|y_1-y_2|}{\min(y_1,y_2)} + (1-x)\frac{|y_1-y_2|}{1-\max(y_1,y_2)}.
\end{equation}
This gives
\begin{align}\label{eq:Term1Bound-Bin}
|\mathcal{I}| & \le \sum_{k_1,k_2}\sum_{(i,j) \in T_{k_1, k_2, k_1, k_2}} \left[ A_{ij}\frac{|\hat{B}_{k_1k_2}-B_{k_1k_2}|}{\min(\hat{B}_{k_1k_2},B_{k_1k_2})} + (1-A_{ij})\frac{|\hat{B}_{k_1k_2}-B_{k_1k_2}|}{1-\max(\hat{B}_{k_1k_2},B_{k_1k_2})}\right]\notag\\
&\le \sum_{k_1,k_2}\sum_{(i,j) \in T_{k_1, k_2, k_1, k_2}} \left[ A_{ij}\frac{|\hat{B}_{k_1k_2}-B_{k_1k_2}|}{1/2B_{k_1k_2}} + (1-A_{ij})\frac{|\hat{B}_{k_1k_2}-B_{k_1k_2}|}{1-3/2B_{k_1k_2}}\right]\notag\\
&\le \sum_{k_1,k_2}\sum_{(i,j) \in T_{k_1, k_2, k_1, k_2}} \left[ 2|\hat{B}_{k_1k_2}-B_{k_1k_2}|\frac{A_{ij}}{B_{k_1k_2}} + 2(1-A_{ij})|\hat{B}_{k_1k_2}-B_{k_1k_2}|\right]\notag\\
& = O_P\left(\frac{1}{n}\right)O_P(n^2) = O_P(n).
\end{align}

Moreover, under the assumption in \eqref{eq:CurrentEvent} and setting $y_2=1$ in \eqref{eq:basiclog}, we have
\begin{align*}
  |\log(\hat{B}_{k_1k_2})| & \le \left|\log \left( \frac{1}{2}B_{k_1k_2}\right)\right| \le \left |\log\left(\frac{1}{2}\rho_nB_{0,\min}\right)\right|\le \frac{1-\frac{1}{2}\rho_nB_{0,\min}}{\frac{1}{2}\rho_nB_{0,\min}} \le c'\frac{1}{\rho_n} , \\
 |\log(B_{k_1k_2})| & \le |\log(\rho_nB_{0,\min})| \le \frac{1-\rho_nB_{0,\min}}{\rho_nB_{0,\min}}\le c'\frac{1}{\rho_n}, \\
 |\log(1-\hat{B}_{k_1k_2})| & \le |\log(1/3)|, |\log(1-B_{k_1k_2})| \le |\log(2/3)|
 \end{align*}
where $B_{0,\min} = \min_{k_1,k_2}B_0.$ Therefore,
\begin{align*}
  |\mathcal{II}| \le & \sum_{(k_1,k_2)\ne (l_1, l_2)}\sum_{T_{k_1, k_2, l_1, l_2}}\left[   A_{ij}(|\log(\hat{B}_{k_1k_2})|+|\log(B_{k_1k_2})|) \right. \\
           &  \left.  + (1-A_{ij})(|\log(1-\hat{B}_{k_1k_2})| + |\log(1-B_{k_1k_2})|)    \right]\\
 \le & \sum_{(k_1,k_2)\ne (l_1, l_2)}\sum_{T_{k_1, k_2, l_1, l_2}}\left[   c' A_{ij}/\rho_n + (1-A_{ij})(|\log(1/3)|+|\log(2/3)|) \right]  = O_P(\frac{n}{\rho_n}).
\end{align*}
Therefore, for binomial deviance we also have 
\begin{equation}\label{seq:BinBound2}
L(A,K) \le L_0(A,K) + O_P(n/\rho_n).
\end{equation}

\noindent Case 2: $K' < K$.  Without loss of generality, assume the $k_1, k_2$ and $l_1, l_2, l_3$ in Lemma~\ref{lem:pigeon} are 1, 2 and 1, 2, 3 respectively. 
We have
\begin{align}\label{Expansion1}
L(A,K') - L_0(A,K) &= \sum_{k_1,k_2, l_1, l_2}\sum_{(i,j) \in T_{k_1, k_2, l_1, l_2}}[\ell(A_{ij}, \hat{P}_{ij}) - \ell(A_{ij}, B_{l_1 l_2})] \notag\\
=& \sum_{(i,j) \in T_{1, 2, 1, 2}}[\ell(A_{ij}, \hat{B}_{12}) - \ell(A_{ij}, B_{12})]  + \sum_{(i,j) \in T_{1, 2, 1,3}}[\ell(A_{ij}, \hat{B}_{12}) - \ell(A_{ij}, B_{13})] \notag\\
&+\sum_{(k_1,k_2, l_1, l_2) \notin \{ (1,2,1,2), (1,2,1,3)\}} \sum_{(i,j) \in T_{k_1, k_2, l_1, l_2}}[\ell(A_{ij}, \hat{P}_{ij}) - \ell(A_{ij}, B_{l_1l_2})]
\end{align}
For both the $L_2$ and the binomial deviance losses and any index set $T$, the function of the form 
$$f(p) = \sum_{(i,j) \in T}\ell(A_{ij}, p)$$
is always minimized when $p = \frac{1}{|T|} \sum_{(i,j)\in T}A_{ij}$.   Applying this in \eqref{Expansion1} and taking the first two terms together, we have
\begin{align}\label{Expansion2}
L(A,K') - L_0(A,K) &\ge   \sum_{(i,j) \in T_{1, 2, 1, 2}}[\ell(A_{ij}, \hat{p}) - \ell(A_{ij}, B_{12})]  + \sum_{(i,j) \in T_{1, 2,1, 3}}[\ell(A_{ij}, \hat{p}) - \ell(A_{ij}, B_{13})] \notag\\
&+\sum_{(k_1,k_2, l_1, l_2) \notin \{ (1,2,1,2), (1,2,1,3)\}} \sum_{(i,j) \in T_{k_1, k_2, l_1, l_2}}[\ell(A_{ij}, \hat{p}_{k_1,k_2, l_1, l_2}) - \ell(A_{ij}, B_{l_1l_2})]\\
&:= \mathcal{III} + \mathcal{IV} + \mathcal{V}  , \notag 
\end{align}
where $\hat{p}$ is the average of $A_{ij}$ over $T_{1, 2, 1, 2} \cup T_{1, 2, 1, 3}$ and $\hat{p}_{k_1,k_2, l_1, l_2}$ is the average of $A_{ij}$ over $T_{k_1, k_2, l_1, l_2}$. Note that $\hat{p} = t\hat{p}_1 + (1-t)\hat{p}_2$, where $\hat{p}_1 = \hat{p}_{1,2,1,2}$ and $\hat{p}_2 = \hat{p}_{1,2,1,3}$ and $t=\frac{|T_{1, 2, 1, 2}|}{|T_{1, 2, 1, 2}| + |T_{1, 2, 1, 3}|}$ for which $|T_{1, 2, 1, 3}| \sim |T_{1, 2, 1,2}| $ by Lemma~\ref{lem:pigeon}. Similarly, $p_1 = B_{12}$ and $p_2 = B_{13}$.

We now proceed to bound term $\mathcal{III}$. Define $f(\bx, p) = \sum_{(i,j) \in T_{1, 2, 1, 2}}\ell(A_{ij}, p)$ for any $p$ in the domain of $y$, where $\bx = \{A_{ij}\}_{T_{1, 2, 1, 2}}$ in any order. We can write $\mathcal{III}$ as 
$$\mathcal{III} = f(\bx, t\hat{p}_1 + (1-t)\hat{p}_2) - f(\bx, p_1).$$
By Lemma~\ref{lem:StrongConvexity}, for the squared loss we have
\begin{equation}\label{eq:strongconvex}
f(\bx, \lambda\hat{p}_1 + (1-t)\hat{p}_2) \ge f(\bx,\hat{p}_1)  + \frac{m}{2}(1-t)^2|\hat{p}_1-\hat{p}_2|^2.
\end{equation}
where $m = 2|T_{1, 2, 1, 2}|$. For the binomial deviance, as long as $p>0$, \eqref{eq:strongconvex} is also true when
$$m= |T_{1, 2, 1, 2}|\left(\frac{\hat{p}_1}{[1-\frac{1}{1+\delta^{1/3}}]^2} + \frac{1-\hat{p}_1}{[\frac{1}{1+\delta^{1/3}}]^2}\right)$$
and $\delta = \frac{\hat{p}_1}{1-\hat{p}_1} $. It is easy to see that $m = \Theta_P(|T_{1, 2, 1, 2}|)$ as well.

From \eqref{eq:strongconvex}, we have
$$f(\bx, t\hat{p}_1 + (1-t)\hat{p}_2) - f(\bx, p_1) \ge  \frac{m}{2}(1-t)^2|\hat{p}_1-\hat{p}_2|^2 + f(\bx,\hat{p}_1) - f(\bx, p_1)$$
and it always holds that  
$$|\hat{p}_1-\hat{p}_2| \ge |p_1-p_2| - |\hat{p}_1-p_1|- |\hat{p}_2-p_2|.$$
From Lemma~\ref{lem:pigeon}, we have $N = |T_{1,2,1,2}| \ge  \tau_{K'} n^2$ for some constant $\tau_{K'}$. Therefore, $ |\hat{p}_1-p_1|$ and $|\hat{p}_2-p_2|$ are upper bounded by $O_{P}(\sqrt{\frac{\rho_n}{n^2}})$ by  Bernstein's inequality,  and we have
$$|\hat{p}_1-\hat{p}_2| \ge |p_1-p_2| - O_{P}(\sqrt{\frac{\rho_n}{n^2}}) \ge c_{K'} \rho_n$$
for some constant $c_{K'}$.

We still need a lower bound on $ f(\bx,\hat{p}_1) - f(\bx, p_1)$. Note that the sameterm in $\mathcal{I}$ from $\eqref{eq:Decomposition0}$  is controlled similarly to \eqref{eq:Term1Bound-L2} and  \eqref{eq:Term1Bound-Bin} under the $L_2$ loss and the binomial deviance,  respectively.  The only difference is the bound on $ |\hat{p}_1-p_1|$ is now $O_{P}(\sqrt{\frac{\rho_n}{n^2}})$. Specifically, for the $L_2$ loss, we have
\begin{equation}\label{eq:DiscrepancyL2}
|f(\bx,\hat{p}_1) - f(\bx, p_1)| = O_P(n\rho_n^{3/2}) , 
\end{equation}
and for the binomial deviance 
\begin{equation}\label{eq:DiscrepancyBin}
|f(\bx,\hat{p}_1) - f(\bx, p_1)| = O_P(n\sqrt{\rho_n}) . 
\end{equation}
Combining all of the above gives 
\begin{align}\label{eq:Term3Bound}
\mathcal{III} &= f(\bx, t\hat{p}_1 + (1-t)\hat{p}_2) - f(\bx, p_1) = \omega_P(n^2\rho_n^2)
\end{align}
for the $L_2$  loss, which also holds for the binomial deviance  as long as $\rho_n^{-1} = o(n^{2/3})$. Term $\mathcal{IV}$ can be bounded in exactly the same way.  

The remaining term $\mathcal{V}$ is negative. For the $L_2$ loss, we have
\begin{align*}
\mathcal{V} &= \sum_{(k_1,k_2, l_1, l_2) \notin \{ (1,2,1,2), (1,2,1,3)\}} \sum_{(i,j) \in T_{k_1, k_2, l_1, l_2}}[(A_{ij}- \hat{p}_{k_1,k_2, l_1, l_2})^2 - (A_{ij}- B_{l_1l_2})^2]\\
& = - \sum_{(k_1,k_2, l_1, l_2) \notin \{ (1,2,1,2), (1,2,1,3)\}}|T_{k_1, k_2, l_1, l_2}|(\hat{p}_{k_1,k_2, l_1, l_2}-B_{l_1l_2})^2   \ge - O_P(\rho_n).
\end{align*}
For the binomial deviance, using the same inequalities as \eqref{eq:Term1Bound-Bin} (but here $|T_{k_1, k_2, l_1, l_2}|$ can be of a lower order than $n^2$), we get
\begin{align*}
\mathcal{V} &\ge - \sum_{(k_1,k_2, l_1, l_2) \notin \{ (1,2,1,2), (1,2,1,3)\}}\sum_{(i,j) \in T_{k_1, k_2, k_1, k_2}} \left[ 2A_{ij}\frac{|\hat{B}_{k_1k_2}-B_{k_1k_2}|}{B_{k_1k_2}} + 2(1-A_{ij})|\hat{B}_{k_1k_2}-B_{k_1k_2}|\right]\\
& \ge O_P(- \sum_{(k_1,k_2, l_1, l_2) \notin \{ (1,2,1,2), (1,2,1,3)\}}\sqrt{\rho_n|T_{k_1, k_2, k_1, k_2}|})   \ge - O_P(\sqrt{\rho_n n^2}).
\end{align*}

Combining the bounds and ignoring smaller order terms, with high probability we have, for the $L_2$ loss
\begin{equation}\label{eq:Bound1}
L(A,K') \ge L_0(A,K) + c n^2\rho_n^2
\end{equation}
where $c$ is a constant.   This bound is also true with the additional requirement of $\rho_n^{-1} = o(n^{2/3})$ for the binomial deviance. 

Finally, combining \eqref{seq:L2Bound2} and \eqref{eq:Bound1},  for the $L_2$  loss we have 
$$L(A,K') \ge L_0(A,K)+ \Theta_P(n^2\rho_n^2) \ge L(A,K) + \Theta_P(n^2\rho_n^2) - O_P(n\rho_n).$$
and therefore $\p(L(A,K') > L(A,K)) \to 1$.

Similarly, by comparing \eqref{seq:BinBound2} and \eqref{eq:Bound1}, for the binomial deviance
$$L(A,K') \ge L_0(A,K)+ \Theta_P(n^2\rho_n^2) \ge L(A,K) + \Theta_P(n^2\rho_n^2) - O_P(n/\rho_n), $$
and the same conclusion holds as long as $\rho_n^{-1} = o(n^{1/3})$.
\end{proof}

\begin{lemma}\label{lem:StrongConvexity}
Let $\bx  = (x_1, x_2, \cdots, x_N)$ be an $N$-dimensional binary vector  and $f(\bx, y) = \sum_{i = 1}^N \ell(x_i, y)$ for a function $\ell: \bR_+^2 \to \bR$. For $\ell(x,y) = (x-y)^2$, we have
$$f(\bx, y) \ge f(\bx, \bar{\bx}) + N(y-\bar{\bx})^2$$
and for $\ell(x,y) = -x\log(y) - (1-x)\log(1-y)$, we have
$$f(\bx, y) \ge f(\bx, \bar{\bx}) + \frac{1}{2}N \left( \frac{\bar{\bx}}{[1-\frac{1}{1+\delta^{1/3}}]^2} + \frac{1-\bar{\bx}}{[\frac{1}{1+\delta^{1/3}}]^2}\right)(y-\bar{\bx})^2$$
where  $\bar{\bx} = \frac{1}{N} \sum_{i=1}^Nx_i$ and $\delta = \frac{\bar{\bx}}{1-\bar{\bx}}$.
\end{lemma}
\begin{proof}[(Proof of Lemma~\ref{lem:StrongConvexity})] 
For both losses, $f$ is strongly convex in $y$ and achieves its minimum at $y = \bar{\bx}$.   Strong convexity implies  (Ch.\ 9 of \cite{boyd2004convex})
\begin{equation}\label{eq:strongconvex2}
f(\bx, y) \ge f(\bx,\bar{\bx})  + \frac{m}{2}(y-\bar{\bx})^2,
\end{equation}
where $m$ is a constant such that 
$$\frac{\partial^2 f(\bx, y)}{\partial y^2} \ge m$$
 for all $y$ in its domain. For the $L_2$ loss, it is easy to see that $m = 2N$. For the binomial deviance, as long as $y>0$, 
$$\frac{\partial^2 f(\bx, y)}{\partial y^2} = N\left(\frac{\bar{\bx}}{y^2} + \frac{1-\bar{\bx}}{(1-y)^2}\right), $$
which is minimized at $y^*$ such that $(\frac{y^*}{1-y^*})^3 = \frac{\bar{\bx}}{1-\bar{\bx}} = \delta$. Thus we have
$$\frac{\partial^2 L(\bx, p)}{\partial p^2} \ge m:= N \left(\frac{\bar{\bx}}{[1-\frac{1}{1+\delta^{1/3}}]^2} + \frac{1-\bar{\bx}}{[\frac{1}{1+\delta^{1/3}}]^2}\right).$$
\end{proof}

\subsection{Proof of Theorem~\ref{thm:RDPG-consistency}}

To prove Theorem~\ref{thm:RDPG-consistency}, we need two results published elsewhere, with slight modifications.   The first is a result on concentration of the spectral embedding adapted from \cite{athreya2017statistical}.  
\begin{lemma}\label{lem:embedding}
Assume the network $A$ is from a random dot product graph model with latent space dimension $K$ satisfying Assumption~\ref{ass:RDPG-parameter} and let the sampling proportion $p$ in ECV  be a fixed constant. Assume all the conditions of Theorem~\ref{thm:simplebound} and that $K$ is a fixed constant. Let $P=U\Sigma U^T$ be the eigen-decomposition of $P$ with eigenvalues in non-increasing order, and let $\hat{U}\hat{\Sigma}\hat{U}^T$ be the eigen-decomposition of $\hat{A}$ defined in \eqref{eq:simpleHS}. Let $X = U\Sigma^{1/2}$ and $\hat{X} = \hat{U}\hat{\Sigma}^{1/2}$. Then there exists an orthogonal transformation $W \in \bR^{n\times n}$ such that 
$$\p\left(\max_{i \in [n]}\norm{\hat{X}_{i\cdot}-WX_{i\cdot}} \le \tilde{C}\frac{\log^2n}{\sqrt{n\rho_n}}\right) \to 1$$
where $X_{i\cdot}$ is the $i$-th row of matrix $X$ and $\tilde{C}$ is a constant depending on $C$ in Lemma~\ref{lemma:concentration}, $K$ and $\psi_1$ and $\psi_2$ in  Assumption~\ref{ass:RDPG-parameter}.
\end{lemma}

\begin{proof}
This follows from Theorem 26 of \cite{athreya2017statistical} by noting the spectrum of $\hat{A}$ overlaps with $\frac{1}{p}P_{\Omega}A$ and concentration in the form of Lemma~\ref{lemma:concentration} holds under the Assumption \ref{ass:RDPG-parameter}.   
\end{proof}
The second result is an entry-wise concentration bound on the empirical eigenvectors obtained by using several tools of \cite{eldridge2017unperturbed}.

\begin{lemma}[Theorem 17 of \cite{eldridge2017unperturbed}]\label{lem:FixTerm}
Let $X$ be an $n\times n$ symmetric random matrix such that $\e X_{ij} = 0$,  $\e|X_{ij}|^p \le \frac{1}{n}$ for all $i, j \in [n]$ and $p\ge 2$, and all of its entries on and above the diagonals are independent. Let $u$ be an $n$-vector with $\norm{u}_{\infty}=1$. For constants $\xi >1$ and $0 < \kappa < 1$, with probability at least $1-n^{-\frac{1}{4}(\log_{\mu}{n})^{\xi-1}(\log_{\mu}{e})^{-\xi}}$, where $\mu = \frac{2}{\kappa+1}$, 
$$\norm{X^pu}_{\infty} < \log ^{p\xi}{n} \text{~~for all~~} p\le \frac{\kappa}{8}\log^{\xi}{n} . $$

\end{lemma}

\begin{lemma}[Corrected version of Theorem 9 of \cite{eldridge2017unperturbed}]\label{lem:deterministicControl}
Assume $A = P+H$ where $P \in \bR^{n\times n}$ and $H \in \bR^{n\times n}$ are symmetric matrices with $\rank{P} = K$. Let $\lambda_1 \ge \lambda_2 \cdots \ge \lambda_K >0$ be the eigenvalues of $P$ and let $\hat{\lambda}_k$ be the eigenvalues of $A$. For any $s \in [n]$, let $\Lambda_s = \{i:\lambda_i=\lambda_s\}$, $d_s = |\Lambda_s|$ and define the gap  as $\delta_s = \min_{i\notin \Lambda_s}|\lambda_s-\lambda_i|$. Let $\Delta^{-1}_{s,t} = \min\{d_i/\delta_i\}_{i \in \{s,t\}}$. Define $\lambda_t^* = |\lambda_t|-\norm{H}$. For any $t\in [n]$, if $\norm{H} < \lambda_t/2$, 
there exist eigenvectors $u_1, \cdots, u_K$ of $P$ corresponding to $\lambda_1, \cdots, \lambda_K$ and eigenvectors $\hat{u}_1, \cdots, \hat{u}_K$ of $A$ corresponding to $\hat{\lambda}_1, \cdots, \hat{\lambda}_K$ such that
 for all $j \in [n]$:
\begin{align*}
|(\hat{u}_t-u_t)_j| \le& |u_{t,j}|\left( 8d_t[\frac{\norm{H}}{\delta_t}]^2+\frac{\norm{H}}{\lambda_t^*}\right) + \left( \frac{|\lambda_t|}{\lambda_t^*}\right)\zeta_j(u_t; H, \lambda_t)\\
& + \frac{2\sqrt{2}\norm{H}}{\lambda_t^*}\sum_{s\ne t}\frac{|\lambda_s|}{\Delta_{s,t}}\left[ |u_{s,j}| + \zeta_j(u_s; H, \lambda_t) \right]
\end{align*}
where $\zeta(u;H,\lambda)$ is a $n$-vector defined for any vector $u\in \bR^n$, symmetric matrix $H \in \bR^{n\times n}$ and scalar $\lambda$ with its $j$-th entry given by 
$$\zeta_j(u;H,\lambda) = \left|\left[   \sum_{p\ge 1}(\frac{2H}{\lambda})^pu  \right]_j\right|.$$
In particular, the eigenvectors $\{u_k\}_{k=1}^K$ and $\{\hat{u}_k\}_{k=1}^K$ are unique up to an orthogonal transformation for those with eigenvalue multiplicity larger than 1.
\end{lemma}
The correction is the new condition $\norm{H} < \lambda_t/2$  in Lemma~\ref{lem:deterministicControl} which was not in the original statement of the result in \cite{eldridge2017unperturbed};   they used $\norm{H} < \lambda_t$ implicitly but that was not enough (see their formula (10)).

We are now ready to introduce  the following lemma about the concentration of adjacency matrix eigenvectors in $\ell_{\infty}$ norm whose proof is given after the proof of Theorem~\ref{thm:RDPG-consistency}. It is an extension of Theorem 3 in \cite{eldridge2017unperturbed} to the random dot product graph setting to give the entrywise concentration of eigenvectors for the random dot product graph. Notice that entrywise bound can also be obtained by directly using   Lemma~\ref{lem:embedding} but Lemma~\ref{lem:KConcentration} gives a better result.

\begin{lemma}\label{lem:KConcentration}
Assume  $A$ is generated from the random dot product graph with latent space dimension $K$ satisfying Assumption~\ref{ass:RDPG-parameter} and $\mu(P) \le t_n^2$. Let $\lambda_1 \ge \lambda_K \cdots, \ge \lambda_K >0$ be the eigenvalues of $P$ and let $\hat{\lambda}_k$ be the eigenvalues of $A$. If for some $\xi > 1$, $\frac{\log^{2\xi}n}{n\rho_n} = o(1)$, then there exist eigenvectors $u_1, \cdots, u_K$ of $P$ corresponding to $\lambda_1, \cdots, \lambda_K$ and eigenvectors $\hat{u}_1, \cdots, \hat{u}_K$ of $A$ corresponding to $\hat{\lambda}_1, \cdots, \hat{\lambda}_K$ as well as positive constants $C_1, C_2$ such that 
\begin{equation}\label{eq:Kconcentration}
\norm{\hat{u}_K - u_K}_{\infty} \le C_1\frac{t_n\log^{\xi}n}{n\sqrt{\rho_n}}
\end{equation}
with probability at least $1-2n^{-\delta}$, where $C_1$ depend on $K$ and $\psi_1$ in Assumption~\ref{ass:RDPG-parameter} while $C$ and $\delta$ are the constants in Lemma~\ref{lemma:concentration}. In particular, such eigenvectors are unique up to an orthogonal transformation for the part with eigenvalue multiplicity larger than 1.
\end{lemma}

Next lemma shows  that the  assumption we make on the eigenvalues of $P$ under the random dot product graph model  ensures the model is sufficiently distinguishable from its lower rank version on a significant proportion of entries.

\begin{lemma}\label{lem:RDPG-identifiable}
Under the conditions of Theorem~\ref{thm:RDPG-consistency}, let $R^{(q)} = P^0-P^0_q$ for $q \le K$ where $P^0_q$ is the rank-$q$ truncated SVD of $P^0$. Then for any $q < K$, there exists constants $\psi_3^{(q)}>0$ and $\kappa>0$ such that
\begin{equation}\label{eq:RDPG-identifiable}
|\Delta_q|:= \left|\left\{(i,j): i\le j, |R^{(q)}_{ij}|\ge \psi_3^{(q)}  \right\}\right| \ge \kappa n^2.
\end{equation}
\end{lemma}
\begin{proof}[(Proof of Lemma~\ref{lem:RDPG-identifiable})]
From the bounded coherence, we have  
\begin{align*}
|R_{ij}^{(q)}| = & |\sum_{k=q+1}^K\lambda_k^0U_{i,(q+1):K}^TU_{j,(q+1):K}|   \le \sum_{k=q+1}^K\lambda_k^0|U_{i,(q+1):K}^TU_{j,(q+1):K}|\\
& \le \sum_{k=q+1}^K\lambda_k^0\norm{U_{i,(q+1):K}}\norm{U_{j,(q+1):K}} \le \psi n \frac{K}{n}a \le \psi K a.
\end{align*}
The lemma follows from the fact that $\sum_{i,j} |R^{(q)}_{ij}|^2 = \norm{R^{(q)}}_F^2 = \sum_{k=q+1}^K(\lambda^{0}_k)^2 \ge \frac{(K-q)n^2}{\psi^2}.$
\end{proof}

\begin{proof}[(Proof of Theorem~\ref{thm:RDPG-consistency})]
We use a strategy similar to the proof of Theorem~\ref{thm:K-Consistency}.   Again, we consider one random ECV split and condition on the training data and the splitting index, and aim to show that  $L(A, K') >  L(A,K)$ with probability tending to 1 for $K' < K$. For simplicity, we also condition on the events of Lemma~\ref{lem:embedding} and Lemma~\ref{lem:KConcentration}, which happen with probability tending to 1. The key argument is again based on independence of training  and validation data.  Define  the loss and the oracle loss respectively as 
$$L(A,K) = \sum_{(i,j) \in \Omega^c}\ell(A_{ij},\hat{P}_{ij}), \ \ L_0(A,K) = \sum_{(i,j) \in \Omega^c}\ell(A_{ij},P_{ij}), $$
where $\ell(x,y) = (x-y)^2$.   

Under the random dot product graph model, we have $P_{ij} = X_{i\cdot}^TX_{j\cdot} = \tilde{X}_{i\cdot}^T\tilde{X}_{j\cdot}$ and $\hat{P}_{ij} = \hat{X}_{i\cdot}^T\hat{X}_{j\cdot}$, where $X$ and $\hat{X}$ are defined in Lemma~\ref{lem:embedding}, have rank $K$ and $K'$ respectivel,  and $\tilde{X} = WX$ with the $W$ from Lemma~\ref{lem:embedding}.

\noindent Case 1: $K' = K$. In this situation, we have
\begin{align}
|\hat{P}_{ij}-P_{ij}| & = |\hat{X}^T_{i\cdot}\hat{X}_{j\cdot} - X_{i\cdot}^TX_{j\cdot}|= |\hat{X}^T_{i\cdot}\hat{X}_{j\cdot} - \tilde{X}^T_{i\cdot}\tilde{X}_{j\cdot}|\notag\\
&\le \norm{\hat{X}_{j\cdot}-\tilde{X}_{j\cdot}}\norm{\hat{X}_{i\cdot}} + \norm{\hat{X}_{i\cdot}-\tilde{X}_{i\cdot}}\norm{\tilde{X}_{j\cdot}}\notag\\
& \le \norm{\hat{X}_{j\cdot}-\tilde{X}_{j\cdot}}(\norm{\tilde{X}_{i\cdot}} + \norm{\hat{X}_{i\cdot}-\tilde{X}_{i\cdot}}) + \norm{\hat{X}_{i\cdot}-\tilde{X}_{i\cdot}}\norm{\tilde{X}_{j\cdot}}\notag\\
& = \norm{\hat{X}_{j\cdot}-\tilde{X}_{j\cdot}}(\norm{X_{i\cdot}} + \norm{\hat{X}_{i\cdot}-\tilde{X}_{i\cdot}}) + \norm{\hat{X}_{i\cdot}-\tilde{X}_{i\cdot}}\norm{X_{j\cdot}}
\end{align}
By Lemma~\ref{lem:embedding} and the coherence assumption, 
\begin{equation}\label{eq:pdiff}
\max_{ij}|\hat{P}_{ij}-P_{ij}|  = O\left(\left(\mu(P^0)\frac{K}{n}\right)^{1/2}\frac{\log^2{n}}{\sqrt{n\rho_n}}\right) = O\left(\frac{\log^2{n}}{n\sqrt{\rho_n}}\right) . 
\end{equation}
Applying Bernstein's inequality to the test entries, we have
\begin{align}
|L(A,K)-& L_0(A,K)| = |\sum_{(i,j)\in \Omega^c}\ell(A_{ij},\hat{P}_{ij})-\ell(A_{ij},P_{ij})|\notag
\le \sum_{(i,j)\in \Omega^c}|\hat{P}_{ij}-P_{ij}|(2A_{ij}+\hat{P}_{ij}+P_{ij})\notag\\
& \le  \sum_{(i,j)\in \Omega^c}\max_{ij}|\hat{P}_{ij}-P_{ij}|(2A_{ij}+2P_{ij} + \max_{ij}|\hat{P}_{ij}-P_{ij}|)\notag\\
& = O_P\left(\frac{\log^2{n}}{n\sqrt{\rho_n}}|\Omega^c|\rho_n\right) = O_P(n\sqrt{\rho_n}\log^2{n}), 
\end{align}
and therefore 
\begin{equation}\label{eq:RDPG-l2-equal}
L(A,K) = L_0(A,K) + O_P(n\sqrt{\rho_n}\log^2{n}).
\end{equation}

%
%

\noindent Case 2: $K'=K-1$.  In a slight abuse of notation, we use $\hat{X}^{(K-1)}$ to denote the $K-1$-dimensional empirical spectral embedding vectors, so that $\hat{P}^{(K-1)} = \hat{X}^{(K-1)}(\hat{X}^{(K-1)})^{T}$, while $\hat{X}$ and $\hat{P} =  \hat{X}^{(K)}(\hat{X}^{(K)})^T$ are the $K$-dimensional embedding and the corresponding estimate of $P$. We have
\begin{align}\label{eq:InitialDiff}
|\hat{P}_{ij}^{(K-1)}-P_{ij}|& = |\hat{X}^{(K-1)T}_{i\cdot}\hat{X}^{(K-1)}_{j\cdot} - X^T_{i\cdot}X_{j\cdot}|  = |\hat{X}^{T}_{i\cdot}\hat{X}_{j\cdot} - X^T_{i\cdot}X_{j\cdot}-\hat{X}_{iK}\hat{X}_{jK}|\notag\\
& \ge |\hat{X}_{iK}\hat{X}_{jK}| - |\hat{X}^{T}_{i\cdot}\hat{X}_{j\cdot} - X^T_{i\cdot}X_{j\cdot}|.
\end{align}
The second term is the same as in \eqref{eq:pdiff}.  To lower bound the first term, note 
\begin{align*}
|\hat{X}_{iK}\hat{X}_{jK}| & \ge |X_{iK}X_{jK}|  - |\hat{X}_{iK}\hat{X}_{jK}-X_{iK}X_{jK}| .  
\end{align*}
We first bound  $|\hat{X}_{iK}\hat{X}_{jK}-X_{iK}X_{jK}|$:  
\begin{align*}
|\hat{X}_{iK}-X_{iK}| & = |\sqrt{\hat{\lambda}_K}\hat{u}_{iK}-\sqrt{\lambda_K}u_{iK}|\le |\sqrt{\hat{\lambda}_K}-\sqrt{\lambda_K}||\hat{u}_{iK}| + \sqrt{\lambda_K}|\hat{u}_{iK}-u_{iK}|\\
& \le |\sqrt{\hat{\lambda}_K}-\sqrt{\lambda_K}|(|\hat{u}_{iK}| + |\hat{u}_{iK}-u_{iK}|)  + \sqrt{\lambda_K}|\hat{u}_{iK}-u_{iK}|\\
& \le \frac{|\hat{\lambda}_K-\lambda_K|}{\sqrt{\lambda_K}}\left(\norm{u_K}_{\infty} + \norm{\hat{u}_K-u_K}_{\infty} \right) + \sqrt{\lambda_K}\norm{\hat{u}_K-u_K}_{\infty}
\end{align*}
where the last inequality follows from
$$|\sqrt{x}-\sqrt{y}| \le \frac{|x-y|}{2\sqrt{\min(x,y)}}$$
and the fact that $\hat{\lambda}_K > \lambda_K/2$ for $n$ sufficiently large by assumption.   We will use Lemma~\ref{lem:KConcentration} to control this term.   Though Lemma~\ref{lem:KConcentration} only gives the result  up to an orthogonal transformation, we only need to bound the inner product between two rows of the eigenvector matrix, which is invariant under such orthogonal transformations. Thus without loss of generality, we can assume the eigenvectors $U$ and $\hat{U}$ are the ones in the conclusion of Lemma~\ref{lem:KConcentration}.    Applying Assumption~\ref{ass:RDPG-parameter} and Lemma~\ref{lem:KConcentration} and dropping higher-order terms, we have 
\begin{equation}\label{eq:embedding-entrywise}
|\hat{X}_{iK}-X_{iK}| \le 2C\psi_1C_1\frac{\sqrt{a}\log^{\xi}{n}}{n\sqrt{\rho_n}} + C_1\sqrt{\psi_1}\frac{\sqrt{a}\log^{\xi}n}{\sqrt{n}} \le 2C_1\sqrt{\psi_1}\sqrt{a}\frac{\log^{\xi}n}{\sqrt{n}}.
\end{equation}
Therefore, 
\begin{align}\label{eq:product-diff}
|\hat{X}_{iK}\hat{X}_{jK}-X_{iK}X_{jK}| & \le  |\hat{X}_{iK}-X_{iK}||\hat{X}_{jK}| + |\hat{X}_{jK}-X_{jK}||X_{iK}|\notag\\
& \le |\hat{X}_{iK}-X_{iK}||X_{jK}| + |\hat{X}_{iK}-X_{iK}||\hat{X}_{jK}-X_{jK}| + |\hat{X}_{jK}-X_{jK}||X_{iK}|\notag\\
& \le 6C_1\sqrt{\psi_1}\sqrt{a}\frac{\log^{\xi}n}{\sqrt{n}}\max_{i}|X_{iK}| \le 6C_1\psi_1\sqrt{K}a\frac{\sqrt{\rho_n}\log^{\xi}n}{\sqrt{n}}.
\end{align}
Now for any $(i,j) \in \Delta_{K-1}$, by Lemma~\ref{lem:RDPG-identifiable}, we have
\begin{equation}\label{eq:DeltaSquare}
|\hat{X}_{iK}\hat{X}_{jK}| \ge |X_{iK}X_{jK}|  - |\hat{X}_{iK}\hat{X}_{jK}-X_{iK}X_{jK}| \ge \frac{\psi_3^{(q)2}\rho_n}{\psi_1}-6C_1\psi_1\sqrt{K}a\frac{\sqrt{\rho_n}\log^{\xi}n}{\sqrt{n}} \ge \frac{\psi_3^{(q)2}\rho_n}{2\psi_1}
\end{equation}
in which the last inequality holds for $n$ sufficiently large when $\log^{2\xi}n = o(n\rho_n)$, as assumed.   Combining \eqref{eq:pdiff} and \eqref{eq:DeltaSquare} in \eqref{eq:InitialDiff}, we have that for any $(i,j) \in \Delta_{K-1}$
\begin{equation}\label{eq:pdifflower}
\min_{ij}|\hat{P}_{ij}^{(K-1)}-P_{ij}| \ge  \Theta(\rho_n-\frac{\log^2{n}}{n\sqrt{\rho_n}})  = \Theta(\rho_n) \ , 
\end{equation}
as long as $n^{1/3}\log^{4/3}{n} = o(n\rho_n)$, and 
$$\e(L(A,K-1)-L_0(A,K)) = \sum_{(i,j)\in \Omega^c}\e(\ell(A_{ij},\hat{P}_{ij}^{(K-1)})-\ell(A_{ij},P_{ij})) = \sum_{(i,j)\in \Omega^c}(\hat{P}_{ij}^{(K-1)}-P_{ij})^2$$
where the expectation is taken over the test entries. Using the same argument as in Lemma~\ref{lem:pigeon} implies that $|\Delta_{K-1} \cap \Omega^c| = \Theta_P(n^2)$. Therefore, by Hoeffding's inequality 
\begin{align}\label{eq:RDPGsmall}
L(A,K-1)-L_0(A,K) &\ge \frac{1}{2}\e(L(A,K-1)-L_0(A,K))   = \frac{1}{2}\sum_{(i,j)\in \Omega^c}(\hat{P}_{ij}^{(K-1)}-P_{ij})^2 \notag\\
& \ge \frac{1}{2}\sum_{(i,j)\in \Omega^c\cap \Delta_{K-1}}(\hat{P}_{ij}^{(K-1)}-P_{ij})^2 \ge \Theta_P(n^2\rho_n^2)
\end{align}
with probability tending to 1. Comparing \eqref{eq:RDPG-l2-equal} and \eqref{eq:RDPGsmall}, we get
$$\p(L(A,K-1) > L(A,K)) \to 1$$
as long as $n^{1/3}\log^{4/3}n = o(n\rho_n)$. 

\noindent Case 3: $K' < K-1$. This case is essentially the same as Case 2. The only difference is that instead of \eqref{eq:InitialDiff},  we now need
\begin{align*}
|\hat{P}_{ij}-P_{ij}|  \ge |\sum_{K' < q \le K}\hat{X}_{iq}\hat{X}_{jq}| - |\hat{X}^{T}_{i\cdot}\hat{X}_{j\cdot} - X^T_{i\cdot}X_{j\cdot}|.
\end{align*}
We have
\begin{align*}
|\sum_{K' < q \le K}\hat{X}_{iq}\hat{X}_{jq}| &\ge |\sum_{K' < q \le K}X_{iq}X_{jq}|  - \sum_{K' < q \le K}|\hat{X}_{iq}\hat{X}_{jq}-X_{iq}X_{jq}|\\
&= |\rho_n R^{(K')}_{ij}|- \sum_{K' < q \le K}|\hat{X}_{iq}\hat{X}_{jq}-X_{iq}X_{jq}|.
\end{align*}
The lower bound for the first term is available from Lemma~\ref{lem:RDPG-identifiable} up to the order of $K-K'$. Each of the remaining terms can be controlled as in \eqref{eq:product-diff}.    Thus we get the  same bound as in Case 2 and$\p(L(A,K') > L(A,K)) \to 1$.    Combining all three cases, we have shown that $\p(\hat{K} < K) \to 0.$
\end{proof}

\begin{proof}[(Proof of Lemma~\ref{lem:KConcentration})]
Let $\{u_k\}$ and $\{\hat{u}_k\}$ be the eigenvectors from the conclusion of Lemma~\ref{lem:deterministicControl}. 
Define
$$Z_k = \norm{\sum_{p\ge 1}(2H/\lambda_K)^pu_k}_{\infty}, k \in [K]$$
if all the series in the definition are finite, and $Z_k = \infty$ if the series diverge for any  component.   Our goal is to bound $Z_k$. For any $1\le k \le K$ and $1 \le l \le n$, 
\begin{align}\label{eq:Z-decomposition}
|(\sum_{p\ge 1}(2H/\lambda_K)^pu_k)_l| &\le \sum_{p\ge 1}|(2H/\lambda_K)^pu_k)_l|    = \sum_{p\ge 1}|(\frac{2\gamma}{\lambda_K}\cdot \frac{H}{\gamma})^pu_k)_l|  \notag\\
& = \sum_{1\le p\le \frac{\kappa}{8}(\log^{\xi}{n})}|[(\frac{2\gamma}{\lambda_K}\cdot \frac{H}{\gamma})^pu_k]_l| + \sum_{p > \frac{\kappa}{8}(\log^{\xi}{n})}|[(\frac{2H}{\lambda_K})^pu_k]_l| \notag \\
&  = \norm{u_k}_{\infty}\sum_{1\le p\le \frac{\kappa}{8}(\log^{\xi}{n})}|[(\frac{2\gamma}{\lambda_K}\cdot \frac{H}{\gamma})^p\frac{u_k}{\norm{u_k}_{\infty}}]_l| + \sum_{p > \frac{\kappa}{8}(\log^{\xi}{n})}|[(\frac{2H}{\lambda_K})^pu_k]_l|
\end{align}
where the constant $\xi$ is defined in Lemma \ref{lem:FixTerm}, $\kappa \in (0,1)$ is a constant, and $\gamma$ depends on $n$ and $\rho_n$. To apply Lemma~\ref{lem:FixTerm}, we need to select a $\gamma$ to ensure $\e |H_{ij}/\gamma|^p \le 1/n$.   Since $H_{ij} \le 1$, as long as $\gamma > 1$, for all $p\ge 2$
$$\e|H_{ij}/\gamma|^p \le \e |H_{ij}/\gamma|^2 = \frac{P_{ij}(1-P_{ij})}{\gamma^2} \le \frac{\max_{ij}P_{ij}}{\gamma^2} \le \frac{\rho_n}{\gamma^2}. $$

Setting $\gamma = \sqrt{n\rho_n}$, we have $\e|H_{ij}/\gamma|^p \le \frac{1}{n}$ for all $i,j \in [n]$ and $p \ge 2$.

Now define events
$$E_k = \left\{\norm{(\frac{H}{\gamma})^p\frac{u_k}{\norm{u_k}_{\infty}}}_{\infty} < (\log{n})^{p\xi}, \text{~~for all~~} p\le \frac{\kappa}{8}(\log^{\xi}{n})\right\}, k = 1, 2, \cdots, K, $$
and
$$E_{0} = \left\{ \norm{H} \le C \sqrt{n\rho_n} \right\}, $$
where $C$ is the constant from Lemma~\ref{lemma:concentration}. By Lemmas~\ref{lem:FixTerm} and~\ref{lemma:concentration}, we have 
$$\p(\cap_{k=0}^KE_k) \ge 1-Kn^{-\frac{1}{4}(\log_{\mu}{n})^{\xi-1}(\log_{\mu}{e})^{-\xi}}-n^{-\delta}.$$

Under the event $\cap_{k=0}^KE_k$, for any $k \in [K]$ and $l \in [n]$, 
\begin{align}
\sum_{1\le p\le \frac{\kappa}{8}(\log^{\xi}{n})}|(\frac{2\gamma}{\lambda_K}\cdot \frac{H}{\gamma})^p\frac{u_k}{\norm{u_k}_{\infty}})_l| & \le \sum_{1\le p\le \frac{\kappa}{8}(\log^{\xi}{n})}(\frac{2\gamma}{\lambda_K}\log^{\xi}{n})^{p} 
= \sum_{1\le p\le \frac{\kappa}{8}(\log^{\xi}{n})}(\frac{2\sqrt{n\rho_n}}{\lambda_K}\log^{\xi}{n})^{p}\notag \\
& \le \sum_{1\le p\le \frac{\kappa}{8}(\log^{\xi}{n})}(\frac{2\sqrt{n\rho_n}}{\frac{n\rho_n}{\psi_1}}\log^{\xi}{n})^{p} 
 = \sum_{1\le p\le \frac{\kappa}{8}(\log^{\xi}{n})}(\frac{2\psi_1}{\sqrt{n\rho_n}}\log^{\xi}{n})^{p} .  \notag
\end{align}
Under the condition $\frac{\log^{2\xi}{n}}{n\rho_n} = o(1)$, for sufficiently large $n$, we have $\frac{2\psi_1}{\sqrt{n\rho_n}}\log^{\xi}{n} < 1/2$. Therefore
\begin{equation}\label{eq:FiniteSumBound}
\norm{u_k}_{\infty}\sum_{1\le p\le \frac{\kappa}{8}(\log^{\xi}{n})}|(\frac{2\gamma}{\lambda_K}\cdot \frac{H}{\gamma})^p\frac{u_k}{\norm{u_k}_{\infty}})_l|  \le 4\psi_1\frac{\log^{\xi}{n}}{\sqrt{n\rho_n}}  \norm{u_k}_{\infty}.
\end{equation}

On the other hand, for the second term in \eqref{eq:Z-decomposition}, we have
\begin{align*}
\sum_{p > \frac{\kappa}{8}(\log^{\xi}{n})}|[(\frac{2H}{\lambda_K})^pu_k]_l| & \le \sum_{p > \frac{\kappa}{8}(\log^{\xi}{n})}\norm{(\frac{2H}{\lambda_K})^pu_k}_{\infty}  \le \sum_{p > \frac{\kappa}{8}(\log^{\xi}{n})}\norm{(\frac{2H}{\lambda_K})^pu_k}_2 \notag\\
& \le  \sum_{p \ge \lceil\frac{\kappa}{8}(\log^{\xi}{n})\rceil}\norm{(\frac{2H}{\lambda_K})^p} = \sum_{p \ge \lceil\frac{\kappa}{8}(\log^{\xi}{n})\rceil}\frac{\norm{2H}^p}{\lambda_K^p} .
\end{align*}
Under the event $\cap_{k=0}^KE_k$, we have $\norm{2H} \le 2C\sqrt{n\rho_n} <  \frac{1}{2}\frac{n\rho_n}{\psi_1}<\frac{1}{2}\lambda_K$ for sufficiently large $n$, which gives 
\begin{equation}\label{eq:TailSumBound}
\sum_{p > \frac{\kappa}{8}(\log^{\xi}{n})}|[(\frac{2H}{\lambda_K})^pu_k]_l|   \le \sum_{p \ge \lceil\frac{\kappa}{8}(\log^{\xi}{n})\rceil}(\frac{2C\psi_1}{\sqrt{n\rho_n}} )^p 
 \le 2(\frac{2C\psi_1}{\sqrt{n\rho_n}}  )^{\lceil \frac{\kappa}{8}(\log^{\xi}{n})\rceil} \le 2(\frac{2C\psi_1}{\sqrt{n\rho_n}}  )^{ \frac{\kappa}{8}(\log^{\xi}{n})}. 
\end{equation}

Combining \eqref{eq:Z-decomposition}, \eqref{eq:FiniteSumBound} and \eqref{eq:TailSumBound}, under the event $\cap_{k=0}^KE_k$, we have 
$$Z_k \le 4\psi_1\frac{\log^{\xi}{n}}{\sqrt{n\rho_n}} \norm{u_k}_{\infty} + 2(\frac{2C\psi_1}{\sqrt{n\rho_n}}  )^{ \frac{\kappa}{8}(\log^{\xi}{n})}, k \in [K] . $$
The second term is dominated by the first, so for sufficiently large $n$
\begin{equation}\label{eq:b2}
Z_k \le 8\psi_1\frac{\log^{\xi}{n}}{\sqrt{n\rho_n}}  \norm{u_k}_{\infty}, \ k \in [K].
\end{equation}

Since $d_s \le K$ and $\delta_s = \Theta(n\rho_n)$ so Assumption~\ref{ass:RDPG-parameter} and Lemma~\ref{lem:deterministicControl} give
\begin{align*}\label{eq:b1}
\norm{\hat{u}_K - u_K}_{\infty} &\le \norm{u_K}_{\infty}\left( 8d_K(\frac{\norm{H}}{\delta_K})^2+\frac{2\norm{H}}{\lambda_K}\right) + Z_K + \frac{4\sqrt{2}\norm{H}}{\lambda_K}\sum_{1\le s < K}\lambda_s\min(\frac{d_s}{\delta_s},\frac{d_K}{\delta_K})\left( \norm{u_s}_{\infty} + Z_s\right)\\
&\le \norm{u_K}_{\infty}\left( 8K(\frac{2C\psi_1\sqrt{n\rho_n}}{n\rho_n})^2+\frac{2C\psi_1\sqrt{n\rho_n}}{n\rho_n}\right) + 16\psi_1\frac{\log^{\xi}{n}}{\sqrt{n\rho_n}}\norm{u_k}_{\infty}\\
&~~~~~~ +  \frac{4\sqrt{2}C\psi_1\sqrt{n\rho_n}}{n\rho_n}(K-1)\left( \frac{2\psi_1^2K n\rho_n}{n\rho_n}( \max_{1\le k \le K} \norm{u_k}_{\infty} + 8\psi_1\frac{\log^{\xi}{n}}{\sqrt{n\rho_n}}    \max_{1\le k \le K} \norm{u_k}_{\infty})\right)\\
& \le \max_{1\le k \le K} \norm{u_k}_{\infty} \cdot \frac{32\psi_1\log^{\xi}{n}}{\sqrt{n\rho_n}}
\end{align*}
for sufficiently large $n$. Therefore, from $\max_{1\le k\le K}\norm{u_k}_{\infty}\le \norm{U}_{2,\infty} \le \sqrt{\frac{K}{n}\mu(P)} \le \frac{\sqrt{K}t_n}{\sqrt{n}}$, we get
$$\norm{\hat{u}_K - u_K}_{\infty} \le \frac{32\psi_1\sqrt{K}t_n\log^{\xi}{n}}{n\sqrt{\rho_n}}.$$
Finally, taking $C_1 = 32\psi_1\sqrt{K}$ gives the result under the event $\cap_{k=0}^KE_k$, which for sufficiently large $n$, happens with probability at least
$$1-Kn^{-\frac{1}{4}(\log_{\mu}{n})^{\xi-1}(\log_{\mu}{e})^{-\xi}}-n^{-\delta} \ge 1-2n^{-\delta}.$$

\end{proof}

\section{Additional discussions of the method and empirical results}

\subsection{Rank estimation for general directed networks} 
\label{sec:lowrank}

Here we test ECV on the task of selecting the
best rank for a directed network model.  There are no obvious
competitors for this task, since the NCV is designed for the block
model family.   Assume $M = XY^T$  where $X, Y \in \bR^{n\times K}$
are such that $M_{ij} \in [0,1]$.  This can be viewed as  a directed
random dot product graph model \citep{young2007random}, with $K$ being
the dimension of its latent space.  We can use the ECV with either the
AUC loss or the SSE loss for model selection here, and with
either of the two stability selection methods (average or mode).    In simulations, we generate two $n \times K$  matrices $S_1$ and $S_2$ with each element drawn independently from the uniform distribution on $(0,1)$, and set $P = S_1S_2^T$.   We then normalize to [0, 1] by setting $M = (\max_{i,j}P_{ij})^{-1}P$ and generate the network adjacency matrix  $A$ with independent Bernoulli edges and $\e A = P$. 

We fix $K=3$ or $K=5$ and vary the number of nodes $n$.
The candidate set is $K \in \{1, 2, \cdots, 8\}$.
Table~\ref{tab:RankEstimation} shows the distribution of estimated
$\hat K$ under various settings.  When the sample size is sufficiently
large, all versions of ECV can estimate $K$ well. The AUC-based ECV is
always more accurate that the SSE-based ECV, and works better for
smaller sample sizes.   The estimation is already quite stable for this task so stability selection does not offer much improvement.

\begin{table}
\centering
{\footnotesize
\caption{Frequency of estimated rank values for directed random dot product graph model in 200 replications. The dashes ``-" indicates the zero occurrence. } 
\label{tab:RankEstimation}
\begin{tabular}{c|c|l|rrrrrrrr}
\hline
$K$ & $n$ & method & $\hat{K}$:  1 & 2 & 3 & 4 & 5 & 6 & 7 & 8 \\ 
  \hline
\multirow{6}{*}{3} &\multirow{6}{*}{600} &
ECV-AUC & 42 &  61 &  97 &  - &  - &  - &  - &  - \\ 
&&ECV-AUC-mode & 40 &  61 &  99 &  - &  - &  - &  - &  - \\ 
&&ECV-AUC-avg & 42 &  61 &  97 &  - &  - &  - &  - &  - \\ 
&&ECV-SSE &144 &  42 &  14 & - &  - &  - &  - &  - \\    
&&ECV-SSE-mode &157 &  39 &  4 & - &  - &  - &  - &  - \\    
&&ECV-SSE-avg &144 &  55 &  1 & - &  - &  - &  - &  - \\    
   \hline
\multirow{6}{*}{3} &\multirow{6}{*}{750} &
ECV-AUC & - &  1 &  199 &  - &  - &  - &  - &  - \\ 
&&ECV-AUC-mode & - &  1 &  199 &  - &  - &  - &  - &  - \\ 
&&ECV-AUC-avg & - &  1 &  199 &  - &  - &  - &  - &  - \\  
&&ECV-SSE &11 &  59 &  130 & - &  - &  - &  - &  - \\    
&&ECV-SSE-mode &6 &  52 &  142 & - &  - &  - &  - &  - \\    
&&ECV-SSE-avg &5 &  67 &  128 & - &  - &  - &  - &  - \\    
   \hline
\multirow{6}{*}{3} &\multirow{6}{*}{900} &
ECV-AUC & - &  - &  3 &  - &  - &  - &  - &  - \\ 
&&ECV-AUC-mode & - &  - &  3 &  - &  - &  - &  - &  - \\ 
&&ECV-AUC-avg & - &  - &  3 &  - &  - &  - &  - &  - \\ 
&&ECV-SSE &- &  4 &  196 & - &  - &  - &  - &  - \\    
&&ECV-SSE-mode &- &  2 &  198 & - &  - &  - &  - &  - \\    
&&ECV-SSE-avg &- & 2 &  198 & - &  - &  - &  - &  - \\    
   \hline
\multirow{6}{*}{5} &\multirow{6}{*}{1500} &
ECV-AUC & 39 &  20 &  26 &  33 &  82 &  - &  - &  - \\ 
&&ECV-AUC-mode & 31 &  20 &  28 &  33 &  88 &  - &  - &  - \\ 
&&ECV-AUC-avg & 39 &  20 &  26 &  33 &  82 &  - &  - &  - \\ 
&&ECV-SSE &  133 &  34 & 20 &  11 & 2 &  - &  - &  - \\    
&&ECV-SSE-mode &134 & 39 &  13 & 10 &  4 &  - &  - &  - \\    
&&ECV-SSE-avg &117 & 52 &  18 & 13 &  - &  - &  - &  - \\    
   \hline
\multirow{6}{*}{5} &\multirow{6}{*}{1800} &
ECV-AUC & - &  - &  1 &  3 &  196 &  - &  - &  - \\ 
&&ECV-AUC-mode & - &  - &  1 &  3 &  196 &  - &  - &  - \\ 
&&ECV-AUC-avg & - &  - &  1 &  3 &  196 &  - &  - &  - \\ 
&&ECV-SSE &  10 &  10 & 29 &  46 & 105 &  - &  - &  - \\    
&&ECV-SSE-mode &9 & 9 &  31 & 28 &  123 &  - &  - \\    
&&ECV-SSE-avg &4 & 13 &  30 & 47 &  106 &  - &  - &  - \\    
   \hline
\multirow{6}{*}{5} &\multirow{6}{*}{2000} &
ECV-AUC & - &  - &  - &  - &  200 &  - &  - &  - \\ 
&&ECV-AUC-mode & - &  - &  - &  - &  200 &  - &  - &  - \\ 
&&ECV-AUC-avg & - &  - &  - &  - &  200 &  - &  - &  - \\ 
&&ECV-SSE &  - &  - & 5 &  14 & 181 &  - &  - &  - \\    
&&ECV-SSE-mode &- & - &  6 & 11 &  183 &  - &  - \\    
&&ECV-SSE-avg &- & - &  5 & 17 &  178 &  - &  - &  - \\    
\hline
\end{tabular}
}
\end{table}

\newpage

\subsection{Tuning regularized spectral clustering}\label{secsec:tuneRSC}

Regularized spectral clustering has been proposed to improve performance of spectral clustering in sparse networks, but regularization itself frequently depends on a tuning parameter that has to be selected correctly in order to achieve the improvement.   Several different regularizations have been proposed and analyzed \citep{chaudhuri2012spectral, amini2013pseudo}.   ECV can be used to tune all of them, but for concreteness here we focus on the proposal by \cite{amini2013pseudo}, which replaces the usual normalized graph Laplacian $L = D^{-1/2}AD^{-1/2}$, where $D$ is the diagonal matrix of node degrees, by the Laplacian computed from the regularized adjacency matrix 
\begin{equation}\label{eq:reg2}
  A_{\tau} = A + \tau\frac{\hat{d}}{n}\mbone\mbone^T
\end{equation}
where $\hat{d}$ is the average node degree  and $\tau$ is a tuning parameter, typically within $[0,1]$.  The scale of the multiplier is motivated by theoretical results under the stochastic block model \citep{gao2015achieving, le2017concentration}.  
This regularization is known to improve concentration \citep{le2017concentration}, but also the larger $\tau$ is, the more noise it adds, and thus we aim to select the best value of $\tau$ that balances these two effects.  \cite{joseph2016impact}  proposed a data-driven way to select $\tau$ by Davis-Kahan estimate under the stochastic block model and the degree corrected model.   Using the ECV is an alternative general data-driven way of selecting $\tau$ which does not rely on model assumptions.   

 Choosing a good $\tau$ is expected to give good clustering accuracy, defined as proportion of correctly clustered nodes under the best cluster matching permutation, 
$$\max_{\hat{\V{c}}^p \in \text{perm}(\hat{\V{c}})}|\{i \in [n], \hat{c}^p_i = c_i\}|/n.$$
We can directly use Algorithm~\ref{algo:genericECV} with the candidate set $\ccal$ being a grid of $\tau$ values and the matrix completion procedure applied to regularized partial adjacency matrices for each $\tau$, as long as we can specify a loss function.     Ideally, we would prefer a model-free loss function, applicable even when the block model does not hold.    In general, choosing a loss function for cross-validation in clustering is difficult. While there is some work in the classical clustering setting \citep{tibshirani2001estimating, sugar2003finding, tibshirani2005cluster}, it has not been discussed much in the network setting, and the loss function we propose next, one of a number of reasonable options,  may be of independent interest.

For any cluster label vector $\V{c}$, the set of node pairs $\vcal\times\vcal$ will be divided into $K(K+1)/2$ classes defined by $H(i,j) = (c_i,c_j)$.   We treat each $H(i,j)$ as an unordered pair, since the network is undirected in spectral clustering.    To compare two vectors of labels $\V{c}_1$ and $\V{c}_2$, we can compare their corresponding partitions $H_1$ and $H_2$ by computing co-clustering difference or normalized mutual information (NMI) between them \citep{yao2003information}.   For instance, the co-clustering matrix for $H_1$ is defined to be the $n^2\times n^2$ matrix $G_1$ such that $G_{1,(j-1)n+i,(q-1)n+p} = \ind\{H_1(i,j)=H_1(p,q)\}$, reflecting whether or not two edges are in the same partition of $H_1$. Then the co-clustering difference between $H_1$ and $H_2$ is defined as the squared Frobenius norm of the difference between the two co-clustering matrices
$$\textsc{CCD}(H_1,H_2) = \norm{G_1-G_2}_F^2/2.$$

We apply this measure to choose the tuning parameter $\tau$ as follows:  for each split $m  = 1, 2, \cdots, N$ of ECV and each candidate value of $\tau$, we complete the adjacency matrix after removing the held-out entries and estimate cluster labels $\hat{c}^{(m)}_{\tau}$ and the corresponding $\hat{H}^{(m)}_{\tau}$ by regularized spectral clustering on the completed matrix with the candidate value of $\tau$.  We also compute $\hat H_{\tau}$, the partition corresponding to regularized spectral clustering on the full adjacency matrix with the same value of $\tau$.   Then we choose $\tau$ by comparing these partitions constrained to the held-out set, 
$$\hat{\tau} = \arg\min_{\tau \in \ccal} \sum_{m =1}^N \textsc{CCD}(\hat{H}^{(m)}_{\tau,\Omega^c_m},\hat H_{\tau,\Omega^c_m}).$$
Intuitively, if $\tau$ is a good value, the label vectors that generate $\hat{H}^{(m)}_{\tau,\Omega^c_m}$ and $\hat{H}_{\tau,\Omega^c_m}$ should both be close to the truth, and so the co-clustering matrices should be similar; if $\tau$ is a bad choice, then both label vectors will contain more errors, likely to be non-matching, and the corresponding co-clustering difference will be larger.

We test the ECV on this task on networks generated from the degree corrected model under the setting described in Section~\ref{secsec:sim-blockmodel}, with $n=600$, $K=3$, a power law distribution for $\theta_i$, balanced community sizes $\pi = (1/3, 1/3, 1/3)$, out-in ratio $\beta=0.2$, and average degree $\lambda=5$, since regularization is generally only relevant when the network is sparse.  The candidate set for the tuning parameter $\tau$ is $\ccal=\{0.1, 0.2, \cdots,1.9, 2\}$.   Without regularization, at this level of sparsity spectral clustering works very poorly.    We use the ECV procedure described in Section~\ref{secsec:tuneRSC} as well as its two stabilized versions to select $\tau$.   We also report the accuracy for each fixed value of $\tau$ in $\ccal$ as well as the the Davis-Kahan estimator (DKest)  of $\tau$ proposed by \cite{joseph2016impact}.

\begin{figure}[H]
\begin{center}
\includegraphics[width=0.8\textwidth]{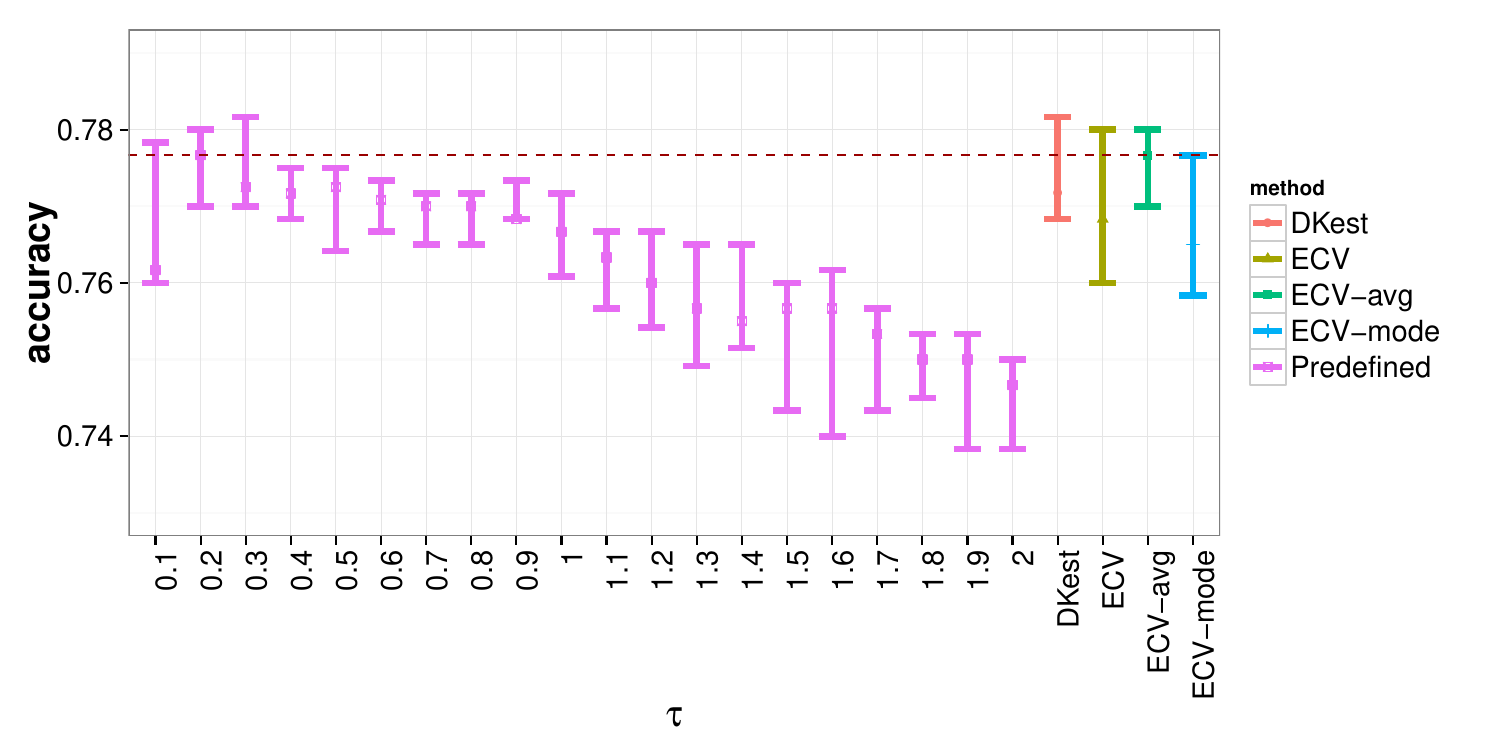}
\caption{The median clustering accuracy for different fixed values of $\tau$ and for DKest and ECV tuning.  The true model is degree corrected block model with $n=600$, $K=3$, $\lambda=5$, $\beta=0.2$ and $t=0$. }
\label{fig:reg-SP}
 \end{center}
\end{figure}

In the sparse setting, spectral clustering may occasionally suffer from bad local optima found by $K$-means. Thus we report the median clustering accuracy out of 200 replications, as well as its $95\%$ confidence interval calculated by bootstrap.  Figure~\ref{fig:reg-SP} shows the confidences intervals for the median accuracy of regularized spectral clustering for all tuning strategies out of 200.   Without regularization, the clustering accuracy is below 0.5 (not shown).   The accuracy jumps up with regularization for small $\tau$ values, and decreases slowly as $\tau$ increases.   All data-driven methods give close to optimal performance, with Davis-Kahan estimator and the ECV with stability selection by average giving the best result, closely followed by the ECV without stability selection and the ECV with stability selection by mode.   Again, considering that the Davis-Kahan estimator is a model-based method designed specifically for this purpose, and ECV is a generic tuning method, this is a good result for the ECV.  

\newpage

\subsection{Additional model selection results under the  block models}\label{appendix:deviance}
Due to the space, we only show the numerical study of the ECV for the block model selection in limited configurations in Section~\ref{secsec:sim-blockmodel} of the paper. In this section, we include more details and additional studies. To be self-contained, we describe the model generating method here again, with additional information.

For the degree corrected block model, we first sample 300 values from the
power law distribution with the lower bound $1$ and scaling parameter
$5$, and then set the node degree parameters $\theta_i$, $i=1,\cdots,
n$ by randomly and independently choosing one of these 300 values.
For the stochastic block model, we set $\theta_i =  1$ for all $i$.  Let $\pi \propto (1,
2^t, \cdots, K^t)$ be the proportions of nodes in the $K$ communities;
$t$ controls the size balance (when $t=0$ the communities have equal
sizes).  Let $B_0 = (1-\beta)  I+ \beta \mbone \mbone^T$ and $B
\propto \Theta B_0 \Theta$, so that $\beta$ is the out-in ratio (the
ratio of between-block probability and within-block probability of
edge).  The scaling is selected so that the average node degree is
$\lambda$.   We consider several combinations of size and the number of communities: $(n=600, K=3)$, $(n=600, K=5)$ and $(n=1200, K=5)$.   For each configuration, we then vary three aspects of the model:
\begin{enumerate}
\item Sparsity:  set the expected average degree $\lambda$ to $15$, $20$, $30$, or $40$, fixing $t=0$ and $\beta = 0.2$.
\item Community size: set $t$ to  $0$, $0.25$, $0.5$, or $1$, fixing $\lambda = 40$ and $\beta = 0.2$.
\item Out-in ratio: set $\beta$ to $0$, $0.25$, or $0.5$,  fixing $\lambda = 40$ and $t = 0$.
\end{enumerate}
 All results are based on 200 replications. The four methods compared on this task are the ECV (Algorithm~\ref{algo:ECV}) with $L_2$ loss and its stable version where the most frequent selection of 20 independent repetitions is returned, and the corresponding versions of the network CV procedure from \citet{chen2014network}. We only show the results from using the $L_2$ loss for model selection since we observed it works better than binomial deviance for both methods.

\begin{table}[H]
\def~{\hphantom{0}}
\tbl{Overall model selection by two cross-validation methods (fraction correct out of 200 replications). The true model is the degree corrected block model.}{%
\begin{tabular}{ccccc|rrrr}
 \\
  \multicolumn{5}{c}{Configurations} 
&  \multicolumn{2}{c}{Proposed method } &  \multicolumn{2}{c}{\citet{chen2014network}  } \\
\hline
$K$ & $n$ & $\lambda$ & t &$\beta$ & $L_2$ & $L_2$+stability  & $L_2$ & $L_2$+stability   \\ 
  \hline
\multirow{4}{*}{3} &\multirow{4}{*}{600} &
15 & 0 & 0.2 & 0.73&  0.87  &  0.00 &  0.00 \\ 
&&20 &0 & 0.2 &0.97 &  0.99 & 0.02 &  0.00 \\    
&&30 &0 & 0.2 &1.00 &  1.00 & 0.43 &  0.40 \\    
&&40 &0 & 0.2 &1.00 &  1.00  &  0.88 &  0.98 \\    
   \hline
\multirow{4}{*}{5} &\multirow{4}{*}{600} &
15 & 0 & 0.2 & 0.49 &  0.58  &  0.00 &  0.00 \\ 
&&20& 0 & 0.2  &0.90 &  0.95 & 0.00 &  0.00 \\    
&&30& 0 & 0.2  &0.99 &  1.00 & 0.05 &  0.01 \\    
&&40 & 0 & 0.2  &0.99 &  1.00  &  0.27 &  0.24 \\    
   \hline
\multirow{4}{*}{5} &\multirow{4}{*}{1200} &
15 & 0 & 0.2 & 0.67 &  0.76  &  0.00 &  0.00 \\ 
&&20& 0 & 0.2  &0.99 &  0.99 & 0.00 &  0.00 \\    
&&30& 0 & 0.2  &1.00 &  1.00 & 0.04 &  0.00\\    
&&40& 0 & 0.2  &1.00 &  1.00  &  0.41 &  0.33 \\    
\hline
\multirow{4}{*}{3} &\multirow{4}{*}{600} &
40 & 0 & 0.2 & 1.00 &  1.00  &  0.88 &  0.98\\ 
&&40 &0.25 & 0.2 &1.00 &  1.00 & 0.90 &  0.97 \\    
&&40 &0.5 & 0.2 &1.00 &  1.00 & 0.92 &  0.97 \\    
&&40 &1 & 0.2 &0.70 &  0.79  &  0.42 &  0.46 \\    
   \hline
\multirow{4}{*}{5} &\multirow{4}{*}{600} &
40 & 0 & 0.2 & 0.99 &  1.00  &  0.27 &  0.24 \\ 
&&40& 0.25 & 0.2  &0.98 &  1.00 & 0.28 &  0.29 \\    
&&40& 0.5 & 0.2  &0.77&  0.79 & 0.18 &  0.17\\    
&&40 & 1& 0.2  &0.11 &  0.06  &  0.05 &  0.00\\    
   \hline
\multirow{4}{*}{5} &\multirow{4}{*}{1200} &
40 & 0 & 0.2 & 1.00 &  1.00  &  0.41 &  0.33 \\ 
&&40& 0.25&0.2  & 1.00  &1.00&   0.44 &  0.39 \\    
&&40& 0.5 &0.2 & 0.81  &0.83  & 0.21 &  0.16\\    
&&40& 1 & 0.2  &0.10 &  0.06  &  0.00 &  0.06 \\    
\hline
\multirow{3}{*}{3} &\multirow{3}{*}{600} &
40& 0&0.1  & 1.00  &1.00&   0.99 &  1.00 \\    
&&40& 0 &0.2 & 1.00 &  1.00  &  0.88 &  0.98\\    
&&40& 0 & 0.5  &0.95 &  0.97  &  0.00 &  0.00 \\    
   \hline
\multirow{3}{*}{5} &\multirow{3}{*}{600} &
40& 0&0.1  & 1.00  &1.00&   0.79 &  0.96 \\    
&&40& 0 &0.2 & 0.99 &  1.00  &  0.27 &  0.24\\    
&&40& 0 & 0.5  &0.00 &  0.00  &  0.00 &  0.00 \\    
   \hline
\multirow{3}{*}{5} &\multirow{3}{*}{1200} &
40& 0&0.1  & 1.00  &1.00&   0.90 &  0.99 \\    
&&40& 0 &0.2 & 1.00 &  1.00  &  0.41 &  0.33\\    
&&40& 0 & 0.5  &0.00 &  0.00  &  0.00 &  0.00 \\    
\hline
\end{tabular}}
\label{tab:DCSBM-compareNCV-2}
\end{table}

Table~\ref{tab:DCSBM-compareNCV-2} shows the fraction of times the correct model was selected when the true
model is the degree corrected model, which is a more complete version of Table~\ref{tab:DCSBM-compareNCV}.   Table~\ref{tab:SBM-compareNCV} shows the counterpart of the results in Table~\ref{tab:DCSBM-compareNCV} under the stochastic block model as the true model. The task is easier under the stochastic block model as the model is simpler, but the general pattern is very similar to the degree corrected model setting. Stability selection clearly improves performance and the ECV performs better overall.

\begin{table}[H]
\def~{\hphantom{0}}
\tbl{Overall model selection by two cross-validation methods (fraction correct out of 200 replications). The true model is the stochastic block model.}{%
\begin{tabular}{ccccc|rrrr}
 \\
  \multicolumn{5}{c}{Configurations} 
&  \multicolumn{2}{c}{Proposed method } &  \multicolumn{2}{c}{\citet{chen2014network}  } \\
\hline
$K$ & $n$ & $\lambda$ & t &$\beta$ & $L_2$ & $L_2$+stability  & $L_2$ & $L_2$+stability   \\ 
  \hline
\multirow{4}{*}{3} &\multirow{4}{*}{600} &
15 & 0 & 0.2 & 1.00 & 1.00 & 0.99 & 1.00 \\ 
&&20& 0 & 0.2  & 1.00 & 1.00 & 1.00 & 1.00 \\ 
 && 30& 0 & 0.2  & 1.00 & 1.00 & 0.99 & 1.00 \\ 
&&40 & 0 & 0.2  & 1.00 & 1.00 & 1.00 & 1.00 \\ 
   \hline
\multirow{4}{*}{5} &\multirow{4}{*}{600} &
15 & 0 & 0.2 & 0.81 & 0.88 & 0.71 & 0.86 \\ 
&&20& 0 & 0.2  & 1.00 & 1.00 & 0.98 & 1.00 \\ 
&&30& 0 & 0.2  &  1.00 & 1.00 & 0.98 & 1.00 \\ 
&&40 & 0 & 0.2  & 0.99 & 1.00 & 0.98 & 1.00 \\ 
   \hline
\multirow{4}{*}{5} &\multirow{4}{*}{1200} &
15 & 0 & 0.2 & 0.98 & 0.98 & 0.91 & 0.96 \\ 
&&20& 0 & 0.2  & 1.00 & 1.00 & 0.99 & 1.00 \\ 
&&30& 0 & 0.2  & 1.00 & 1.00 & 0.96 & 1.00 \\ 
&&40 & 0 & 0.2  & 1.00 & 1.00 & 0.97 & 1.00 \\ 
  \hline
\multirow{4}{*}{3} &\multirow{4}{*}{600} &
40 & 0 & 0.2 & 1.00 & 1.00 & 1.00 & 1.00 \\ 
&&40& 0.25 & 0.2  & 1.00 & 1.00 & 0.99 & 1.00 \\ 
&&40& 0.5 & 0.2  & 1.00 & 1.00 & 0.97 & 1.00 \\ 
&&40 & 1& 0.2  & 0.73 & 0.76 & 0.34 & 0.43 \\ 
     \hline
\multirow{4}{*}{5} &\multirow{4}{*}{600} &
40 & 0 & 0.2 &   0.99 & 1.00 & 0.98 & 1.00 \\ 
&&40& 0.25 & 0.2  &  1.00 & 1.00 & 0.95 & 1.00 \\ 
&&40& 0.5 & 0.2  & 0.82 & 0.86 & 0.64 & 0.78 \\ 
&&40 & 1& 0.2  & 0.06 & 0.01 & 0.17 & 0.10 \\ 
   \hline
\multirow{4}{*}{5} &\multirow{4}{*}{1200} &
40 & 0 & 0.2 & 1.00 & 1.00 & 0.97 & 1.00 \\ 
&&40& 0.25 & 0.2  & 1.00 & 1.00 & 0.98 & 1.00 \\ 
&&40& 0.5 & 0.2  & 0.93 & 0.94 & 0.73 & 0.89 \\ 
&&40 & 1& 0.2  & 0.01 & 0.01 & 0.21 & 0.06 \\
\hline
\multirow{3}{*}{3} &\multirow{3}{*}{600} &
40& 0&0.1  & 1.00 & 1.00 & 0.98 & 1.00 \\ 
&&40& 0 &0.2 & 1.00 & 1.00 & 1.00 & 1.00 \\ 
&&40& 0 & 0.5 & 0.92 & 0.96 & 0.83 & 0.96 \\ 
\hline
\multirow{3}{*}{5} &\multirow{3}{*}{600} &
40& 0&0.1  & 0.99 & 1.00 & 0.97 & 1.00 \\ 
&&40& 0 &0.2 &  0.99 & 1.00 & 0.98 & 1.00 \\ 
&&40& 0 & 0.5 & 0.00 & 0.00 & 0.00 & 0.00 \\ 
     \hline
\multirow{3}{*}{5} &\multirow{3}{*}{1200} &
40& 0&0.1  & 1.00 & 1.00 & 0.98 & 1.00 \\ 
&&40& 0 &0.2 & 1.00 & 1.00 & 0.97 & 1.00 \\ 
&&40& 0 & 0.5 & 0.00 & 0.00 & 0.00 & 0.00 \\
\hline
\end{tabular}}
\label{tab:SBM-compareNCV}
\end{table}

\begin{table}[H]
\centering
{\small
\caption{Overall correct selection rate by two cross-validation methods in 200 replications when binomial deviance is used as the loss function. The true model is the degree corrected block model.} 
\label{tab:DCSBM-compareNCV-dev}
\begin{tabular}{ccccc|rrrr}
  \multicolumn{5}{c}{Configurations} 
&  \multicolumn{2}{c}{Proposed method } &  \multicolumn{2}{c}{\citet{chen2014network}  } \\
\hline
$K$ & $n$ & $\lambda$ & t &$\beta$ & deviance &deviance+stability  & deviance &deviance+stability   \\ 
  \hline
\multirow{4}{*}{3} &\multirow{4}{*}{600} &
15 & 0 & 0.2 & 0.47 & 0.45 & 0.00 & 0.00 \\ 
&&20& 0 & 0.2  & 0.89 & 0.96 & 0.00 & 0.00 \\ 
&&30& 0 & 0.2  & 1.00 & 1.00 & 0.26 & 0.14 \\ 
&&40 & 0 & 0.2  & 1.00 & 1.00 & 0.84 & 0.96 \\ 
   \hline
\multirow{4}{*}{5} &\multirow{4}{*}{600} &
15 & 0 & 0.2 & 0.34 & 0.40 & 0.00 & 0.00 \\ 
&&20& 0 & 0.2  & 0.82 & 0.93 & 0.00 & 0.00 \\ 
&&30& 0 & 0.2  & 0.97 & 1.00 & 0.01 & 0.00 \\ 
&&40 & 0 & 0.2  & 0.99 & 1.00 & 0.13 & 0.10 \\ 
   \hline
\multirow{4}{*}{5} &\multirow{4}{*}{1200} &
15 & 0 & 0.2 & 0.45 & 0.53 & 0.00 & 0.00 \\ 
&&20& 0 & 0.2  & 0.94 & 0.98 & 0.00 & 0.00 \\ 
&&30& 0 & 0.2  & 1.00 & 1.00 & 0.00 & 0.00 \\ 
&&40 & 0 & 0.2  & 1.00 & 1.00 & 0.23 & 0.15 \\ \hline
\hline

\multirow{4}{*}{3} &\multirow{4}{*}{600} &
40 & 0 & 0.2 & 1.00 & 1.00 & 0.86 & 0.95 \\ 
&&40& 0.25 & 0.2  & 1.00 & 1.00 & 0.89 & 0.94 \\ 
&&40& 0.5 & 0.2  & 1.00 & 1.00 & 0.89 & 0.95 \\ 
&&40 & 1& 0.2  & 0.64 & 0.71 & 0.29 & 0.41 \\ 
   \hline
\multirow{4}{*}{5} &\multirow{4}{*}{600} &
40 & 0 & 0.2 & 0.97 & 1.00 & 0.16 & 0.10 \\ 
&&40& 0.25 & 0.2  & 0.98 & 1.00 & 0.12 & 0.09 \\ 
&&40& 0.5 & 0.2  & 0.72 & 0.79 & 0.07 & 0.04 \\ 
&&40 & 1& 0.2  &  0.12 & 0.07 & 0.07 & 0.03 \\    \hline
\multirow{4}{*}{5} &\multirow{4}{*}{1200} &
40 & 0 & 0.2 & 1.00 & 1.00 & 0.21 & 0.17 \\ 
&&40& 0.25 & 0.2  & 1.00 & 1.00 & 0.21 & 0.15 \\ 
&&40& 0.5 & 0.2  & 0.81 & 0.82 & 0.09 & 0.04 \\ 
&&40 & 1& 0.2  & 0.09 & 0.07 & 0.02 & 0.01 \\
\hline
\multirow{3}{*}{3} &\multirow{3}{*}{600} &
40& 0&0.1  & 1.00 & 1.00 & 0.99 & 1.00 \\ 
&&40& 0 &0.2 & 1.00 & 1.00 & 0.85 & 0.96 \\ 
&&40& 0 & 0.5  & 0.96 & 0.97 & 0.00 & 0.00 \\ 
     \hline
\multirow{3}{*}{5} &\multirow{3}{*}{600} &
40& 0&0.1  & 0.99 & 1.00 & 0.59 & 0.82 \\ 
&&40& 0 &0.2 & 0.98 & 1.00 & 0.18 & 0.10 \\ 
&&40& 0 & 0.5 & 0.00 & 0.00 & 0.00 & 0.00 \\ 
     \hline
\multirow{3}{*}{5} &\multirow{3}{*}{1200} &
40& 0&0.1  & 1.00 & 1.00 & 0.93 & 1.00 \\ 
&&40& 0 &0.2 & 0.99 & 1.00 & 0.22 & 0.15 \\ 
&&40& 0 & 0.5 &  0.00 & 0.00 & 0.00 & 0.00 \\ 
\hline
\end{tabular}
}
\end{table}

\begin{table}[H]
\centering
{\small
  \caption{Overall correct selection rate of two cross-validation methods in 200 replications when binomial deviance is used as the loss function. The true model is the stochastic block model.} 
\label{tab:SBM-compareNCV-dev}
\begin{tabular}{ccccc|rrrr}
  \multicolumn{5}{c}{Configurations} 
&  \multicolumn{2}{c}{Proposed method } &  \multicolumn{2}{c}{\citet{chen2014network}  } \\
\hline
$K$ & $n$ & $\lambda$ & t &$\beta$ & deviance &deviance+stability  & deviance &deviance+stability   \\ 
  \hline
\multirow{4}{*}{3} &\multirow{4}{*}{600} &
15 & 0 & 0.2 & 1.00 & 1.00 & 0.98 & 1.00 \\ 
&&20& 0 & 0.2  & 1.00 & 1.00 & 0.99 & 1.00 \\ 
&&30& 0 & 0.2  & 1.00 & 1.00 & 0.99 & 1.00 \\ 
&&40 & 0 & 0.2  & 1.00 & 1.00 & 0.99 & 1.00 \\ 
 \hline
\multirow{4}{*}{5} &\multirow{4}{*}{600} &
15 & 0 & 0.2 & 0.82 & 0.89 & 0.71 & 0.87 \\ 
&&20& 0 & 0.2  & 0.99 & 1.00 & 0.97 & 1.00 \\ 
&&30& 0 & 0.2  & 0.99 & 1.00 & 0.98 & 1.00 \\ 
&&40 & 0 & 0.2  & 1.00 & 1.00 & 0.97 & 1.00 \\ 
     \hline
\multirow{4}{*}{5} &\multirow{4}{*}{1200} &
15 & 0 & 0.2 & 0.97 & 0.98 & 0.92 & 0.96 \\ 
&&20& 0 & 0.2  & 1.00 & 1.00 & 1.00 & 1.00 \\ 
&&30& 0 & 0.2  & 1.00 & 1.00 & 0.96 & 1.00 \\ 
&&40 & 0 & 0.2  & 1.00 & 1.00 & 0.96 & 1.00 \\ 
\hline
\multirow{4}{*}{3} &\multirow{4}{*}{600} &
40 & 0 & 0.2 & 1.00 & 1.00 & 1.00 & 1.00 \\ 
&&40& 0.25 & 0.2  & 1.00 & 1.00 & 0.99 & 1.00 \\ 
&&40& 0.5 & 0.2  & 1.00 & 1.00 & 0.95 & 1.00 \\ 
&&40 & 1& 0.2  &  0.67 & 0.73 & 0.26 & 0.38 \\ 
     \hline
\multirow{4}{*}{5} &\multirow{4}{*}{600} &
40 & 0 & 0.2 &0.99 & 1.00 & 0.94 & 1.00 \\ 
&&40& 0.25 & 0.2  & 1.00 & 1.00 & 0.95 & 1.00 \\ 
&&40& 0.5 & 0.2  & 0.80 & 0.86 & 0.58 & 0.74 \\ 
&&40 & 1& 0.2  &   0.12 & 0.04 & 0.21 & 0.12 \\   
 \hline
\multirow{4}{*}{5} &\multirow{4}{*}{1200} &
40 & 0 & 0.2 & 1.00 & 1.00 & 0.97 & 1.00 \\ 
&&40& 0.25 & 0.2  & 1.00 & 1.00 & 0.96 & 1.00 \\ 
&&40& 0.5 & 0.2  & 0.93 & 0.94 & 0.68 & 0.84 \\ 
&&40 & 1& 0.2  &  0.05 & 0.01 & 0.23 & 0.11 \\ 
 \hline
\multirow{3}{*}{3} &\multirow{3}{*}{600} &
40& 0&0.1  & 1.00 & 1.00 & 0.98 & 1.00 \\ 
&&40& 0 &0.2 & 1.00 & 1.00 & 0.99 & 1.00 \\ 
&&40& 0 & 0.5 & 0.94 & 0.97 & 0.85 & 0.97 \\ 
     \hline
\multirow{3}{*}{5} &\multirow{3}{*}{600} &
40& 0&0.1  & 0.99 & 1.00 & 0.96 & 1.00 \\ 
&&40& 0 &0.2 & 0.99 & 1.00 & 0.93 & 1.00 \\ 
&&40& 0 & 0.5 & 0.00 & 0.00 & 0.00 & 0.00 \\
     \hline
\multirow{3}{*}{5} &\multirow{3}{*}{1200} &
40& 0&0.1  & 1.00 & 1.00 & 0.97 & 1.00 \\ 
&&40& 0 &0.2 & 1.00 & 1.00 & 0.97 & 1.00 \\ 
&&40& 0 & 0.5 & 0.00 & 0.00 & 0.00 & 0.00 \\ 
\hline
\end{tabular}
}
\end{table}

\newpage

\subsection{Selecting the number of communities}\label{appendix:KselectionSBM}

When the type of a block model (stochastic block model or the degree corrected model) is known or assumed, there are multiple
methods available for selecting the number of communities $K$, which
can be benchmarked against cross-validation methods.  For this task, we compare the previously mentioned two CV methods, as well as the model-free ECV with the sum of squared error and the AUC as loss functions,
 described in Section \ref{secsec:ECV-AUC}, with two versions of stability selection (average or
 mode).     
We include two additional methods designed specifically for choosing
$K$ under block models, which we would expect to be at least as
accurate as cross-validation considering that they use the true model
and cross-validation does not.   The method of
\cite{wang2015likelihood} is a BIC-type criterion based on an
asymptotic analysis of the likelihood ratio statistic.    Another
BIC-type method proposed by \cite{saldana2014how} is based on the
composite likelihood but it is computationally infeasible for
networks with more than 1000 nodes (using the implementation on the
authors' website) and it was less accurate than LR-BIC in our
experiments on smaller networks, so we omitted it from comparisons.
From the class of eigenvalues-based methods proposed by
\cite{le2015estimating}, we include the best-performing variant based
on the Bethe-Hessian matrix with moment correction.  Due to the
large number of methods, we first compare just the cross-validation
methods, cross-validation methods, and their stabilized
versions, and then we compare the best of the cross-validation methods
with the other two.  Only the results under the degree corrected model as the true model are included but the results are similar when the true model is the stochastic block model.
 
 Table~\ref{tab:K-Selection-CV-Lambda} shows the comparison between
 the cross-validation methods when we vary the average network degree
 with fixed $\beta=0.2$ and balanced communities. Both stability
 selection methods improve the accuracy of  ECV,  and the average is
 better for all versions of the ECV.    On the other hand, for \citet{chen2014network} the
 most frequently selected $K$ is typically more accurate than the
 rounded average.    Further,  all the variants of the ECV work as
 well as or better than \citet{chen2014network} in all configurations. 
 
Tables~\ref{tab:K-Selection-CV-Pi} and \ref{tab:K-Selection-CV-Beta} compare the same methods when we vary $t$ fixing $\lambda$ and $\beta$, and vary $\beta$ while fixing $\lambda$ and $t$, respectively. The pattern is very similar to Table~\ref{tab:K-Selection-CV-Lambda}.

\begin{table}[H]
\centering
{\small
\caption{The rate of correctly estimating the number of communities (out of 200 replications) when varying the network average degree and fixing $t=0$, $\beta=0.2$. The true model is the degree corrected block model.} 
\label{tab:K-Selection-CV-Lambda}
\begin{tabular}{ccc|rrr|rrr|rrr|rrr}
  \hline
 \multicolumn{3}{c|}{} & \multicolumn{9}{c|}{Proposed Method} & \multicolumn{3}{c}{\citet{chen2014network}} \\
\hline
 \multicolumn{3}{c|}{Configurations} & \multicolumn{3}{c|}{$L2$} & \multicolumn{3}{c|}{AUC} & \multicolumn{3}{c|}{sum. sq. err.} & \multicolumn{3}{c}{$L_2$} \\
\hline
$K$ & $n$ & $\lambda$ &  & mode  & avg & & mode & avg  & & mode & avg & & mode & avg\\ 
  \hline
\multirow{4}{*}{3} &\multirow{4}{*}{600} &
15 & 0.99 & 1.00 & 1.00 & 0.99 & 1.00 & 1.00 & 0.99 & 1.00 & 1.00& 0.82 & 0.99 & 0.94  \\ 
& &20 & 1.00 & 1.00 & 1.00 & 1.00 & 1.00 & 1.00 & 1.00 & 1.00 & 1.00& 0.98 & 1.00 & 1.00  \\ 
 & &30 & 1.00 & 1.00 & 1.00 & 1.00 & 1.00 & 1.00 & 1.00 & 1.00 & 1.00& 0.99 & 1.00 & 1.00  \\ 
 & &40 & 1.00 & 1.00 & 1.00 & 1.00 & 1.00 & 1.00 & 1.00 & 1.00 & 1.00& 0.99 & 1.00 & 1.00  \\ 
  \hline
\multirow{4}{*}{5} &\multirow{4}{*}{600} &
15 & 0.57 & 0.60 & 0.72 & 0.55 & 0.59 & 0.68 & 0.33 & 0.34 & 0.68& 0.01 & 0.00 & 0.00  \\ 
&&20 & 0.92 & 0.95 & 0.96 & 0.93 & 0.96 & 0.99 & 0.86 & 0.91 & 0.99& 0.43 & 0.67 & 0.36  \\ 
 && 30 & 0.99 & 1.00 & 1.00  & 1.00 & 1.00 & 1.00 & 1.00 & 1.00 & 1.00& 0.76 & 0.99 & 0.91 \\ 
 & &40 & 0.99 & 1.00 & 1.00  & 1.00 & 1.00 & 1.00 & 1.00 & 1.00 & 1.00& 0.76 & 0.98 & 0.89 \\ 
   \hline
\multirow{4}{*}{5} &\multirow{4}{*}{1200} &
15  & 0.74 & 0.79 & 0.85& 0.73 & 0.79 & 0.83 & 0.22 & 0.26 & 0.83 & 0.01 & 0.00 & 0.00  \\ 
& &20 & 0.99 & 0.99 & 1.00 & 0.98 & 0.99 & 0.99 & 0.94 & 0.97 & 0.99& 0.76 & 0.95 & 0.67  \\ 
 & &30 & 1.00 & 1.00 & 1.00 & 1.00 & 1.00 & 1.00 & 1.00 & 1.00 & 1.00& 0.98 & 1.00 & 1.00  \\ 
 & &40 & 1.00 & 1.00 & 1.00 & 1.00 & 1.00 & 1.00 & 1.00 & 1.00 & 1.00& 0.96 & 1.00 & 1.00  \\ 
   \hline
   \end{tabular}
}
\end{table}

\begin{table}[H]
\centering
{\small
\caption{The rate of correctly estimating the number of communities (out of 200 replications) when varying $t$ and fixing $\lambda=40$, $\beta=0.2$. The true model is the degree corrected block model.} 
\label{tab:K-Selection-CV-Pi}
\begin{tabular}{ccc|rrr|rrr|rrr|rrr}
  \hline
 \multicolumn{3}{c|}{} & \multicolumn{9}{c|}{Proposed Method} & \multicolumn{3}{c}{\citet{chen2014network}} \\
\hline
 \multicolumn{3}{c|}{Configurations} & \multicolumn{3}{c|}{$L2$} & \multicolumn{3}{c|}{AUC} & \multicolumn{3}{c|}{sum. sq. err.} & \multicolumn{3}{c}{$L_2$} \\
\hline
$K$ & $n$ & $t$ &  & mode  & avg & & mode & avg  & & mode & avg & & mode & avg\\ 
  \hline
\multirow{4}{*}{3} &\multirow{4}{*}{600} &
0 & 1.00 & 1.00 & 1.00 & 1.00 & 1.00 & 1.00 & 1.00 & 1.00 & 1.00& 0.99 & 1.00 & 1.00  \\ 
&& 0.25 & 1.00 & 1.00 & 1.00 & 1.00 & 1.00 & 1.00 & 1.00 & 1.00 & 1.00& 0.99 & 1.00 & 1.00  \\ 
&& 0.5 & 1.00 & 1.00 & 1.00 & 1.00 & 1.00 & 1.00 & 1.00 & 1.00 & 1.00 & 0.99 & 1.00 & 1.00 \\ 
&& 1 & 0.70 & 0.79 & 0.67 & 0.85 & 0.86 & 0.90 & 0.82 & 0.83 & 0.90 & 0.49 & 0.48 & 0.49  \\ 
   \hline
\multirow{4}{*}{5} &\multirow{4}{*}{600} &
0 & 0.99 & 1.00 & 1.00 & 1.00 & 1.00 & 1.00 & 1.00 & 1.00 & 1.00& 0.76 & 0.98 & 0.89  \\ 
&& 0.25 & 0.98 & 1.00 & 1.00  & 1.00 & 1.00 & 1.00 & 1.00 & 1.00 & 1.00& 0.64 & 0.95 & 0.84 \\ 
&& 0.5 & 0.77 & 0.80 & 0.80 & 0.80 & 0.80 & 0.83 & 0.74 & 0.77 & 0.83 & 0.36 & 0.55 & 0.70 \\ 
&&1 & 0.11 & 0.06 & 0.07 & 0.03 & 0.01 & 0.01 & 0.01 & 0.00 & 0.01& 0.06 & 0.01 & 0.01  \\ 
   \hline
\multirow{4}{*}{5} &\multirow{4}{*}{1200} &
0 & 1.00 & 1.00 & 1.00 & 1.00 & 1.00 & 1.00 & 1.00 & 1.00 & 1.00& 0.96 & 1.00 & 1.00 \\ 
&&0.25 & 1.00 & 1.00 & 1.00 & 1.00 & 1.00 & 1.00 & 1.00 & 1.00 & 1.00& 0.97 & 1.00 & 1.00  \\ 
&&0.5 & 0.81 & 0.83 & 0.83 & 0.86 & 0.89 & 0.91 & 0.74 & 0.74 & 0.91 & 0.60 & 0.64 & 0.66 \\ 
&&1 & 0.10 & 0.06 & 0.07 & 0.04 & 0.01 & 0.01 & 0.00 & 0.00 & 0.01& 0.01 & 0.00 & 0.00  \\ 
   \hline
   \end{tabular}
}
\end{table}

\begin{table}[H]
\centering
{\small
\caption{The rate of correctly estimating the number of communities (out of 200 replications) when varying $\beta$ and fixing $\lambda=40$, $t=0$. The true model is the degree corrected block model.} 
\label{tab:K-Selection-CV-Beta}
\begin{tabular}{ccc|rrr|rrr|rrr|rrr}
  \hline
 \multicolumn{3}{c|}{} & \multicolumn{9}{c|}{Proposed Method} & \multicolumn{3}{c}{\citet{chen2014network}} \\
\hline
 \multicolumn{3}{c|}{Configurations} & \multicolumn{3}{c|}{$L2$} & \multicolumn{3}{c|}{AUC} & \multicolumn{3}{c|}{sum. sq. err.} & \multicolumn{3}{c}{$L_2$} \\
\hline
$K$ & $n$ & $\beta$ &  & mode  & avg & & mode & avg  & & mode & avg & & mode & avg\\ 
  \hline
\multirow{3}{*}{3} &\multirow{3}{*}{600} &
0.1 & 1.00 & 1.00 & 1.00 & 1.00 & 1.00 & 1.00 & 1.00 & 1.00 & 1.00& 1.00 & 1.00 & 1.00  \\ 
 && 0.2 & 1.00 & 1.00 & 1.00 & 1.00 & 1.00 & 1.00 & 1.00 & 1.00 & 1.00 & 0.99 & 1.00 & 1.00 \\ 
  && 0.5 & 0.96 & 1.00 & 1.00  & 0.96 & 1.00 & 1.00 & 0.12 & 0.16 & 1.00 & 0.00 & 0.00 & 0.00\\ 
   \hline
\multirow{3}{*}{5} &\multirow{3}{*}{600} &
0.1 & 1.00 & 1.00 & 1.00  & 0.98 & 1.00 & 0.99 & 1.00 & 1.00 & 0.99& 0.85 & 1.00 & 0.99 \\ 
 & &0.2 & 0.99 & 1.00 & 1.00 & 1.00 & 1.00 & 1.00 & 1.00 & 1.00 & 1.00& 0.76 & 0.98 & 0.89  \\ 
  &&0.5 & 0.00 & 0.00 & 0.00 & 0.00 & 0.00 & 0.00 & 0.00 & 0.00 & 0.00 & 0.00 & 0.00 & 0.00 \\ 
   \hline
\multirow{3}{*}{5} &\multirow{3}{*}{1200} &
0.1 & 1.00 & 1.00 & 1.00 & 0.99 & 1.00 & 1.00 & 1.00 & 1.00 & 1.00& 0.95 & 1.00 & 1.00  \\ 
 && 0.2 & 1.00 & 1.00 & 1.00  & 1.00 & 1.00 & 1.00 & 1.00 & 1.00 & 1.00& 0.96 & 1.00 & 1.00 \\ 
 && 0.5 & 0.00 & 0.00 & 0.00 & 0.00 & 0.00 & 0.00 & 0.00 & 0.00 & 0.00 & 0.00 & 0.00 & 0.00 \\ 
   \hline
   \end{tabular}
}
\end{table}

Next, we compare the best variants of the two cross-validation methods with the model-based methods with results shown in Table~\ref{tab:DCSBM-KSelection-All}.   The methods of \citet{wang2015likelihood} and \citet{le2015estimating} perform perfectly most of the time, and outperform cross-validation when $K$ is large and the network is sparse (harder settings).   This is expected since cross-validation is a general method and the other two rely on the true model; they also cannot be applied to any other tasks.   It is also not clear how they behave under model misspecification (important given that in the real world not many networks follow exactly the stochastic block model or the degree corrected model), while cross-validation can still be expected to give reasonable results;   in particular, the ECV selection can be interpreted as the optimal model from the block model family in terms of link prediction for the observed network.

\begin{table}[H]
\centering
{\footnotesize
\caption{The rate of correctly estimating the number of communities (out of 200 replications) for the best variant of each method. The true model is the degree corrected block model.} 
\label{tab:DCSBM-KSelection-All}
\begin{tabular}{c|c|c|c|c|rrrrr}
  \hline
 \multicolumn{5}{c}{Configurations} & \multicolumn{2}{c}{Proposed Method} & \citet{chen2014network}& \citet{wang2015likelihood}& \citet{le2015estimating}\\
\hline
$K$ & $n$ & $\lambda$ & t &$\beta$ & $L_2$ + avg.stable  & AUC+avg.stable & $L_2$+ mode.stable &  &   \\ 
  \hline
\multirow{4}{*}{3} &\multirow{4}{*}{600} &
15 &0 & 0.2& 1.00  & 1.00 & 0.99& 1.00 & 1.00 \\ 
 && 20&0 & 0.2 & 1.00 & 1.00 & 1.00 & 1.00 & 1.00 \\ 
 &&30&0 & 0.2 & 1.00 & 1.00 & 1.00 & 1.00 & 1.00 \\ 
 && 40&0 & 0.2 & 1.00 & 1.00 & 1.00 & 1.00 & 1.00 \\ 
   \hline
\multirow{4}{*}{5} &\multirow{4}{*}{600} &
15&0 & 0.2 & 0.72  & 0.59 & 0.00& 1.00 & 1.00 \\ 
 && 20&0 & 0.2 & 0.96  & 0.96 & 0.67& 1.00 & 1.00 \\ 
&&30&0 & 0.2 & 1.00  & 1.00 & 0.99& 1.00 & 1.00 \\ 
&&40&0 & 0.2 & 1.00  & 1.00 & 0.98& 1.00 & 1.00 \\ 
   \hline
\multirow{4}{*}{5} &\multirow{4}{*}{1200} &
15 &0 & 0.2& 0.85  & 0.79 & 0.00& 1.00 & 1.00 \\ 
&&20&0 & 0.2 & 1.00  & 0.99 & 0.95& 1.00 & 1.00 \\ 
 &&30 &0 & 0.2& 1.00  & 1.00 & 1.00& 1.00 & 1.00 \\ 
&&40 &0 & 0.2& 1.00 & 1.00 & 1.00 & 1.00 & 1.00 \\ 
   \hline
\multirow{4}{*}{3} &\multirow{4}{*}{600} &
40 & 0 & 0.2 & 1.00 & 1.00 & 1.00 & 1.00 & 1.00 \\ 
&& 40 & 0.25 & 0.2 & 1.00 & 1.00 & 1.00 & 1.00 & 1.00 \\ 
&& 40 & 0.5 & 0.2 & 1.00 & 1.00 & 1.00 & 1.00 & 1.00 \\ 
&& 40 & 1& 0.2 & 0.67  & 0.86 & 0.48& 1.00 & 1.00 \\ 
   \hline
\multirow{4}{*}{5} &\multirow{4}{*}{600} &
40 & 0 & 0.2 & 1.00  & 1.00 & 0.98& 1.00 & 1.00 \\ 
 && 40 & 0.25 & 0.2  & 1.00 & 1.00 & 0.95 & 1.00 & 1.00 \\ 
&& 40 & 0.5 & 0.2  & 0.80  & 0.80 & 0.55& 1.00 & 0.99 \\ 
&& 40 & 1& 0.2 & 0.07 & 0.01 & 0.01 & 0.46 & 0.10 \\ 
   \hline  
\multirow{4}{*}{5} &\multirow{4}{*}{1200} &
40 & 0 & 0.2 & 1.00 & 1.00 & 1.00 & 1.00 & 1.00 \\ 
 && 40 & 0.25 & 0.2  & 1.00 & 1.00 & 1.00 & 1.00 & 1.00 \\ 
&& 40 & 0.5 & 0.2 & 0.83  & 0.89& 0.64 & 1.00 & 1.00 \\ 
&& 40 & 1& 0.2  & 0.07  & 0.01 & 0.00& 0.45 & 0.12 \\ 
   \hline
\multirow{3}{*}{3} &\multirow{3}{*}{600} &
 40 & 0 & 0.1  & 1.00 & 1.00 & 1.00 & 1.00 & 1.00 \\ 
&& 40 & 0 & 0.2  & 1.00 & 1.00 & 1.00 & 1.00 & 1.00 \\ 
&& 40 & 0 & 0.5  & 0.98  & 0.96 & 0.00& 1.00 & 1.00 \\ 
   \hline         
\multirow{3}{*}{5} &\multirow{3}{*}{600} &
 40 & 0 & 0.1   & 1.00 & 1.00 & 1.00 & 1.00 & 1.00 \\ 
&& 40 & 0 & 0.2   & 1.00  & 1.00 & 0.98& 1.00 & 1.00 \\ 
&& 40 & 0 & 0.5  & 0.00 & 0.00 & 0.00 & 0.00 & 0.00 \\ 
   \hline
\multirow{3}{*}{5} &\multirow{3}{*}{1200} &
 40 & 0 & 0.1  & 1.00 & 1.00 & 1.00 & 1.00 & 1.00 \\ 
&& 40 & 0 & 0.2  & 1.00 & 1.00 & 1.00 & 1.00 & 1.00 \\ 
&& 40 & 0 & 0.5 & 0.00 & 0.00 & 0.00 & 0.00 & 0.00 \\ 
   \hline
   
      \end{tabular}
   }
\end{table}

Table~\ref{tab:DCSBM-KSelection-All} in the paper shows the accuracy of selecting $K$ from multiple methods under the degree corrected model, and results under the stochastic block model are given in Table~\ref{tab:SBM-KSelection-All} below. The pattern is similar, except the ECV variant with AUC as the loss has a problem with perfectly separated communities ($\beta=0$, an unrealistic scenario, presumably due to many ties affecting the AUC). 

\begin{table}[htb]
\centering
{\footnotesize
\caption{The correct rate for estimating the number of communities in 200 replications from the best variant of each method. The underlying true model is the stochastic block model.} 
\label{tab:SBM-KSelection-All}
\begin{tabular}{c|c|c|c|c|rrrrr}
  \hline
 \multicolumn{5}{c}{Configurations} & \multicolumn{2}{c}{Proposed Method} & \citet{chen2014network}& \citet{wang2015likelihood}& \citet{le2015estimating}\\
\hline
$K$ & $n$ & $\lambda$ & t &$\beta$ & $L_2$ + avg.stable  & AUC+avg.stable & $L_2$+ mode.stable &  &   \\ 
  \hline
\multirow{4}{*}{3} &\multirow{4}{*}{600} &
15&0 & 0.2 &1.00 & 1.00 & 1.00 & 1.00 & 1.00 \\ 
&& 20&0 & 0.2 & 1.00 & 1.00 & 1.00 & 1.00 & 0.99 \\ 
&&30&0 & 0.2 & 1.00 & 1.00 & 1.00 & 1.00 & 1.00 \\ 
&&40&0 & 0.2 & 1.00 & 1.00 & 1.00 & 1.00 & 1.00 \\ 
       \hline
\multirow{4}{*}{5} &\multirow{4}{*}{600} &
15 &0 & 0.2& 0.89 & 0.89 & 0.86 & 0.99 & 1.00 \\ 
&&20&0 & 0.2 & 1.00 & 1.00 & 1.00 & 1.00 & 1.00 \\ 
&&30 &0 & 0.2& 1.00 & 1.00 & 1.00 & 1.00 & 1.00 \\ 
&&40 &0 & 0.2& 1.00 & 0.99  & 1.00& 1.00 & 1.00 \\ 
     \hline
\multirow{4}{*}{5} &\multirow{4}{*}{1200} &
15 &0 & 0.2& 0.98  & 0.99 & 0.96& 1.00 & 1.00 \\ 
&&20&0 & 0.2 & 1.00 & 1.00 & 1.00 & 1.00 & 1.00 \\ 
&&30 &0 & 0.2& 1.00 & 1.00 & 1.00 & 1.00 & 1.00 \\ 
&&40 &0 & 0.2& 1.00 & 1.00 & 1.00 & 1.00 & 1.00 \\ 
   \hline
  \hline
\multirow{4}{*}{3} &\multirow{4}{*}{600} &
40 & 0 & 0.2 & 1.00 & 1.00 & 1.00 & 1.00 & 1.00 \\ 
&& 40 & 0.25 & 0.2  &1.00 & 1.00 & 1.00 & 1.00 & 1.00 \\ 
&& 40 & 0.5 & 0.2  & 1.00 & 1.00 & 1.00 & 1.00 & 1.00 \\ 
&& 40 & 1& 0.2 & 0.72 & 0.85 & 0.44 & 1.00 & 1.00 \\ 
   \hline
\multirow{4}{*}{5} &\multirow{4}{*}{600} &
40 & 0 & 0.2 & 1.00 & 1.00 & 0.98 & 1.00 & 1.00 \\ 
&& 40 & 0.25 & 0.2  & 1.00 & 1.00 & 1.00 & 1.00 & 1.00 \\ 
&& 40 & 0.5 & 0.2  & 0.86  & 0.88 & 0.78& 1.00 & 0.99 \\ 
&& 40 & 1& 0.2 & 0.05 & 0.01 & 0.10 & 0.72 & 0.05 \\
   \hline  
\multirow{4}{*}{5} &\multirow{4}{*}{1200} &
40 & 0 & 0.2 & 1.00 & 1.00 & 1.00 & 1.00 & 1.00 \\ 
&& 40 & 0.25 & 0.2  & 1.00 & 1.00 & 1.00 & 1.00 & 1.00 \\ 
&& 40 & 0.5 & 0.2  & 0.95  & 0.96 & 0.89& 1.00 & 1.00 \\ 
&& 40 & 1& 0.2 & 0.03  & 0.01 & 0.06& 0.79 & 0.07 \\  
   \hline
  \hline
\multirow{3}{*}{3} &\multirow{3}{*}{600} &
 40 & 0 & 0.1   & 1.00 & 1.00 & 0.93 & 1.00 & 1.00 \\ 
&& 40 & 0 & 0.2   & 1.00 & 1.00 & 1.00 & 1.00 & 1.00 \\ 
&& 40 & 0 & 0.5  & 0.95  & 1.00 & 0.96& 0.82 & 1.00 \\  
   \hline         
\multirow{3}{*}{5} &\multirow{3}{*}{600} &
 40 & 0 & 0.1   & 1.00  & 0.76 & 1.00& 1.00 & 1.00 \\ 
&& 40 & 0 & 0.2   & 1.00 & 0.99 & 1.00 & 1.00 & 1.00 \\ 
&& 40 & 0 & 0.5  & 0.00  & 0.05 & 0.00& 0.00 & 0.00 \\ 
   \hline
\multirow{3}{*}{5} &\multirow{3}{*}{1200} &
 40 & 0 & 0.1   & 1.00  & 0.88 & 1.00& 1.00 & 1.00 \\ 
&& 40 & 0 & 0.2   & 1.00 & 1.00 & 1.00 & 1.00 & 1.00 \\ 
&& 40 & 0 & 0.5  & 0.00 & 0.04 & 0.00 & 0.00 & 0.00 \\
   \hline
   
      \end{tabular}
   }
\end{table}

\newpage

\subsection{The impact of training proportion $p$ and replication number $N$}\label{appendix:robustness}

\begin{figure}[H]
\centering
\begin{subfigure}{.45\textwidth}
  \centering
  \includegraphics[width=\linewidth]{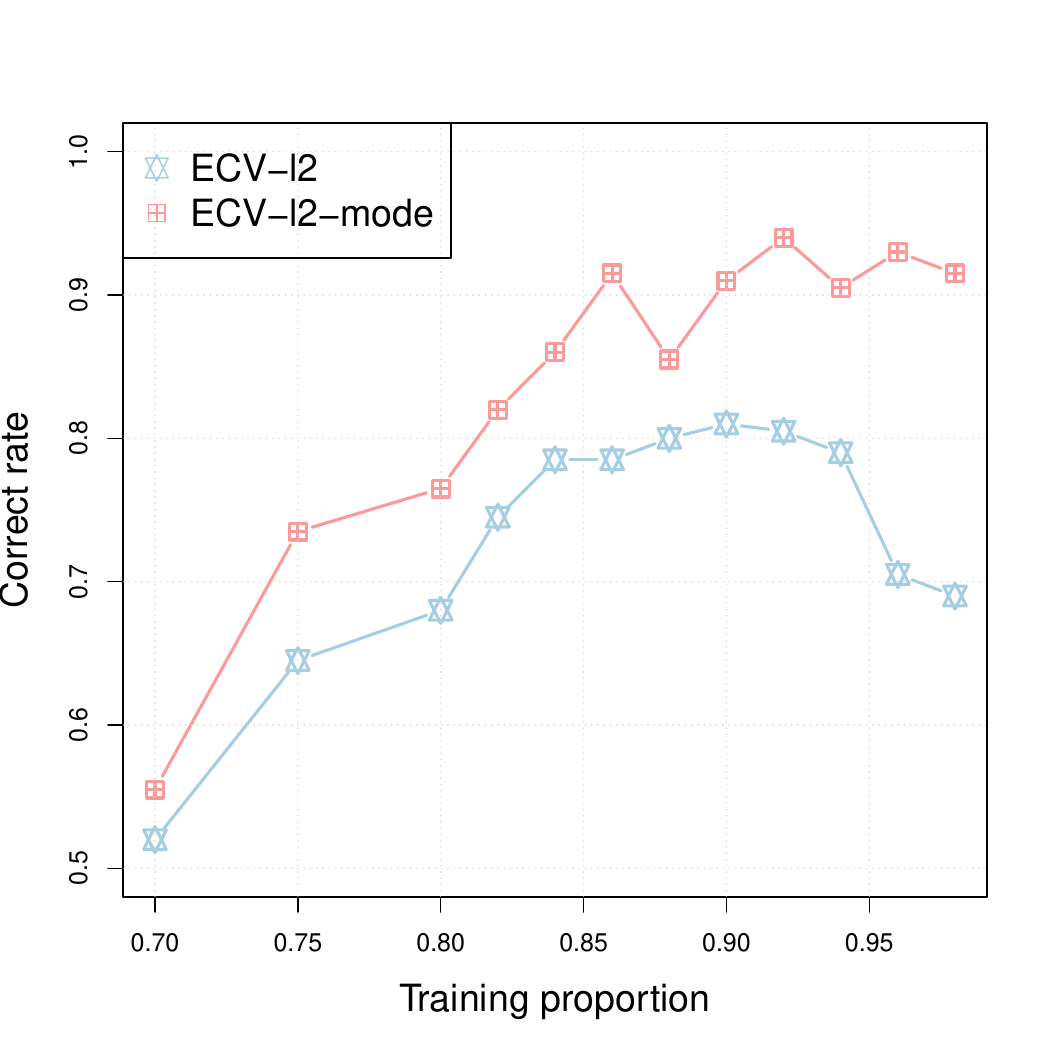}
  \caption{Varying $p$, $N=3$.}
\end{subfigure}%
\begin{subfigure}{.45\textwidth}
  \centering
  \includegraphics[width=\linewidth]{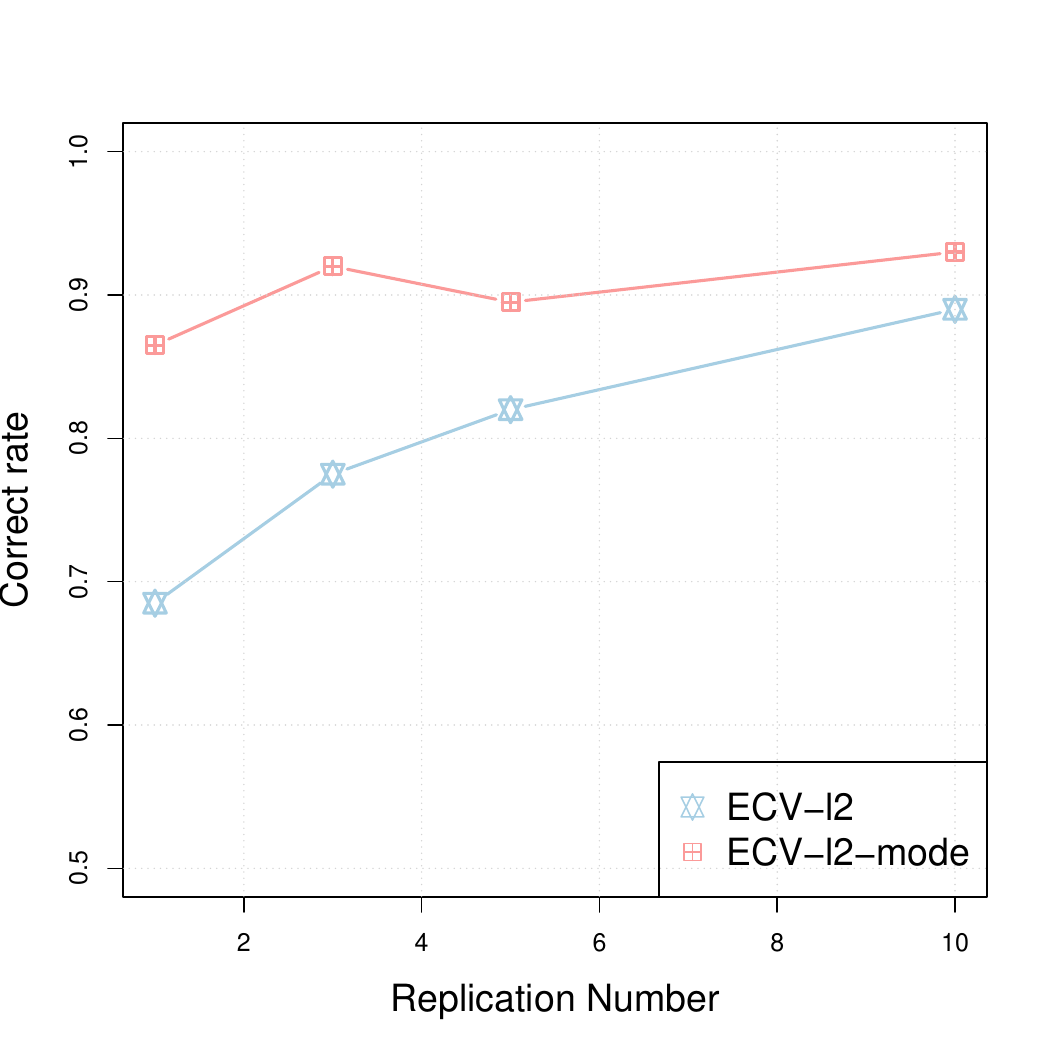}
  \caption{Varying $N$, $p=0.9$. } 
\end{subfigure}
\caption{The rate of correctly selecting between the stochastic block model and the degree corrected model as a function of $p$ and $N$. }
\label{fig:RobustnessTest}
\end{figure}

\begin{figure}[H]
\centering
\begin{subfigure}{.45\textwidth}
  \centering
  \includegraphics[width=\linewidth]{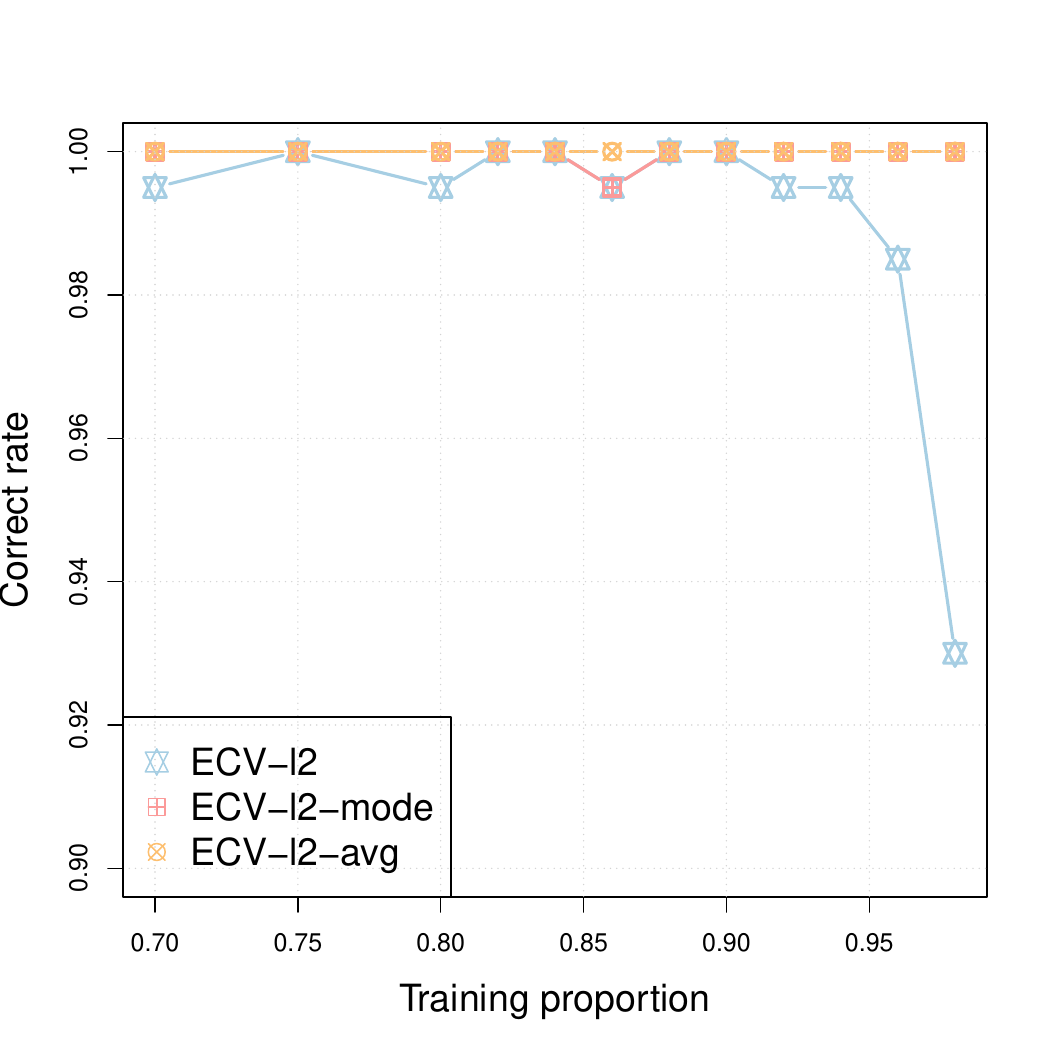}
  \caption{Varying $p$, $N=3$.}
\end{subfigure}%
\begin{subfigure}{.45\textwidth}
  \centering
  \includegraphics[width=\linewidth]{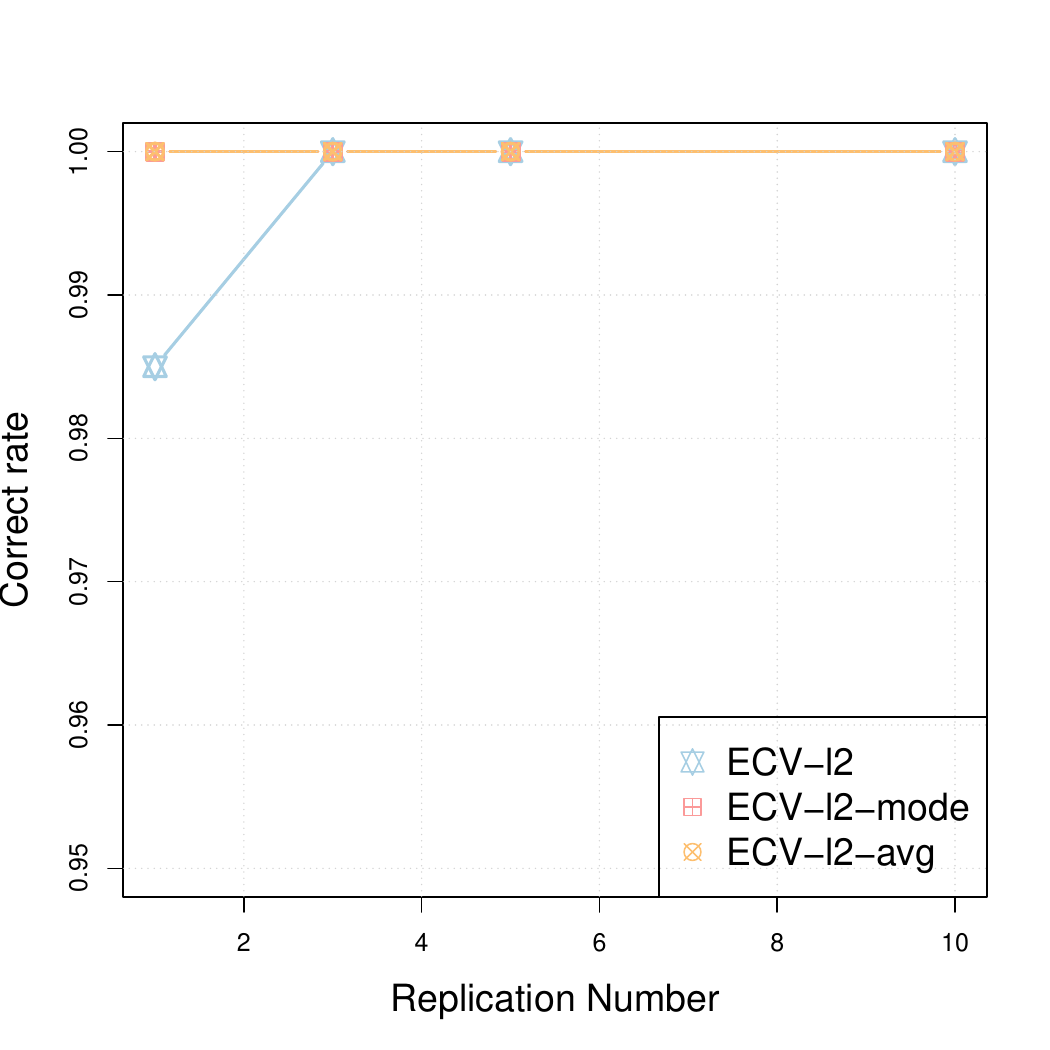}
  \caption{Varying $N$, $p=0.9$. }
\end{subfigure}
\caption{The rate of correctly selecting $K$ under the true model as a function of $p$ and $N$.}
\label{fig:Robustness-K}
\end{figure}

This simulation study illustrates the impact of $p$ and $N$ on the performance of ECV for the block model selection task in Section~\ref{secsec:sim-blockmodel}.    The true model is the degree corrected model with $K=3$ equal-sized communities, $n=600$, average degree 15,  and the out-in ratio 0.2.  The results are averaged over 200 replications. Figures~\ref{fig:RobustnessTest}~and~\ref{fig:Robustness-K} show the effects of varying $p$ and $N$ on model selection and estimation of $K$, respectively.  Clearly, a small $p$ will not produce enough data to fit the model accurately.  A very large $p$ is also not ideal since the test set will be very small.  The larger the number of replications $N$, the better in general.   The stability selection step makes our procedure much more robust to the choice of $p$ and $N$, with similar performance for $p > 0.85$ and  all values of $N$ considered. In all our examples in the paper, we use $p=0.9, N=3$.

\newpage

\subsection{Comparison of different matrix completion algorithms}\label{secsec:compareSoftImpute}

A natural question to ask about ECV is whether the choice of matrix completion algorithm affects its performance.    Here we empirically compare the simple matrix completion algorithm \eqref{eq:simpleHS} we use with a more advanced algorithm called hardImpute  \citep{mazumder2010spectral}. The softImpute algorithm intends to solve the nuclear norm regularized least squares problem,  and the hardImpute solves the rank regularized version of the same problem. They are generally believed to be more accurate on matrix completio compared to the one-step thresholding method \citep{chi2019matrix}.   However, our main interest is their effect on model selection performed by ECV.  

We compare these two methods, our standard ECV and the hardImpute version (ECV-h) on the same tasks: selecting between the block models and selecting the number of communities.  The underlying true model is the stochastic block model with $5$ communities and average degree 40, as in Table~\ref{tab:SBM-compareNCV} and Table~\ref{tab:SBM-KSelection-All}. We set the out-in ratio $\beta$ to be $0.2, 0.3, 0.4$.     Table~\ref{tab:Compare-softImpute} reports the results on model selection as well as the matrix competion error,  using the correct rank, measured in Frobenius norm, and average running time. The hardImpute algorithm is implemented in \cite{softImpute}.    Due to potential implementation differences, the timing comparison may not be fair and we only include it for reference;  it likely benefits hardImpute because it is implemented in Fortran  while the algorithm \eqref{eq:simpleHS} is implemented in R.
We see that the hardImpute algorithms does a better job on matrix completion, at the cost of a longer comuting time;  however, the results on model selection are essentially the same.  This suggests that a good enough matrix completion is sufficient for model selection.  
\begin{table}[H]
\centering
{\footnotesize
\caption{The rates of correct overall block model selection, correct number of communities selection, matrix imputation errors and running time for two versions of ECV:  ECV uses matrix completion in  \eqref{eq:simpleHS}  and  ``ECV-h" uses the hardImpute algorithm.} 
\label{tab:Compare-softImpute}
\begin{tabular}{c|rr|rr|rr|rr}
  \hline
& \multicolumn{2}{c|}{overall selection} & \multicolumn{2}{c|}{$K$ selection} & \multicolumn{2}{c|}{matrix error} & \multicolumn{2}{c}{time (sec.)}\\
\hline
$\beta$ & ECV& ECV-h & ECV& ECV-h & ECV& ECV-h & ECV & ECV-h\\
  \hline
0.2 & 1.00  &1.00 & 1.00  & 1.00  & 0.26 & 0.20 & 0.06 & 0.27 \\ 
 0.3 & 0.96  & 0.96  & 0.98  & 0.96  & 0.37 & 0.29 & 0.06 & 0.27 \\
 0.4& 0.00  & 0.00  & 0.00  & 0.00 & 0.55 & 0.44 & 0.06 & 0.27 \\
   \hline
   \end{tabular}
}
\end{table}

\newpage

\subsection{Simulation example of networks with dependent edges}

Our framework relies on the assumption that given the matrix $M$, the upper triangular entries of $A$ are independently generated. In this example, we run a small  experiment to investigate robustness of ECV to the violation of this independence assumption.    The model is based on the stochastic block model as in Table~\ref{tab:SBM-compareNCV} and Table~\ref{tab:SBM-KSelection-All}, with $K=5, n=600, \beta = 0.2, \lambda=30$. To generate the network, we first order the nodes so that the first 120 nodes are from community 1, the next 120 nodes are from community 2, and so on.
 Then we take 10 consecutive nodes as a group, resulting in 60 non-overlapping groups of 10.  Within each group, we generate the $45$ binary edges between these 10 nodes by truncating a multivariate Gaussian random vector $x \sim N(0, \Sigma)$, where $\Sigma$  has ones on the diagonal, $\rho$ on the first sub-diagonals, and zero everywhere else.  The truncation threshold is set to ensure that the marginal distribution matches matches the original block model with no dependence, but we can now control the degree of dependence between edges by varying  $\rho$.   We vary $\rho$ from 0 to 0.5 (for larger values $\Sigma$ is not positive semi-definite) and evaluate the performance of both overall model selection and estimating the number of communities, shown in Tables~\ref{tab:SBM-compareNCV-dependent} and~\ref{tab:SBM-KSelection-All-dependent}.   As $\rho$ increases, only the method  of \citet{chen2014network} with no stability selection shows some degradation in performance on overall model selection, while our method and both methods combined with stability selection stay the same.

\begin{table}[H]
\centering
{\small
\caption{Overall correct model selection rate by two cross-validation methods (out of 200 replications). The true model is the stochastic block model with dependent edges by truncating correlated multivariate Gaussian. Here we have $n=600, K=5, \beta=0.2$ and $\lambda = 30$. When $\rho = 0$, it coincides with the setting of Table~\ref{tab:SBM-compareNCV}.} 
\label{tab:SBM-compareNCV-dependent}
\begin{tabular}{c|rrrr}
 &  \multicolumn{2}{c}{Proposed Method}  &  \multicolumn{2}{c}{\citet{chen2014network}}  \\ 
\hline
$\rho$ & $L_2$ & $L_2$+stable  & $L_2$ & $L_2$+stable  \\ 
  \hline
0  & 0.99 & 1.00 & 0.98 & 1.00 \\ 
0.1& 0.99 & 1.00 & 0.98 & 1.00 \\ 
  0.2 & 1.00 & 1.00 & 0.96 & 1.00 \\ 
  0.3& 1.00 & 1.00 & 0.97 & 1.00 \\ 
  0.5  & 1.00 & 1.00 & 0.95 & 1.00 \\ 
\hline

\end{tabular}
}
\end{table}

\begin{table}[H]
\centering
{\footnotesize
\caption{The correct rate for estimating the number of communities in 200 replications.  The true model is the stochastic block model with dependent edges by truncating correlated multivariate Gaussian. Here we have $n=600, K=5, \beta=0.2$ and $\lambda = 30$. When $\rho = 0$, it coincides with the setting of Table~\ref{tab:SBM-KSelection-All}.} 
\label{tab:SBM-KSelection-All-dependent}
\begin{tabular}{c|rrrrr}
  \hline
 & \multicolumn{2}{c}{Proposed Method} & \citet{chen2014network}& \citet{wang2015likelihood}& \citet{le2015estimating}\\
\hline
$\rho$ & $L_2$ + avg.stable  & AUC+avg.stable & $L_2$+ mode.stable &  &   \\ 
  \hline
0& 1.00 & 1.00 & 1.00 & 1.00 & 1.00 \\
0.1 & 1.00 & 0.99 & 1.00 & 1.00 & 1.00 \\
 0.2 & 1.00 & 1.00 & 1.00 & 1.00 & 1.00 \\
  0.3 & 1.00 & 1.00 & 1.00 & 1.00 & 1.00 \\ 
  0.5 & 1.00 & 1.00 & 1.00 & 1.00 & 1.00 \\ 
   \hline   
      \end{tabular}
   }
\end{table}

\end{appendix}

\end{document}